\documentclass[twocolumn]{aastex63}
\usepackage{graphicx}
\usepackage{amsmath,amssymb}
\usepackage{xcolor}
\usepackage{multirow} 
\usepackage{comment}
\usepackage{bigdelim} 
\usepackage{enumerate}

\def \sn{SN\,2012au}

\usepackage{colortbl}

\newlength{\bibitemsep}
\newlength{\bibparskip}
\let\oldthebibliography\thebibliography
\renewcommand\thebibliography[1]{%
  \oldthebibliography{#1}%
  \setlength{\parskip}{\bibitemsep}%
  \setlength{\itemsep}{\bibparskip}%
}

\shorttitle{\sn{} Radio re-brightening}
\shortauthors{Eli Wiston and Friends}

\graphicspath{{./}{figures/}}

\begin{document}

\title{Old and Bright: The Remarkable Radio Brightening of the Engine-driven \sn{} Several Years After Explosion Signals the Birth of a PWN}

\author[0009-0002-4843-2913]{Eli Wiston}
\affiliation{Department of Astronomy, University of California, Berkeley, CA 94720-3411, USA}
\affiliation{Berkeley Center for Multi-messenger Research on Astrophysical Transients and Outreach (Multi-RAPTOR), University of California, Berkeley, CA 94720-3411, USA}

\author[0000-0003-4768-7586]{Raffaella Margutti}
\affiliation{Department of Astronomy, University of California, Berkeley, CA 94720-3411, USA}
\affiliation{Department of Physics, University of California, 366 Physics North MC 7300, Berkeley, CA 94720, USA}
\affiliation{Berkeley Center for Multi-messenger Research on Astrophysical Transients and Outreach (Multi-RAPTOR), University of California, Berkeley, CA 94720-3411, USA}

\author[0000-0002-8070-5400]{Nayana A. J.}
\affiliation{Department of Astronomy, University of California, Berkeley, CA 94720-3411, USA}
\affiliation{Berkeley Center for Multi-messenger Research on Astrophysical Transients and Outreach (Multi-RAPTOR), University of California, Berkeley, CA 94720-3411, USA}

\author[0000-0002-4670-7509]{Brian D. Metzger}
\affiliation{Department of Physics and Columbia Astrophysics Laboratory, Columbia University, New York, NY 10027, USA}

\author[0000-0002-5358-5642]{Kohta Murase}
\affiliation{Center for Gravitational Physics and Quantum Information, Yukawa Institute for Theoretical Physics, Kyoto, Kyoto 606-8502 Japan}
\affiliation{Department of Physics, Department of Astronomy and Astrophysics, Center for Multimessenger Astrophysics, Institute for Gravitation and the Cosmos, The Pennsylvania State University, University Park, PA 16802, USA}

\author[0000-0002-0763-3885]{Dan Milisavljevic} 
\affiliation{Department of Physics and Astronomy, Purdue University, 525 Northwestern Avenue, West Lafayette, IN 47907, USA}

\author[0000-0003-0466-3779]{Itai Sfaradi}
\affiliation{Department of Astronomy, University of California, Berkeley, CA 94720-3411, USA}
\affiliation{Berkeley Center for Multi-messenger Research on Astrophysical Transients and Outreach (Multi-RAPTOR), University of California, Berkeley, CA 94720-3411, USA}

\author[0000-0002-7706-5668]{Ryan Chornock}
\affiliation{Department of Astronomy, University of California, Berkeley, CA 94720-3411, USA}
\affiliation{Berkeley Center for Multi-messenger Research on Astrophysical Transients and Outreach (Multi-RAPTOR), University of California, Berkeley, CA 94720-3411, USA}

\author[0000-0001-5126-6237]{Deanne L. Coppejans} \affiliation{Department of Physics, University of Warwick, Gibbet Hill Road, Coventry, CV4 7AL, UK}

\author[0000-0002-7735-5796]{Joe Bright}
\affiliation{Astrophysics, Department of Physics, University of Oxford, Keble Road, Oxford, OX1 3RH, UK}

\author[0000-0002-3490-146X]{Garrett K. Keating}
\affiliation{Center for Astrophysics | Harvard \& Smithsonian, 60 Garden Street, Cambridge, MA 02138-1516, USA}

\author[0000-0003-0794-5982]{Giacomo Terreran}
\affiliation{Adler Planetarium, 1300 S DuSable Lake Shore Dr, Chicago, IL 60605, USA}

\author[0000-0002-5857-4264]{Mattias Lazda}
\affiliation{Dunlap Institute for Astronomy \& Astrophysics, University of Toronto, 50 St. George Street, Toronto, ON M5S 3H4, Canada}
\affiliation{David A. Dunlap Department of Astronomy \& Astrophysics, University of Toronto, 50 St. George Street, Toronto, ON M5S 3H4, Canada}

\author[0000-0001-7081-0082]{Maria R. Drout}
\affiliation{David A. Dunlap Department of Astronomy \& Astrophysics, University of Toronto, 50 St. George St., Toronto, ON M5S 3H4, Canada}

\author[0000-0002-3019-4577]{Michael Stroh}
\affiliation{National Radio Astronomy Observatory; 5651 Balloon Fiesta Pkwy. NE; Albuquerque, NM 87113}

\author[0000-0003-2705-4941]{Lauren Rhodes}
\affiliation{Trottier Space Institute at McGill, 3550 Rue University, Montreal, Quebec H3A 2A7, Canada}
\affiliation{Department of Physics, McGill University, 3600 Rue University, Montreal, Quebec H3A 2T8, Canada}

\author[0000-0001-8405-2649]{Ben Margalit}
\affiliation{School of Physics and Astronomy, University of Minnesota, Minneapolis, MN 55455, USA}

\author[0000-0001-8530-8941]{Jonathan Granot}
\affiliation{Department of Natural Sciences, The Open University of Israel, PO Box 808, Ra’anana 4353701, Israel} \affiliation{Astrophysics Research Center of the Open university (ARCO), The Open University of Israel, PO Box 808, Ra’anana 4353701, Israel} 
\affiliation{Department of Physics, The George Washington University, 725 21st Street NW, Washington, DC 20052, USA}

\author[0000-0002-3137-4633]{Fabio De Colle} \affiliation{Instituto de Ciencias Nucleares, Universidad Nacional Aut{\'o}noma de M{\'e}xico, A. P. 70-543 04510 D. F. Mexico}

\author[0000-0002-0592-4152]{Michael Bietenholz}
\affiliation{Department of Physics and Astronomy, York University, Toronto, M3J~1P3, Ontario, Canada}
\affiliation{Hartebeesthoek Radio Astronomy Observatory, PO Box 443, Krugersdorp, 1740, South Africa}

\author[0000-0002-6347-3089]{Daichi Tsuna}
\affiliation{Center for Astrophysics | Harvard \& Smithsonian, 60 Garden Street, Cambridge, MA 02138-1516, USA}

\author[0000-0003-2872-5153]{Samantha Wu}
\affiliation{The Observatories of the Carnegie Institution for Science, Pasadena, CA 91101, USA}
\affiliation{Center for Interdisciplinary Exploration \& Research in Astrophysics (CIERA), Physics \& Astronomy, Northwestern University,
Evanston, IL 60202, USA}

\author[0000-0003-1792-2338]{Tanmoy Laskar}
\affiliation{Department of Physics \& Astronomy, University of Utah, Salt Lake City, UT 84112, USA}

\author[0000-0002-9392-9681]{Edo Berger}
\affiliation{Center for Astrophysics | Harvard \& Smithsonian, 60 Garden Street, Cambridge, MA 02138-1516, USA}

\author[0000-0002-7507-8115]{Daniel Patnaude}
\affiliation{Center for Astrophysics | Harvard and Smithsonian, 60 Garden St, Cambridge, MA 02138}

\author[0000-0003-0528-202X]{Collin~T.~Christy}
\affiliation{Department of Astronomy and Steward Observatory, University of Arizona, 933 North Cherry Avenue, Tucson, AZ 85721-0065, USA}


\begin{abstract}
We present the results from an extensive broad-band (radio to X-rays) observing campaign of the engine-driven Type Ib \sn{} in the first 13 years of evolution. 
The early-time (${\delta}t\leq{190}$\,d) radio and X-ray evolution is well-described by conventional models of a forward shock interacting with a wind-like circumstellar medium ($\rho_{\rm{CSM}}\propto{r}^{-2}$).  However, starting at $\delta{t}\approx{6.7}$\,yr, we detect a significant radio re-brightening. This late-time emission is dominated by a luminous component  characterized by a broad and rapidly evolving spectral peak and a shallow optically thin spectral slope, $F_{\nu}\propto{\nu}^{-0.31\pm0.02}$. These properties imply a compact emitting region ($R\lesssim{10}^{16}$\,cm) expanding at a remarkably slow velocity ($\lesssim{500}$\,km/s) into a high-density environment ( $\geq{10}^4\,\rm{cm}^{-3}$), accompanied by a hard electron power-law index  $p\approx{1.6}$. No soft or hard X-ray emission is detected at any epoch, indicating that high-energy radiation is either strongly absorbed or intrinsically absent.
In the context of aspherical shock-CSM interaction models, these observations imply extreme properties of the CSM (geometry, density, total mass) that lack clear astrophysical  motivation. Instead, we show that the emergence of radiation from a newborn Pulsar Wind Nebula (PWN) naturally explains the radio spectral evolution and high-energy limits, where the emission is governed by the adiabatic expansion of a relic pair plasma. We conclude that \sn{} represents the most compelling candidate for a young, newborn PWN discovered to date, a scenario that can be directly tested with pending Very Long Baseline Interferometry (VLBI) observations.
\end{abstract}


\section{Introduction} \label{sec:intro} 
Core-collapse supernovae (CCSNe) are known to produce synchrotron emission that is observable at radio frequencies. This emission typically originates  from the shock interaction with the circumstellar medium (CSM), where high-velocity supernova (SN)  ejecta plow through dense material, built up from mass lost by the stellar progenitor before explosion  (e.g., \citealt{Chevalier1982a,Chevalier1982b,Chevalier1989,chevalier98,Weiler2002,Chevalier&Franson_2006};  see \citealt{Chevalier2017} for a review). Depending on the structure of the CSM, the radio emission in some cases may persist for long periods of time or re-brighten after many years, as observed in SN 1987A \citep{Staveley-Smith1993}. Late-emerging radio emission from CCSNe has also been theorized as possible manifestation of off-axis Gamma-Ray Burst (GRB)  jets\footnote{At the time of writing the afterglow of the neutron-star merger GW\,170817 is  the best observed example of off-axis jet emission entering the observer's line of sight at late times, e.g., \cite{Margutti21}, their Fig. 7.}  (e.g., \citealt{Rhoads_1997, Granot2002}), or the emergence of radiation powered by young Pulsar Wind Nebulae (PWNe) (e.g., \citealt{Reynolds_Chevalier1984,Slane2017}). 

Historically, the literature on late-time SN radio emission has largely been dominated by single-object studies of notable explosions, 
including e.g.,   SN\,1986J \citep{Sukumar_Allen1989,Bietenhol2008,Bietenholz2010,Bietenhol2017a,Bietenholz_2017}; SN\,1987A \citep{Staveley-Smith1993};  and SN\,1993J \citealt{Fransson1996,Fransson1998,Perez-Torres2002, Bartel_Bietenholz2007, Marti-Vidal2011,Kundu2019}). Other approaches focused on following up samples of known radio SNe, like the recent works by \cite{Bietenholz2021}, \cite{Matsuoka2025}, and \cite{Sfaradi2025}. These studies have found that detected radio SNe can typically be characterized by light curves with a monotonic rise and fall (as predicted by \citealt{chevalier98}) and that Type Ib/c and Type II SNe have distinct inferred distributions of progenitor mass-loss rates. The Very Large Array Sky Survey \citep[VLASS;][]{Lacy2020} and the Variables and Slow Transients Pilot Survey for the Australian Square Kilometre Array Pathfinder (VAST; \citealt{Murphy2013}) provided the first unbiased radio surveys of the northern and (portions of the) southern sky, respectively, in the modern era. These surveys revealed a population of radio bright SNe at late times ($\delta t \gg 1$\,yr), including SNe that were \emph{not} radio-detected at early times \citep{Stroh_2021,Rose2024}. These studies revealed a population of SNe that display significantly brighter radio emission at late times than predicted by standard models. Here we present the study of one such SN, \sn{}, which was independently re-discovered by VLASS \citep{Stroh_2021}.

On 14 March 2012, \sn{} was discovered by the Catalina Real-Time Transient Survey SNHunt project \citep{OpticalDiscovery}. Optical, near-infrared (NIR), and ultraviolet (UV) follow-up of the source revealed Type-Ib SN spectral features, alongside a luminosity in the top $\sim$10\% \citep{Drout_2011} of H-stripped CCSNe \citep{Milisavljevic2013,Takaki2013}. Modeling of the early-time ($\delta t < 60$\,d post-explosion)  UV-optical-NIR light curve indicated a large kinetic energy $E_{\rm k} \approx 10^{52}$\, erg \citep{Milisavljevic2013,Takaki2013}, significantly larger than the $\approx 10^{51}$\,erg that is expected from neutrino-driven explosions \citep{Burrows2000, Wang2022}.
This inference suggested the presence of a central engine driving the explosion. Spectroscopic observations of the event $\sim$1 yr post-explosion further revealed a number of extraordinary features: a ``plateau'' of iron emission lines, intermediate-width  ($\sim2000\,\rm{km~s^{-1}}$)
OI $\lambda$7774 emission, persistent P-Cygni absorption features,  in addition to a slow optical decay. These features, combined with the large $E_{k}$, were similar to those exhibited by SLSNe, such as SN\,2007bi, as well as engine-driven ``hypernovae'', such as SNe 1998bw and 1997dq \citep{Milisavljevic2013,Takaki2013}.  Despite this exceptional optical spectral and temporal behavior, the radio and X-ray properties of \sn{} at early times were unremarkable and similar to those of ordinary CCSNe, consistent with a shock wave expanding through the progenitor's stellar wind \citep{Kamble2014}.

Six years post-explosion, \cite{Milisavljevic2018} acquired spectroscopic observations of \sn{} using the Magellan telescope. The mere persistence of detectable optical emission at this late epoch robustly ruled out radioactive decay as the primary power source. Moreover, the spectrum had evolved significantly since $\delta t \sim 1$\, yr, transitioning to being heavily dominated by forbidden oxygen and sulfur emission lines. These features imply the presence of an ionizing source. Because the spectrum lacked narrow H Balmer lines, the common scenario of interaction with a hydrogen-rich CSM was disfavored. Consequently, \cite{Milisavljevic2018} proposed that \sn{} is powered by a pulsar or magnetar wind nebula, though interaction with an H-poor CSM remained a viable alternative. This central-engine hypothesis has been supported by semi-analytic \citep{Pandey21} and numerical \citep{Omand2023, Dessart2024} modeling, which showed that the optical data are well-reproduced by a nebula powered by a millisecond magnetar. Further supporting this interpretation are spectropolarimetric measurements tracking the supernova's evolution from early times into the nebular phase \citep{DeSoto_2025}. These measurements showed strong polarization oriented along one axis during the photospheric phase, then a new dominant axis, orthogonal to the first, appearing during the nebular phase. \citet{DeSoto_2025} suggested that this dual-axis structure was similar to that predicted for the ejecta of magnetar progenitors. 

In this work we present the analysis of multi-frequency radio (spanning from the meter-wave to the mm-wave bands), and X-ray (including soft and hard X-rays) emission of  \sn{} in the first 13 years of evolution, and we complement  this data set with one epoch of optical photometry (\S\ref{Sec:data}). For consistency, we re-model  the early-time shock dynamics and the resulting synchrotron radiation (originally presented in \citealt{Kamble2014}) in \S\ref{sec:shock_dynamics}. In \S\ref{Sec:TwoComponent}, a model-agnostic fit reveals that the radio data are best described by two distinct emitting components. To interpret these physically, we explore models of CSM interaction (\S\ref{Sec:CSMinteraction}) and PWN (\S\ref{Sec:PWN}), before concluding and discussing future prospects in \S\ref{Sec:Conclusion}. Throughout this analysis, we draw comparisons to the recent results from  very long baseline interferometry (VLBI) measurements of \sn{} from \cite{Lazda_VLBI}: while \citet{Lazda_VLBI} also favor a PWN model, they are unable to rule out the CSM interaction scenario.

We follow \cite{Milisavljevic2013} and \citet{Lazda_VLBI} and adopt 2012 March 3.5 (corresponding to MJD 55989.5) as time of explosion. Times are in the observer frame and are referred to the time of explosion unless otherwise stated. We adopt a distance of $D = 23.5 \pm 3.1$\,Mpc, based on association with the host galaxy NGC 4790 \citep{Theureau2007}.

\section{Observations}\label{Sec:data}
\begin{figure*}[t!]
\includegraphics[width=1.3\columnwidth]{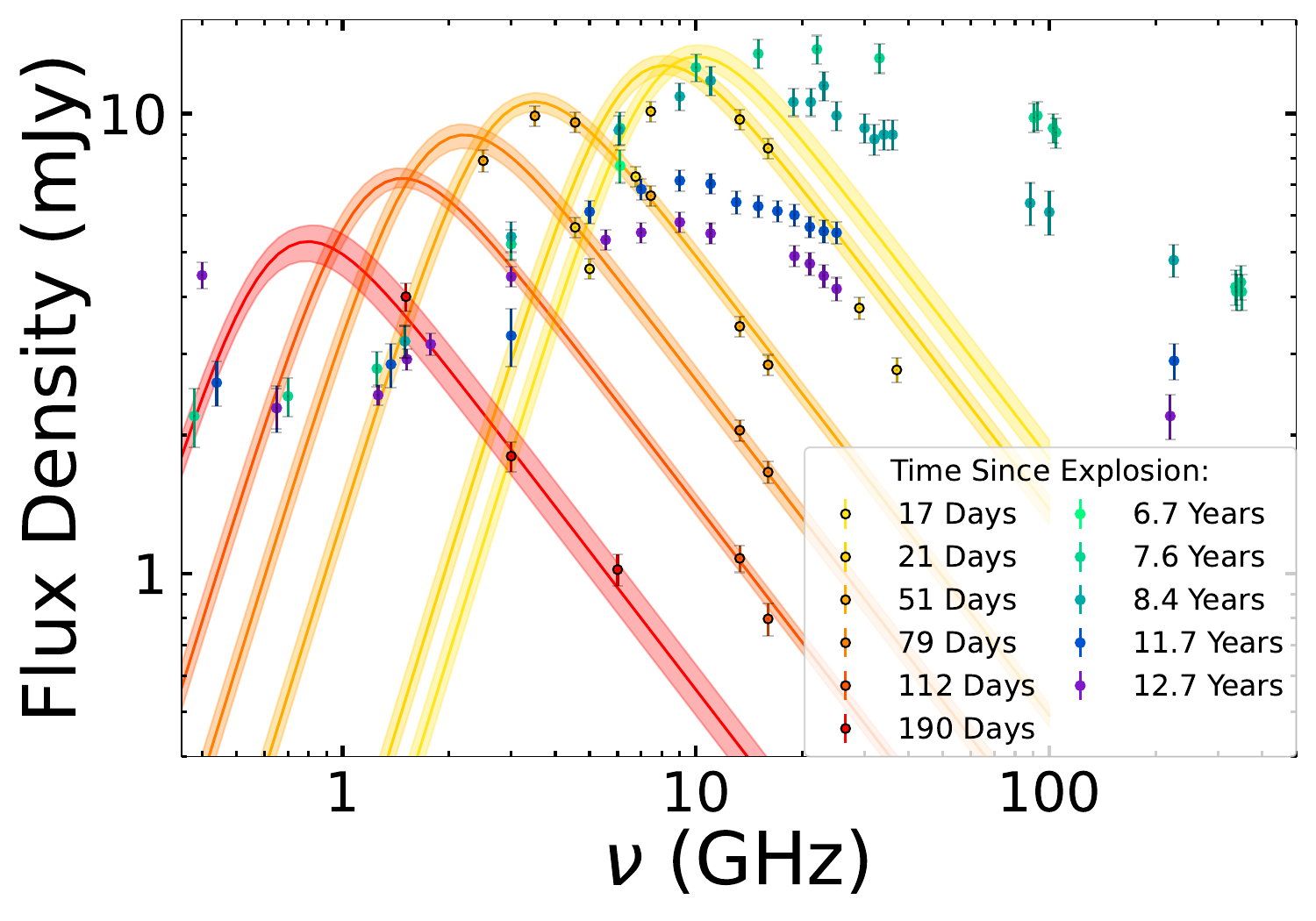}
\includegraphics[width = 0.7\columnwidth]{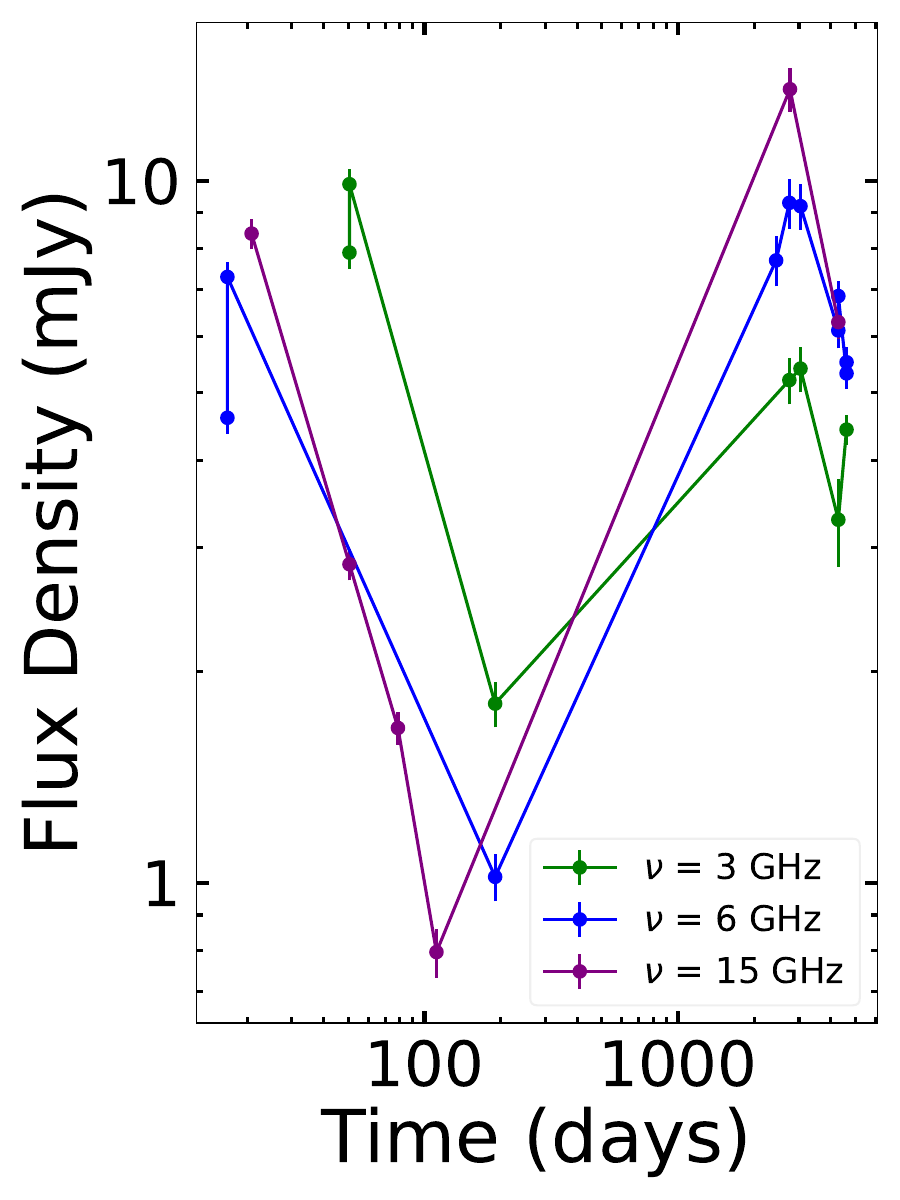}
    \caption{\emph{Left:} Complete \sn{} radio (i.e., meter-wave to  mm-wave) dataset spanning $\approx13$ years of evolution. The early-time data ($\lesssim 6$ months) originate from \cite{Kamble2014}, to which we add one unpublished epoch  at $\delta t = 190$\,d. Solid lines (yellow to red): best-fitting model of synchrotron emission  powered by the deceleration of the explosion's forward shock into a broken power law CSM-density model with a continuous transition between density indices, $s_1$ and $s_2$ from our updated model described in \S\ref{SubSec:thinshellmodel}. For each epoch, we show the 3$\sigma$ credible interval as the shaded region around the best fitting model. The peak of the radio emission at these early times clearly evolves to lower frequencies, as expected for a shock propagating through a medium with decreasing optical depth.  
    The late-time data at $\delta t= 6.7-12.7$\,yr that constitute the focus of this work notably deviate from this early-time trend, indicating the onset of a new phase of radio evolution of \sn{}, which is not a simple extrapolation of the previous behavior. 
    \emph{Right}: Light curves of \sn{} at selected well-sampled frequencies (3 GHz, 6 GHz, and 15 GHz). At early times, the \sn{} follows the typical rise and fall behavior associated with SSA. At late times, the source re-brightens, before dimming again.}
    \label{fig:all_data}
\end{figure*}

Multi-wavelength radio and X-ray observations of \sn{} with the NSF's Karl G. Jansky Very Large Array (VLA) and \emph{Swift}-XRT in the first 6 months of evolution were published by \cite{Kamble2014}. Despite many unusual optical properties, the radio emission was consistent with synchrotron radiation from electrons accelerated by an explosion shock propagating into a typical wind-like density medium (i.e., $\rho_{\rm{CSM}}\propto r^{-2}$) , indicative of a steady mass-loss rate in the final years before core-collapse. The early X-ray data were consistent with inverse-Compton emission from the same medium. We show the early-time radio data  in Fig. \ref{fig:all_data}, with one additional epoch at $\delta t = 190$\,d that was not published by \cite{Kamble2014}. 

We started an extensive multi-wavelength monitoring campaign of the late-time radio and X-ray emission from \sn{} in 2018 November covering the time interval $\delta t= 2438-4730$\,d. 
Our campaign includes observations acquired with the VLA (\S\ref{SubSec:VLA}), the Atacama Large Millimeter Array (ALMA,  \S\ref{SubSec:ALMA}), upgraded Giant Metrewave Radio Telescope (GMRT, \S\ref{SubSec:GMRT}), the Submillimeter Array (SMA, \S\ref{SubSec:SMA}), the Northern Extended Millimeter Array (NOEMA, \S\ref{SubSec:NOEMA}), the Chandra X-ray Observatory (CXO, \S\ref{SubSec:CXO}) and the Nuclear Spectroscopic Telescope Array (NuSTAR, \S\ref{SubSec:NuSTAR}). The full late-time radio dataset, with the early data for comparison, is displayed in Fig. \ref{fig:all_data}, with flux densities listed in Table \ref{Tab:radio}. Uncertainties presented in the table are the 1$\sigma$ map rms values. We add 5\% uncertainty in quadrature on all radio observations at the time of modeling to capture residual systematic flux errors.

\subsection{VLA (1--36 GHz)}\label{SubSec:VLA}
We observed \sn{} with the VLA over five epochs in November 2018, September 2019, July 2020, December 2023, and October 2024, which spanned $\delta t=2438-4621$\,d (Fig. \ref{fig:all_data}).
A log of our radio observations can be found in Table \ref{Tab:radio}. When possible, we split the observed bandwidth into sub-bands to obtain higher spectral resolution. We used both 3C286 and 3C147 as bandpass and flux-density calibrators for different epochs and performed the observations in standard phase referencing mode with complex gain calibrators J1246-0730 (L, S, C), J1239-1023 (S, C, X), J1305-1033 (C, X, Ku), J1305-1033 (Ku, K), J1256-0547 (K), and 3C279 (K, Ka). In addition to this dataset, we reduced one epoch of previously unpublished data from VLA/12A-458 (PI: Kamble) that was acquired on 9 September 2012 ($\delta t = 190$\,d).
3C286 was used as a bandpass and flux-density calibrator, and J130533.0-103319 (C-band) and J1246-0730 (L and S-band) were used for complex gain calibration.

We calibrated both datasets using the VLA pipeline in the Common Astronomy Software Applications package \citep[\textsc{casa},][]{McMullin07} using v4.5.1 for the first 4 epochs, and v6.5.4.9 for the last 2 epochs, with additional flagging. For imaging, we used the \texttt{CLEAN} algorithm in \textsc{casa} using Briggs weighting with a robust parameter of 1 for the first 4 epochs and natural weighting for the last 2 epochs. Where necessary we performed phase-only self-calibration. We used PyBDSM (Python Blob Detection and Source Measurement, \citealt{Mohan15}) to fit the source in the image plane with a Gaussian fixed to the dimensions of the CLEAN beam.

We conducted these observations in the A and D array configurations. In A configuration, the SN was always clearly separated from the host; however, in D configuration at lower frequencies, the resolution was too low to fully avoid host contamination ( $\approx 10\arcsec$ away from radio bright host emission). In these cases, we used our existing observations of the field to subtract the underlying host-galaxy flux.

\subsection{GMRT (300--1300 MHz)}\label{SubSec:GMRT}
We observed \sn{} with the GMRT over four epochs in December 2019, July 2020, December 2023, and November 2024($\delta t = 2847.0 -4643.5$\,d). 
We used 3C286 or 3C147 as flux-density and bandpass calibrators, and QSO J1246-0730, J1248-199, or 3C275 as complex gain calibrators. The measured flux densities and other details of these observations are provided in Table \ref{Tab:radio}.

We reduced and imaged the first 2 epochs of Band~3 (250-500~MHz), 4 (550-850~MHz) and 5 (1000-1400~MHz) observations in \textsc{casa}~v6.1.2 following standard procedures. In each case, we performed a few rounds of phase-only self-calibration, followed by one round of amplitude and phase self-calibration. For imaging we used Briggs weighting with a robust parameter of 1, and used two Taylor terms for the large fractional bandwidth. We measured the flux density in the image-plane using PyBDSM with an elliptical Gaussian fixed to the dimensions of the synthesized beam.

For Band~2 (120-250~MHz) data, direction dependent calibration was necessary. We divided the Band~2 data into three subbands of 25~MHz, selecting only the usable parts of the 
recorded bandwidth. These subbands are centered on 148, 204, and 229~MHz. Each subband was processed independently with the SPAM pipeline \citep{Intema2009,Intema2017} in the default mode, which includes direction-dependent calibration. The processed output visibilities of the pipeline for the three subbands were then combined to create a final combined image using WSClean \citep{Offringa2014}. 
We reduced and imaged the last two epochs of GMRT data in the Astronomical Image Processing System (AIPS) \cite{AIPS}. We do not anticipate any systematic differences between reductions carried out in AIPS compared to those calibrated in \textsc{casa}. We performed several rounds of phase-only self-calibration when needed. The imaging was carried out using a Gaussian fit with the task \texttt{JMFIT}. For all bands, the source positions in our GMRT and VLA images are consistent.

\subsection{ALMA (90--350 GHz)}\label{SubSec:ALMA}
We observed \sn{} with ALMA in October 2019 ($\delta t = 2796$\,d). The observations were taken at 104.5 GHz (Band 3) and 350 GHz (Band 7). The data were calibrated using the ALMA pipeline in \textsc{casa} 5.6.1-8. For imaging, the data were divided into 8 sub-bands, each of bandwith 2 GHz with central frequencies of 90.5, 92.4, 102.5, 104.5, 336.5, 338.4, 348.5 and 350.5 GHz. We imaged the data using the \texttt{tclean} algorithm in \textsc{casa} v4.5.1, using Briggs weighting with a robust parameter of 1. For Band 3 we did phase-only self-calibration. Similarly to the VLA data, we fitted the source in the image plane with a Gaussian fixed to the dimensions of the clean beam with PyBDSM.
The flux densities and observation details are shown in Table \ref{Tab:radio}.

\subsection{SMA (210--240 GHz)}\label{SubSec:SMA}
\sn{} was observed with the Submillimeter Array (SMA) 
on 2020 July 14 ($\delta t = 3055$\,d), 2023 November 16 ($\delta t = 4275$\,d), and 2025 February 14 ($\delta t = 4730$\,d) under the 
 SMA Large Program, POETS (Pursuit of Extragalactic Transients with the SMA).
The data were reduced and calibrated via two different paths which produced equivalent results, a standard user-driven calibration using the MIR calibration suite (\url{https://www.cfa.harvard.edu/~cqi/mircook.html}) and a new pipeline calibration process called COMPASS. Complex gains versus time were calibrated using the nearby blazar 3C279, which also provided a strong continuum source for bandpass calibration. The flux density scale was set using observations of Callisto and Titan, using the brightness temperature models described in NRAO Memo 594\footnote{Butler, B.J.; https://library.nrao.edu/public/memos/alma/memo594.pdf}. The data were spectrally averaged to provide a mean continuum data set for further analysis.  The source flux density was measured using both visibility fitting (assuming a point source) and aperture fitting after imaging within AIPS and a proprietary imaging routine; all results were consistent with regard to position and flux density. The flux densities and observation details are shown in Table \ref{Tab:radio}.

\subsection{NOEMA (88--100 GHz)}\label{SubSec:NOEMA}
We observed the field of \sn{} for a single epoch on 2020 November 17 ($\delta t=$3181\,d) with the NOEMA interferometer.
The observations were carried out between UT1 and UT2 at a central frequency of 94 GHz. We reduced and measured flux density values for the lower side band (LSB), which has a central frequency of $86$ GHz and the upper side band (USB), which has a central frequency of $100$ GHz.
We calibrated the data using \textsc{clic} within the \textsc{gildas}\footnote{\url{https://www.iram.fr/IRAMFR/GILDAS}} software package to perform bandpass, amplitude and phase calibration. We measured the flux density of the target by performing a fit in the \textit{uv}-plane. With such high signal-to-noise measurements, we report flux density measurements in both sub-bands in Table \ref{Tab:radio}.

\subsection{CXO (0.3--10 keV)}\label{SubSec:CXO}
A first deep epoch of CXO observations was obtained on  2018 August 2  ($\delta t=$2343\,d; observation ID 21660; PI Patnaude) and published in \cite{Milisavljevic2018}. We acquired two additional epochs of CXO ACIS-S observations covering the time period $\delta t=$3060--3677\,d. A log of the CXO observations can be found in Table \ref{Tab:Xraydata}. We analyzed the ACIS-S data following standard practice with \texttt{CIAO v4.15} and corresponding calibration files. We re-analyzed the first epoch of CXO data for consistency. No X-ray source is detected at the location of the SN in any epoch, either blindly or with a targeted detection algorithm. Assuming Poissonian statistics, we inferred the $3\,\sigma$ count-rate limits reported in Table \ref{Tab:Xraydata}. We converted the count-rate limits into flux limits adopting an absorbed power-law spectrum with  photon index $\Gamma=2$ and a Galactic neutral hydrogen absorption column in the direction of the transient of   $\rm{NH_{MW}}=3.7\times 10^{20}\,\rm{cm^{-2}}$ \citep{HI4PI}. For each observation, a count-to-flux factor was derived by accounting for the instrumental response at the location of the transient. 
\subsection{NuSTAR (3--80 keV)}\label{SubSec:NuSTAR}
We obtained one epoch of hard X-ray observations with NuSTAR in coordination with the CXO at $\delta t\approx$3035\,d (Obs ID 90601521002; PI Margutti). NuSTAR data were processed and analyzed with the \texttt{nupipeline} within the the NuSTAR Data Analysis Software (v1.9.7) and corresponding calibration files. We removed the time intervals affected by high background with custom scripts. No source is detected at the location of the SN. We report in Table \ref{Tab:Xraydata} the inferred count-rate limits in the 8--20 keV energy band (which corresponds to the peak of the NuSTAR effective area), combining the signal from both NuSTAR modules. As for the CXO observations, we assume a  power-law spectrum with photon index $\Gamma=2$ for the flux calibration. 

\subsection{Optical photometry}\label{SubSec:NIRphot}
We obtained optical imaging of \sn{} on 2020 June 23 ($\delta t = 3034$\,d, using the Low Resolution Imaging Spectrometer \citep[LRIS;][]{Oke1995} mounted on the 10~m Keck I telescope (PI Terreran, Program 2020A\_O285). We used \textit{BgRI}-bands, exposing for 720~s per band. We reduced the images using standard overscan, bias, and flatfielding within \textsc{iraf}.\footnote{\textsc{IRAF} is distributed by the National Optical Astronomy Observatory, which is operated by the Association of Universities for Research in Astronomy (AURA) under a cooperative agreement with the National Science Foundation. http://iraf.noao.edu/} We then extracted the photometry data using the \textsc{SNOoPY}\footnote{Cappellaro, E. (2014). \textsc{SNOoPY}: a package for SN photometry, \url{http://sngroup.oapd.inaf.it/snoopy.html}} package. We performed point-spread function photometry with \textsc{DAOPHOT} \citep{Stetson1987}. We estimated the zero-points and color term based on the magnitudes of field stars present in the Sloan Digital Sky Survey\footnote{\url{http://www.sdss.org}} \citep[SDSS;][]{York2000} catalog (DR9). We converted the SDSS \textit{ugriz} magnitudes to Johnson/Cousins \textit{UBVRI} filters following \cite{Chonis2008}. We quantified the uncertainty on the instrumental magnitude injecting artificial stars \citep[e.g.,][]{Hu2011}. The resulting uncertainty was then added in quadrature to the fit uncertainties returned by \textsc{DAOPHOT} and the uncertainties from the photometric calibration to obtain the total uncertainty on the photometry. We measure  $23.0\pm0.2$~mag, $22.0\pm0.2$~mag, $21.6\pm0.1$~mag and $21.2\pm0.1$~mag for \textit{B}, \textit{g}, \textit{R} and \textit{I}, respectively (reported in Vega magnitudes).

\section{Shock Dynamics and Deviation from a Pure Wind-like Density Model}
\label{sec:shock_dynamics}

\subsection{Radio emission from SNe in the ``thin-shell'' approximation}\label{SubSec:thinshellmodel}

\subsubsection{Dynamics}  \label{subsub:thinshelldynamics}
At early times ($\delta t \leq$190\,d), the radio spectral energy distributions (SEDs) follow a fairly standard evolution, peaking at successively lower frequencies over time (Fig.\ref{fig:all_data}). \cite{Kamble2014} showed that the radio emission in this time range is well explained by synchrotron radiation from electrons accelerated at the forward shock (FS) expanding in a wind-like circumstellar medium (CSM) with density $\rho_{\rm{CSM}}\propto r^{-2}$, as expected from massive stars that have been losing mass at a constant $\dot M/v_{w}$ (where $\dot M$ is the mass-loss rate, and $v_w$ is the wind velocity).
In the following we generalize the approach of \cite{Kamble2014}. 

We first solve for the SN shock dynamics in a given CSM, and then predict the resulting synchrotron emission from the FS. In the ``thin-shell'' approximation (e.g., \citealt{Chevalier1982a}), the conservation of momentum of the shocked shell of material of negligible thickness can be written as: 
\begin{equation}
\label{eq:thin_shell}
    M_{\rm s} \frac{dv_{\rm s}}{dt} = 4\pi R_{\rm s}^2[\rho_{\textrm{ej}}(v_{\textrm{ej}} - v_{\rm s})^2 - \rho_{\textrm{CSM}}v_{\rm s}^2]
\end{equation}
(e.g., \citealt{Chevalier2017}), where $M_{\rm s}$ is the total mass of the thin shell comprising both shocked ejecta and shocked CSM at the shock radius $R_{\rm s}$, propagating at the shock velocity $v_{\rm s} = dR_{\rm s}/dt$; $v_{\textrm{ej}}$ is the ejecta velocity at the reverse shock; $\rho_{\textrm{ej}}$ and $\rho_{\textrm{CSM}}$ are the density profiles of the outer layers of the star and the CSM, respectively. We assume that the velocity of the CSM, $v_w$, is much smaller than $v_s$, such that $v_s -v_w \approx v_s$.

\cite{Chevalier_1987A} derive a density profile for SN ejecta in homologous expansion. The expansion of the stellar envelope and corresponding rarefaction wave is modeled with a combination of self-similar solutions and numerical computation. The resulting profile comprises an outer steep density power law, and an inner flat density profile, with a transition velocity $v_t$ marking the boundary between the two regimes:  
\begin{equation}
\rho_{\textrm{ej}} = 
\left\{
    \begin{array}{lr}
    \label{eq:rho_ej_norm}
        A(t-t_0)^{-3}, &  v_{\textrm{ej}} < v_t\\
        A(t-t_0)^{-3}\left(v_{\textrm{ej}}/v_t\right)^{-n}, & v_{\textrm{ej}} \geq v_t
    \end{array}
\right.
\end{equation}
where:
\begin{equation}
\label{eq:rho_ej_norm2}
    A \equiv \frac{1}{4\pi n} \frac{[3(n-3)M_{\textrm{ej}}]^{5/2}}{[10(n-2)E_{\rm{k}}]^{3/2}},
\end{equation}
\begin{equation}
\label{eq:v_t}
    v_t \equiv \biggl[\frac{10(n-5)E_{\rm{k}}}{3(n-3)M_{\textrm{ej}}}\biggl]^{1/2},
\end{equation}
and $M_\textrm{ej}$ and $E_{\rm{k}}$ are the total explosion ejecta mass and kinetic energy, respectively.  The outer ejecta density index, $n$, varies based on stellar progenitor. For  Wolf-Rayet stars that are possible progenitors of H-stripped  CCSNe, values of $n \approx 10$ are commonly used in the literature (\citealt{Chevalier&Franson_2006}, \citealt{Liu_Tauris_2015}), based on analytic approximations of the simulations from \cite{Matzner_McKee_1999}. Recent results by \cite{Matsuoka2025} that are derived from fitting radio SNe potentially indicate that lower values of $n \approx5$ are possible and  maybe even common. However, in our modeling $n$ is degenerate with other parameters, and  assumptions have to be  made (see below).  We  adopt $n=10$ in what follows, and we note that none of our major conclusions depends on this assumption.

Other radio-SN studies (e.g., \citealt{Matsuoka2025, Soria25} for recent examples) also used the thin-shell approximation to model the dynamics of SN ejecta interacting with a dense CSM. In these studies, the \cite{Chevalier1982a} self-similar solutions for a single power law outer ejecta profile ($\rho_{\rm{ej}} \propto r^{-n}$) interacting with a single power law CSM profile ($\rho_{\rm{CSM}} = Ar^{-s}$) are used. The solution for the temporal evolution of the shock radius is: 
\begin{equation}
\label{Eq:Rsselfsimilar}
    R_s(t) = \Bigg[\frac{(3-s)(4-s)}{(n-3)(n-4)}\frac{U_c^n}{A}   \Bigg]^{\frac{1}{n-s}} t^{\frac{n-3}{n-s}},
\end{equation}
where
\begin{equation}
    U_c \equiv \Big(\frac{3M_{\rm{ej}}}{4\pi v_t^{3-n}} \frac{n-3}{n}\Big)^{1/n}.
\end{equation}
\noindent 
This solution is valid for $s<3$ and $n>5$. Scenarios with $s>3$ in the thin-shell approximation would result in the unphysical situation of an accelerating contact discontinuity with non-accelerating ejecta (a self-similar treatment of this problem can be found in \citealt{Waxman1993}). Similarly, scenarios with $n<5$ would result in a separation between the contact discontinuity and the outer shock wave \citep{Chevalier1982a}. Within the regime of validity ($s<3$ and $n>5$), the solution of Eq. \ref{Eq:Rsselfsimilar} only applies when the two interacting density profiles are power laws. For the present study, we allow for the possibility of more complex CSM profiles by forgoing the analytic solution and instead integrating numerically Eq. \ref{eq:thin_shell} --- similar to the treatment of the X-ray analysis in \cite{Ibik25}. 
Given the explosion parameters and density profiles for the ejecta and CSM, we can numerically solve Equation \ref{eq:thin_shell} to compute the shock radius $R_s(t)$ and velocity $v_s(t)$ evolution. 

\subsubsection{Synchrotron Radiation} \label{subsub:SynchRadiation}
Following \cite{Chevalier&Franson_2006}, we assume that a fraction $\epsilon_B$ of the post-shock internal energy density goes into the magnetic energy density: 
\begin{equation}
\label{eq:B_field}
    \frac{B^2}{8\pi}=\epsilon_{B}\rho_{\rm CSM} v_{\rm s}^2,
\end{equation}
where $\rho_{\rm CSM}$ is the \emph{unshocked} CSM density. An additional parameter that will be relevant is $\epsilon_e$, the fraction of the post-shock energy density that goes into the energy density of relativistic electrons: 
\begin{equation}
    \int_{\gamma_{\rm{min}}}^{\gamma_{\rm{max}}} n(E)EdE=\epsilon_e\rho_{\rm{CSM}}v_{\rm s}^2,
\end{equation}
where $N(E)$ is the number density of electrons per unit electron energy $E$ above a minimum electron  Lorentz factor $\gamma_{\rm{min}} =1$, as in \citep{chevalier98}. We will assume $\gamma_{\rm{max}}=\infty$ unless otherwise specified. We define a system in equipartition when $\epsilon_B = \epsilon_e$, i.e.,  when relativistic electrons and magnetic field carry equal fractions of post-shock energy density.

The combination of Eq.\,\ref{eq:thin_shell}--\ref{eq:v_t} and Eq.\,\ref{eq:B_field} thus give us the $R_s(t)$ and $B(t)$ evolution for a given $\rho_{\rm{CSM}}(r)$ and explosion parameters $E_{\rm{k}}$ and $M_{\rm{ej}}$, from which the synchrotron spectrum can be computed, given values for $\epsilon_B$ and $\epsilon_e$ and the distribution of accelerated electron energies.
As the FS propagates through the CSM, electrons are accelerated at the shock front. Following the formalism of \cite{chevalier98}, we consider a power-law distribution of accelerated electron energies, $N(E) = n_0E^{-p}$, 
as supported by  Particle In Cell (PIC) simulations (e.g., \citealt{Caprioli23} and references therein for a recent review). The spectral shape of the resulting emission will be determined by the relative values of $\nu_{m}$, the characteristic synchrotron frequency, $\nu{_{sa}}$, the synchrotron self-absorption (SSA) frequency, and $\nu_{c}$, the cooling frequency. Different orderings of these break frequencies will result in different slopes and different locations of the spectral peak flux \citep{Granot&Sari}. In order to model all these possible combinations of frequency orderings, we apply the methods described in Appendix C of \cite{Itai_tvd}, self-consistently calculating the  break frequencies and their corresponding SEDs. 

We find that the observed synchrotron emission of \sn{}  consistently falls into the regime of $\nu_{m}<\nu{_{sa}}<\nu_{c}$. In this regime, the flux density at the asymptotic intersection of the optically thin $F_{\nu}\propto \nu^{-(p-1)/2}$ and optically thick $F_{\nu}\propto \nu^{5/2}$ power-law segment is \citep{chevalier98}:
\begin{equation}
\label{eq:F_brk}
    F_{\rm{\nu,brk}} = \frac{\pi R_s^2}{D^2} \frac{c_5}{c_6} (B\,\rm{sin}\theta)^{-1/2} \Big(\frac{\nu_{\rm{brk}}}{2c_1}\Big)^{5/2}
\end{equation}
and the spectral peak occurs at $\nu\approx \nu_{brk}$:
\begin{align}
\label{eq:nu_brk}
    \nu_{\rm{brk}} = 2c_1\Big(R_s f\Big(\frac{\epsilon_e}{\epsilon_B}\Big)\frac{p-2}{6\pi}\Big)^\frac{2}{p+4} \Big(\frac{m_e}{c^2}\Big)^{\frac{2(p-2)}{p+4}}B^\frac{p+6}{p+4}\nonumber \\ c_6^\frac{2}{p+4} (\rm{sin}
    \theta)^\frac{p+2}{p+4},
\end{align}
\noindent
Above, $c_1$, $c_5(p)$, and $c_6(p)$ are functions of physical constants and parameter $p$ defined by \cite{Pacholczyk1970} and fully described in Appendix \ref{app:constants}. $f_V$ is the volumetric emission filling factor, defined as in \cite{chevalier98} such that the total volume of the synchrotron emitting electrons is: 
\begin{equation}
\label{eq:equivalent volume}
    V \equiv \frac{4\pi}{3}R_s^3f,
\end{equation}
and $\theta$ is the pitch angle of accelerated electrons. We note that the alternative choice of an isotropic distribution of pitch angles (as in \citealt{Chevalier2017}) would have no impact on our major conclusions.  Following \cite{chevalier98}, we assume $\rm{sin}(\theta) =1$.
For a discussion of anisotropic pitch angle distributions, see \cite{pitch_angles}. 
\subsection{Application to SN 2012au}
\label{SubSec:12ausingle}
\subsubsection{\texorpdfstring{Early-time radio emission at $\delta t<190$\,d}{Early-time radio emission at delta t < 190 d}}
\label{Subsub:12auearlytime}

We approximate the resulting synchrotron SEDs using a smoothed broken power law model of the form:
\begin{equation}
\label{eq:broken_PL}
    F_\nu = F_{\nu,{\rm brk}} \left[ \left(\frac{\nu}{\nu_{\textrm{brk}}}\right)^{-w(1-p)/2} + \left(\frac{\nu}{\nu_{\textrm{brk}}}\right)^{-w\alpha}\right]^{-1/w},
\end{equation}
where $F_{\nu,{\rm brk}}$ and $\nu_{\rm{brk}}$ are related to the shock properties via Eq. \ref{eq:F_brk} and \ref{eq:nu_brk}, respectively;  $p$ is the power law index for the distribution of emitting electron energies, $N(E) \propto E^{-p}$, $\alpha$ is the optically thick spectral index ($F_{\nu} \propto \nu^{\alpha}$), and $w$ is a smoothing parameter for the transition between optically thick and optically thin power laws. $F_{\nu,{\rm brk}}$, the flux at the intersection of the power laws is related to the observed peak flux density in the light curve by the relation: $F_{\nu,{\rm pk}} = 2^{-1/w}F_{\nu,{\rm brk}}$. Following \cite{Granot&Sari}, when the spectra fall in the $\nu_{m}<\nu{_{sa}}<\nu_{c}$ regime that is relevant to our analysis below, we can parameterize $w$ as a function 
of $p$:
\begin{equation}
\label{eq:smooth}
    w = 1.47 - 0.21p.
\end{equation}

Next, Eq. \ref{eq:thin_shell} depends on the total kinetic energy $E_{\rm{k}}$ and ejecta mass $M_{\rm{ej}}$ of the explosion. Ideally, one would either have these parameters as free in the radio modeling, or use independent estimates of these parameters from other wavelengths.
For example, estimates of $M_{\rm{ej}}$ and $E_{\rm{k}}$ can be obtained from the UV-optical-NIR emission. 
\cite{Milisavljevic2013} model a ``pseudo-bolometric'' (i.e., UV+optical) light curve of \sn\, and estimate $M_{\rm{ej}} \approx (3-5)\,\rm M_\odot$ and $E_{\rm{k}} \approx 1\times10^{52} $ erg. \cite{Takaki2013}  use spectroscopic measurements to estimate an expansion velocity $v$, and a rise-time $t_r$. Applying scaling relationships of these two parameters using  another Type Ib SN\,2008D as a template, \cite{Takaki2013}  estimate  $M_{\rm{ej}} = (5-7)\, \rm{M_\odot}$ and  $E_{\rm{k}} = (7-18)\times10^{51} $ erg. \cite{Pandey21} estimate $M_{\rm{ej}} = (4.7-8.5)\,\rm{M_\odot}$ and $E_{\rm{k}} =(4.8-5.4)\times10^{51} $\,erg from the modeling of the bolometric light curve (where the NIR has been estimated based on the extrapolation of the observed optical emission).

In the radio modeling $E_{\rm{k}}$ and $M_{\rm{ej}}$ are highly degenerate. Both parameters affect the initial ejecta velocity and the transition velocity in the ejecta density profile, $v_t$. At early times, the outermost ejecta, that produce synchrotron emission, are expected to always be moving faster than $v_t$, and therefore, for the purpose of the early radio emission, $M_{\rm{ej}}$ and $E_{\rm{k}}$ only impact the initial velocity of the ejecta. Mathematically, $M_{\rm{ej}}$ and $E_{\rm{k}}$ only enter the calculation through Eq. \ref{eq:rho_ej_norm} and \ref{eq:rho_ej_norm2} as a ratio that determines the normalization of $\rho_{\rm{ej}}\propto \frac{M_{\rm{ej}}^5/2}{E_{\rm{k}^{3/2}}}$. We will thus fix one of these parameters ($M_{\rm{ej}}$) and fit for the other ($E_{\rm{k}}$). We assume  $M_{\rm{ej}} = 5 M_\odot$ to remain within the inferred mass ranges of all three optical studies above.

For the early-time ($\delta t \leq 190$\,d) radio data,
we further assume an ejecta density index of $n = 10$ and
microphysical parameters of $\epsilon_e = \epsilon_B = 0.3$, to allow a direct comparison with the results from \cite{Kamble2014}. We find that there is large variance in the optically thin slope across epochs --- corresponding to values of $p = 2.9-3.5$. Given that the value of $p$ is poorly determined, we choose to assume $p=3$, the standard value in synchrotron modeling of SNe \citep{chevalier98,Chevalier&Franson_2006}.

The free parameters of our model
characterize the CSM density profile, $\rho_{\rm{CSM,0}}(r)$.
We first assumed a single power-law density profile, but failed to adequately fit the data. In particular, the last early-time epoch at $\delta t= 190$\,d was consistently and significantly over-predicted by the model. We thus allow for an extra degree of freedom, and adopt a broken power-law CSM density profile:
\begin{equation}\label{eq:rhoCSMBPL}
\rho_{\textrm{CSM,0}}(r) = 
\left\{
    \begin{array}{lr}
        \rho_{\rm 0} \left(\frac{r}{R_{\rm{brk}}}\right)^{-s_1}, &  r < R_{\rm brk}\\
        \rho_{\rm 0} \left(\frac{r}{R_{\rm brk}}\right)^{-s_2}, &  r > R_{\rm brk}
    \end{array}
\right.
\end{equation}
with free parameters: $\rho_0$, the normalization on the density, $R_{\rm brk}$, a break radius where the density index may change, and $s_1$ and $s_2$, the density indices at radii smaller and larger than $R_{\rm brk}$, respectively. We use a Markov Chain Monte Carlo (MCMC) fit with the \textsc{emcee} Python package \citep{Foreman_Mackey_2013}. We initialize 25 walkers which each iterate the model calculations 1000 times. We discard the first 200 iterations for each walker to only consider samples after the burn-in phase. We report the results of convergence tests $\hat R$ and ESS statistics in Appendix \ref{app:early_corner}. We iterate over the parameters $\rho_0$, $E_{\rm{k}}$, and $R_{\rm{brk}}$ in logarithmic space, while we iterate over the parameters $s_1$ and $s_2$ in linear space. The initial parameter values and 
priors are summarized in Table \ref{tab:early_tab}. The prior on $R_{\rm{brk}}$ is informed by our single power-law CSM density fit above. The corner plot of the resulting MCMC fit can be found in Appendix \ref{app:early_corner}, Fig. \ref{fig:early_shock_fit}, and the best-fitting parameters are reported in Table \ref{tab:early_tab}.
\begin{deluxetable}{c|ccc}
\tablecaption{Early radio data ($\delta t<190$\,d) best-fitting parameters. Equipartition is assumed. \label{tab:early_tab}}
\tablehead{
    \colhead{Parameter} & \colhead{Priors}  & \colhead{Best Fit Values} 
}
\startdata \label{tab:early_best_fit}
$\log_{10}\big ( \frac{E_{\rm{k}}}{\rm{erg}}\big )$ & [50, 53] & $52.54\pm{0.01}$\\
$\log_{10}\big ( \frac{\rho_0}{\rm{g\,cm^{-3}}}\big )$ & [-15, 23]  & $-22.11\pm0.05$\\
$\log_{10}\big (\frac{R_{\rm{brk}}}{\rm{cm}}\big )$ & [16.5, 17]  & $16.76\pm{+0.02}$\\
$s_1$ & [1, 4]  & $1.85\pm0.02$\\
$s_2$ & [1, 4]  & $2.25\pm{0.03}$\\
\enddata
\end{deluxetable}

As seen in Figure \ref{fig:all_data}, the early-time radio data ($\delta t \leq 190$\,d) are well reproduced by $\log_{10}\big ( \frac{E_{\rm{k}}}{\rm{erg}}\big ) = 52.54\pm0.01$, with a broken power law CSM density profile. 
The inferred explosion energy is a factor of a few higher than inferred by \cite{Milisavljevic2013}, \cite{Takaki2013}, and \cite{Pandey21}, which is not surprising given the orthogonal approaches, using completely different datasets. Our modeling results in a higher initial shock velocity than the radio analysis of \cite{Kamble2014}. At $\delta t_0 =$ 25 d (reference time of the analysis of \citealt{Kamble2014}), our shock radius is $R_s(\delta t_0) = 2.5\pm0.1\times10^{16}$\,cm, corresponding to an average velocity of $\approx 0.38$c, compared to the velocity of $\approx 0.2$c from \cite{Kamble2014}. This discrepancy arises from the fact that our modeling self-consistently accounts for shock deceleration from CSM interaction, while \cite{Kamble2014} assumes a constant shock velocity.

At $r<R_{\rm{brk}}$, where $\log_{10}\big (\frac{R_{\rm brk}}{\rm cm}\big)=16.76\pm0.02$,
we find  $\rho_{\rm{CSM,0}} \propto r^{-1.85\pm0.02}$. To match the emission at $\delta t = 190$\,d, which was not included in \cite{Kamble2014}, the CSM density profile steepens to   $\rho_{\rm{CSM,0}} \propto r^{-2.25\pm{0.03}}$ at $r>R_{\rm{brk}}$. 
Our results are reasonably close to a wind-like density profile   ($\rho_{\rm{CSM,0}}\propto r^{-2}$) that was \emph{assumed} by \cite{Kamble2014}, but indicate some important deviations. With this caveat in mind, our \emph{effective} mass-loss rates range between $(4-5)\times10^{-6}\,\rm{M_{\odot}\,yr^{-1}}$ over the period $\delta t =17-190$\,d, assuming $v_w =1000\,\rm{km\,s^{-1}}$ (typical of Wolf-Rayet progenitor stars).
\cite{Kamble2014} reported a slightly lower constant mass-loss rate 
$\dot M = 3.6\times10^{-6} \rm{M_\odot\,yr^{-1}}$. 
Despite indicating higher mass-loss rates, the faster shock velocity of our best-fitting model leads to lower densities at a given time vs. \cite{Kamble2014}. For example, our best-fitting electron number density at $t_0$,
$n_e(t_0) \approx 200\,\rm{ cm}^{-3}$, compared with $n_e(t_0) \approx 900\, \rm{ cm}^{-3}$ of \cite{Kamble2014}.\footnote{Electron number densities are derived adopting a mean molecular weight $\mu_e=1$ for the CSM, that is typically assumed in the literature.} The combination of lower density but larger radii at a given time results in similar inferred $B$ fields (Eq. \ref{eq:B_field}): $B(t_0) = 0.49\pm0.02$\,G for our analysis, roughly consistent with the reported $B(t_0) = 0.54$\,G 
from \cite{Kamble2014}, all assuming equipartition.

\subsubsection{\texorpdfstring{Late-time radio emission at $\delta t>190$\,d}{Late-time radio emission at delta t>190 d}} 
\label{subsec:12aulatetime}

\begin{figure*}[t!]
	\centering 
	\includegraphics[width=1.4\columnwidth]{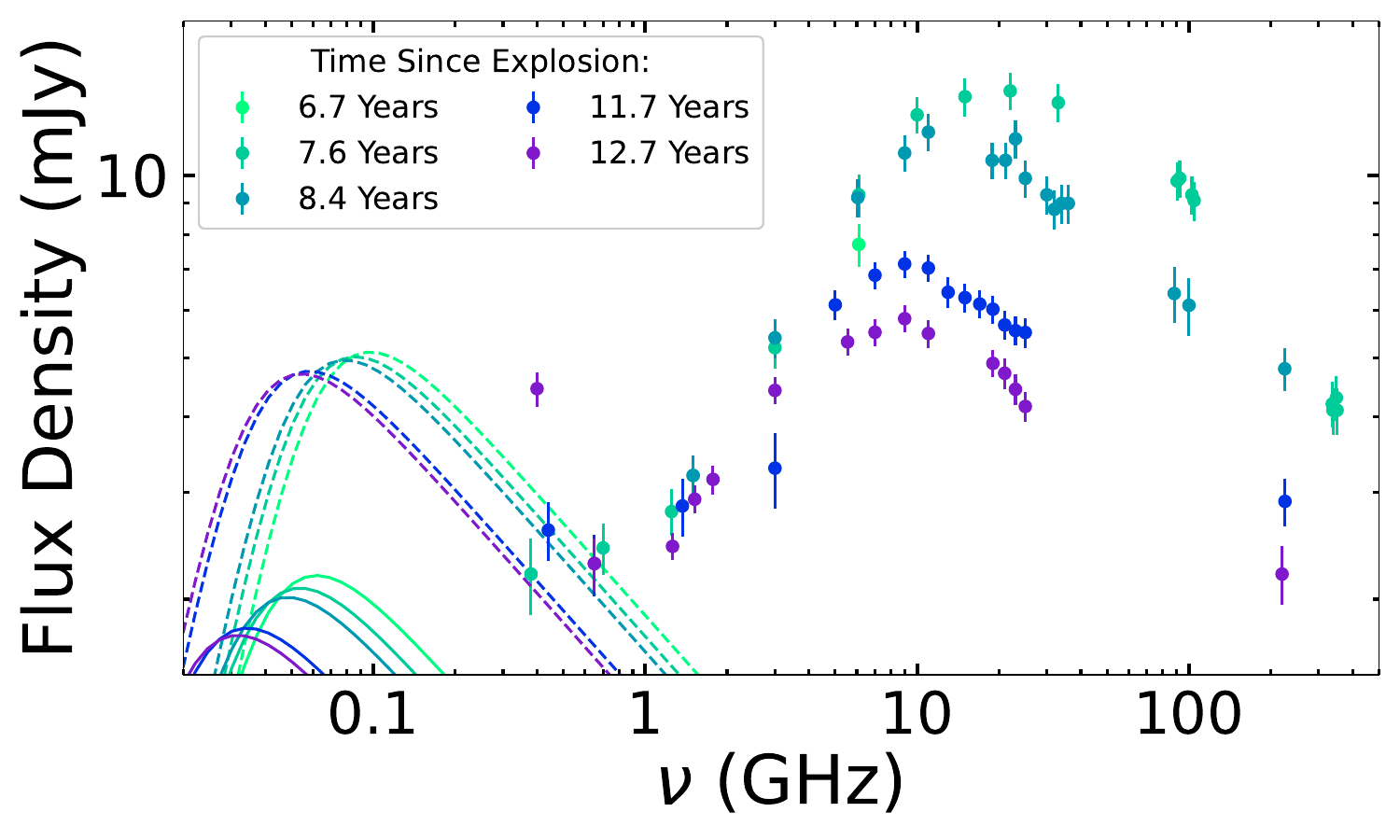}
    \hskip -0.5 cm
    \vskip -0.1 cm
    \caption{The extrapolation of our best-fitting model(s) of \S\ref{subsec:12aulatetime} to late times $\delta t \geq 6$\,yr significantly under-predicts the detected emission (filled circles).  Solid lines: extrapolation of the early-time ($\delta t \leq 190 $d) best-fitting model using a standard wind CSM density index $s=2$, and electron energy distribution index $p=3$. The dashed lines show how a model with shallower $p=2.01$, and a slightly less steep density index $s=1.75$ can match the flux density at low frequencies $\nu \lesssim 1$\,GHz at late-times, 
    while still requiring some additional component to explain the dominant emission at higher frequencies ($\nu \geq 5$\,GHz). This two-component model is explored in \S\ref{Sec:TwoComponent}.}
    \label{fig:low_freq_extrap}
\end{figure*}

We extrapolate our early-time solution to later times.
We assume the CSM density index is a standard wind index $s = 2$ and compute the shock dynamics self-consistently with Eq. \ref{eq:thin_shell}; finally we predict the expected radio SEDs, using Eq. \ref{eq:F_brk}--\ref{eq:broken_PL} (thick solid lines in Fig.  \ref{fig:low_freq_extrap}). By computing the characteristic synchrotron frequencies at the times of interest, we verify that the break frequencies still satisfy $\nu_{m}<\nu{_{sa}}<\nu_{c}$, which implies that Eq. \ref{eq:F_brk} and \ref{eq:nu_brk} are still valid.

As shown by Fig. \ref{fig:low_freq_extrap}, the extrapolation of the early-time model significantly under-predicts the observed flux densities, and 
the overall SED peak is shifted to significantly lower frequencies. 
However, while the emission near the observed SED peak is completely unaccounted for by the extrapolated model, we note that flux densities at $\nu\lesssim1$\,GHz are better matched.  With just a small deviation from the a wind density profile (from $\rho_{\rm{CSM,0}} \propto r^{-2}$ to $\rho_{\rm{CSM,0}} \propto r^{-1.75}$), and a decrease in the electron energy distribution index (from $p = 3$ to $p = 2.01$) the predicted flux density increases  to a level comparable with the observed $F_{\nu}$ at $\nu \lesssim 5$\,GHz
--- see the dashed line extrapolations in Fig. \ref{fig:low_freq_extrap}.

This exercise suggests that slight modifications of the extended CSM density profile at $r>10^{16.8}$\,cm, and some evolution of the shock parameters can reasonably account for
the low-frequency portion of the late-time emission. However, the higher-frequency flux densities at $\gtrapprox$ 5 GHz indicate a clear deviation from this simple scenario of continued shock interaction with the same CSM, suggesting the presence of a different component of emission. 
Thus motivated, we consider in \S\ref{Sec:TwoComponent} a multiple-component model to account for the bright peak of radio emission at $\delta t>6\,\rm{yr}$.

\section{A Two Component Model} \label{Sec:TwoComponent} 
We start with a few broad considerations about the shape of the spectral peak of the  late-time SEDs.
In addition to not being fully explained by extrapolations of the early-time shock, the late-time radio emission from \sn{} exhibits other notably unusual features:
(i) The optically thick slope around $\nu_{\rm{pk}}$ is unusually shallow: F$_{\nu} \propto \nu^{0.5}$, compared to the expected $F_{\nu}\propto \nu^{5/2}$ or $F_{\nu}\propto \nu^{2}$ of the SSA regime (or even steeper for a free-free absorption (FFA)+SSA scenario of a single population of radiating electrons). (ii) The optically thin slope is also unusually shallow, scaling like $F_{\nu} \propto \nu^{-0.3}$, compared to a typical value for young SNe of $F_{\nu} \propto \nu^{-1}$ \citep{Chevalier&Franson_2006} or at the most extreme  end F$_{\nu} \propto \nu^{-0.5}$ \citep{Urosevic2019}. (iii) Overall, the radio SEDs are unusually broad. 

\begin{figure}
    \centering
    \includegraphics[width=\linewidth]{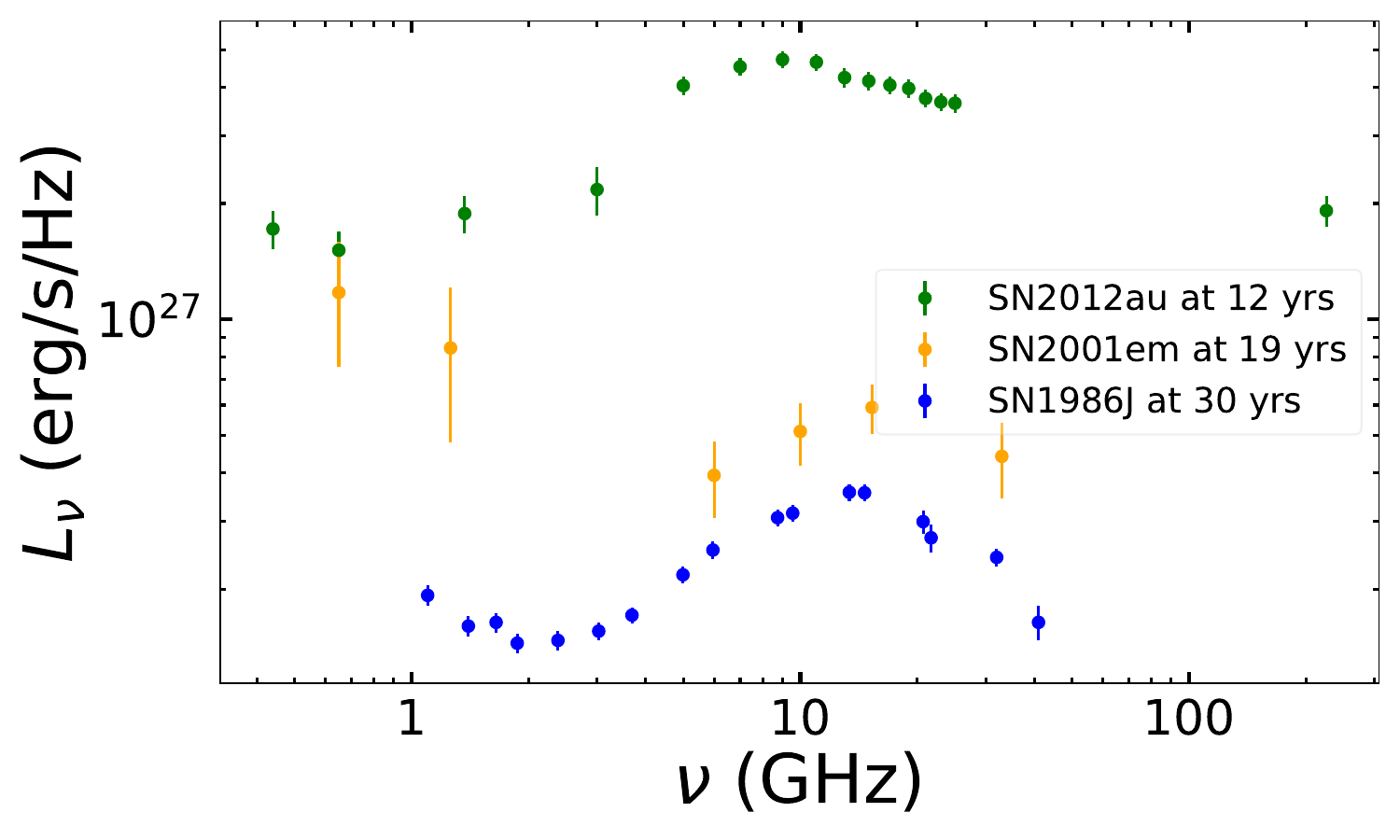}
    \caption{Late-time radio observations of \sn\, (this paper), SN\,1986J \citep{Bietenholz_2017}, and SN\,2001em \citep{01em} show a  qualitatively similar behavior, with an ``inverted'' SED at low frequencies. }
    \label{fig:comparison}
\end{figure}

The radio spectral shape of \sn\, many years post-explosion bears many qualitative similarities with that of the late-time radio emission from two other SNe:
 1986J and 2001em (see Figure \ref{fig:comparison}). The SEDs of the three SNe observed at $\delta t>12$\,yr peak at $\nu \approx(10-20)$\,GHz, and show an ``inversion'' of the optically thick slope around $\nu\approx$\,a few GHz. They also have similarly shallow optically thick slopes close to the peak and, in some epochs, display similarly broad peaks. The analysis of VLBI observations of SN\,1986J revealed that the observations are best explained by two distinct components contributing to the spectrum: a spatially extended component and a compact component dominating the low and high frequency range of emission, respectively \citep{Bietenholz_2017}. A similar explanation was favored for SN\,2001em \citep{01em}, in large part due to its observed similarities with SN\,1986J. VLBI  observations of \sn{}\, from \cite{Lazda_VLBI} suggest a similar two-component scenario, which is directly supported by our radio SED modeling and that we explore in detail below.

\subsection{A Model-Agnostic Approach} \label{SubSec:modelagnostic}
\begin{figure*}[t!]
    \centering
    \includegraphics[trim=1cm 4cm 0.5cm 4cm, width=1.95\columnwidth]{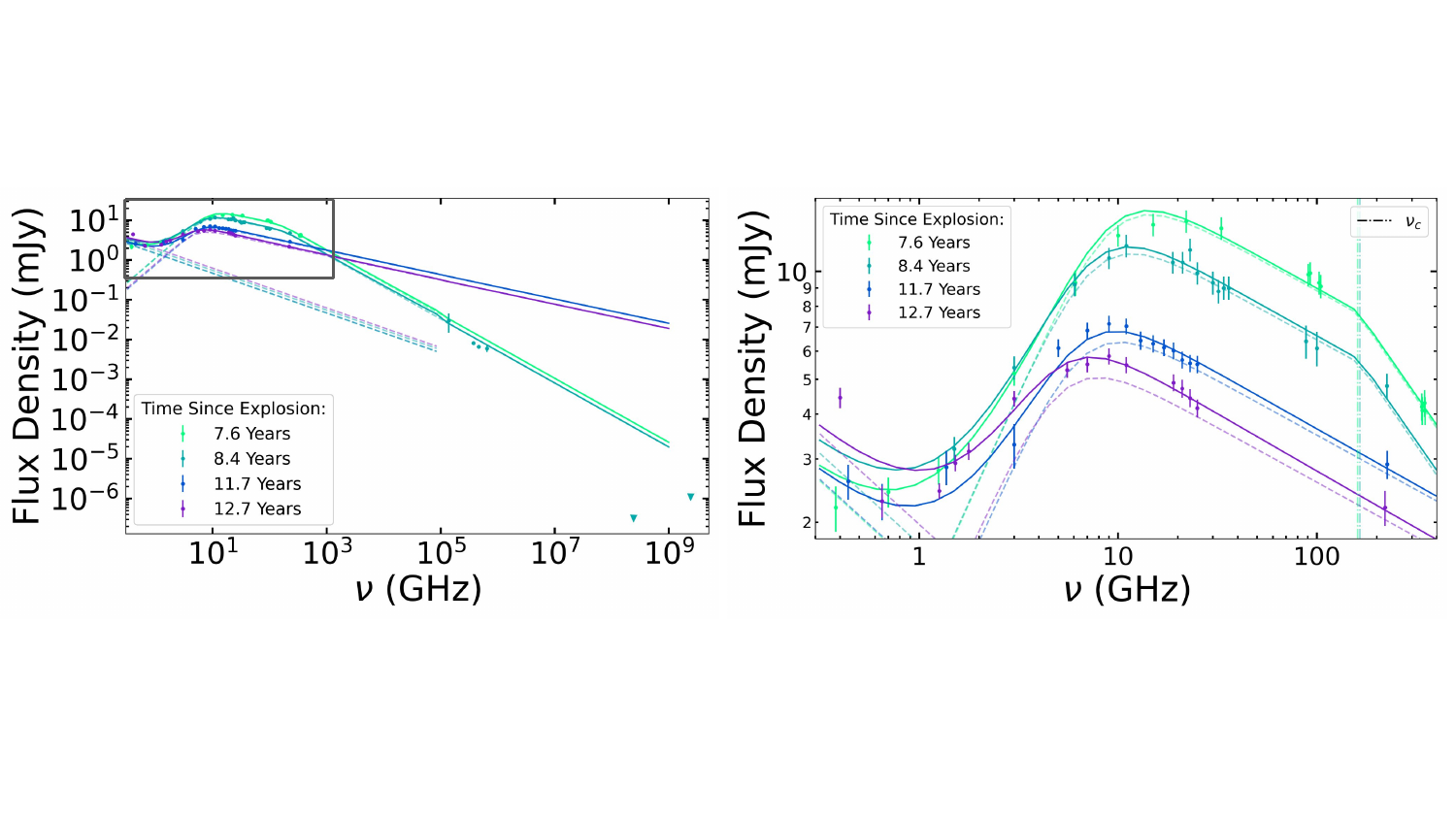}
    \caption{A model-agnostic, two-component fit to the late-time ($\delta t > 6$\,yr) \sn{} data. The model is sum of a low-frequency, single power-law component, and a high-frequency, broken power-law component, with a spectral break that may represent a cooling break. Dashed lines: the best fit individual components for each epoch. Solid lines: the total emission for each epoch. The low frequency model is truncated at $\nu = 10^{5}$\, GHz; while we cannot constrain any spectral break in this component, it is unlikely to extend this far without cooling effects.
    \emph{Left:} The full radio to X-ray SED late-time dataset, including broadband radio data, optical data at $\delta t = 8.4$\,yr and X-ray upper limits at $\delta t = 8.4$\,yr. \emph{Right:} A zoom-in on the broadband radio portion of the dataset - the boxed region in the left plot, with the cooling break frequencies shown by vertical dot-dashed lines for the 7.6 and 8.4 yr epochs.}
    \label{fig:model_agnostic_fit}
\end{figure*}

First, we perform a two-component fit of the late-time radio SEDs adopting a model-agnostic perspective. The goal is to capture the key observed features of the SEDs, while making as few physical assumptions as possible. Our model has a low-frequency component that will capture the behavior of the expanding blastwave from early times, and a higher-frequency component that captures the remaining emission. 

Based on our extrapolations shown in Figure \ref{fig:low_freq_extrap}, it is likely that our observations of the low-frequency component will fall entirely in the optically thin regime. As such, we parametrize this component as a single power law spectrum. With observations showing very flat spectral indices, we fix the power-law of this low-frequency component to $F_\nu \propto \nu^{-1/2}$, which is the shallowest spectral index allowable within the typical range of values for $p$: $2<p<4$ (e.g., \citealt{Chevalier&Franson_2006,Chevalier2017} and references therein) for young (i.e., pre-remnant phase) SNe. We allow the power-law normalization to vary freely between epochs.

We model the high-frequency component as a broken power law as in Eq.\,\,\ref{eq:broken_PL}. However, instead of assuming an optically thick slope based on SSA ($F_\nu \propto \nu^{5/2}$), we treat this slope as a free parameter, $\alpha$. We do not assume any $w(p)$ smoothing parameterization, as we do not assume to know how the characteristic frequencies of the spectrum are ordered, and therefore cannot use the formalism of \cite{Granot&Sari} to describe the smoothing parameter of the broken power law. While $w$, $\alpha$,  and $p$ will be able to vary freely, we will assume that they will stay constant across late-time epochs. 

We find that the optically thin side of the high-frequency component is poorly fit by a single power law segment, particularly at the $\delta t = 7.6$\,yr epoch. To allow the model to better fit the observed SED, we add a cooling-break frequency, $\nu_{\rm{c}}$, at each epoch as a free parameter. At $\nu > \nu_{\rm{c}}$, the spectral index steepens from $F_\nu \propto \nu^{-\frac{(p-1)}{2}}$ to $F_\nu \propto \nu^{-\frac{p}{2}}$, consistent with spectral steepening at an inverse-Compton or synchrotron cooling break \citep{Chevalier2017}. The change in slope at $\nu_{\rm{c}}$ is the only physical prescription added to this model.

The results of this model-agnostic fit can be found in Figure \ref{fig:model_agnostic_fit}, with the corresponding corner plots in Appendix \ref{app:early_corner}. We find that the best-fitting parameters are: $p= 1.61^{+0.05}_{-0.04}$, $\alpha = 1.21^{+0.16}_{-0.13}$, and $w = 0.44^{+0.13}_{-0.10}$. $F_{\nu,\rm{brk}}(t)$ and $\nu_{\rm{brk}}(t)$ follow a consistent power-law evolution in time, as shown in Figure \ref{fig:model_agnostic_F_nu}: $F_{\nu,\rm{brk}} \propto t^{-1.98\pm0.17}$, and $\nu_{\rm{brk}} \propto t^{-0.94 \pm 0.32}$. While fixing the low-frequency optically thin slope to $F_\nu \propto \nu^{-1/2}$ (compared to the early-time value of $F_\nu \propto \nu^{-1}$)  did produce a closer fit to the data, the inferred values of $F_{\nu,\rm{brk}}(t)$ and $\nu_{\rm{brk}}(t)$ are not highly sensitive to our specific choice of spectral slope of the low-frequency component. 

We find statistical evidence for a cooling break in the radio data at $\delta t = 7.6$\,yr, with  $\rm{log}_{10}(\nu_{\rm{c}}/\rm{Hz}) = 11.21^{+0.06}_{-0.06}$.  The broad-band (radio to X-ray) SED at $\delta t = 8.4$\,yr also implies the presence of a cooling break, with $\rm{log}_{10}(\nu_{\rm{c}}/\rm{Hz}) \approx 11.2$ required to maintain consistency with the optical data (Fig. \ref{fig:model_agnostic_fit}). The overestimated emission in the optical regime can also be explained by a distribution of electron energies that is not characterized by a single power law and breaks to a steeper power law at high energies.

More interestingly, we find that even including a cooling break, the extrapolation of the radio SED  leads to a significant overestimate of the X-ray flux at $\delta t = 8.4$\,yr, implying photoelectric absorption of the X-rays. We derive a lower limit on the neutral hydrogen equivalent column density needed to reconcile the model predictions with the deep Chandra and NuSTAR non-detections of \S\ref{SubSec:CXO}  and \S\ref{SubSec:NuSTAR} by considering the contributions from the high-frequency component, only, where the high-frequency component has a cooling break. We consider two chemical compositions of the absorbing material: a Solar-like composition; and a ``SN ejecta''-like composition with $\approx 35$\% of O by mass (e.g., \citealt{Chugai06}). We calculate the respective photoelectric cross sections with \texttt{vphabs} within \texttt{Xspec}.  We find\footnote{The Chandra and the NuSTAR observations translate into similar constraints, and we list here the most constraining lower limit  obtained between the two measurements.}  $\rm{NH_{int}}>2.1\times 10^{22}\,\rm{cm^{-2}}$ for Solar composition, and $\rm{NH_{int}}>5.2\times 10^{20}\,\rm{cm^{-2}}$  for the Oxygen-rich, ejecta-like composition. These are conservative lower limits, as: (i) we are not considering the possible contribution from the low-frequency radio component; (ii) we are not accounting for other sources of X-ray emission, like a reverse shock.  
This photo-electrically absorbed two-component model fits the data well, with a broken power law capturing the high-frequency emission, and an additional optically thin power law accounting for the low-frequency flux densities and hence the spectral index flattening towards lower frequencies.  

\subsection{Physical Inferences} 
\label{SubSec:PhysInfModelAgn}
\begin{deluxetable}{c|cc|ccc} \label{Tab:modelagnostic}
\tablecaption{Fitted and inferred parameters for the high-frequency component of the two-component model-agnostic fit (\S\ref{SubSec:modelagnostic}-\S\ref{SubSec:PhysInfModelAgn}). Inferred parameters are based on the assumptions of equipartition at $\epsilon_e = \epsilon_B = 0.1$, a characteristic frequencies ordering of $\nu_{\rm{m}} < \nu_{\rm{sa}}<\nu_{\rm{c}}$, and a filling factor of
$f_V=0.5$. Note that these values are dependent on our best fit value of the smoothing parameter, $w = 2.2$. For a more typical value of $w=1$, the inferred $R$ are negligibly larger, $B$ fields are a factor of $\approx2$ smaller, and $n_e$ is on the order of $10^{7}$ cm$^{-3}$ instead of  $10^{8}$ cm$^{-3}$. \label{tab:agnostic_fit} }

\tablewidth{0pt}
\tablehead{
\hline
\colhead{\shortstack{\rule{0pt}{3ex}$\delta t$ \\ (yr)}} &
\colhead{\shortstack{\rule{0pt}{3ex}$F_{\rm pk}$ \\ (mJy)}} &
\colhead{\shortstack{\rule{0pt}{3ex}$\nu_{\rm pk}$ \\ (GHz)}} &
\colhead{\shortstack{\rule{0pt}{3ex}$R$ \\ ($10^{16}$ cm)}} &
\colhead{\shortstack{\rule{0pt}{3ex}$B$ \\ (G)}} &
\colhead{\shortstack{\rule{0pt}{3ex}$n_{\rm e}$ \\ ($10^{8}$ cm$^{-3}$)}}
}
\startdata
$7.64$ &$13.2^{+1.1}_{-0.9}$ & $9.6^{+1.3}_{-1.2}$ & $1.1^{+0.2}_{-0.1}$ & $1.6^{+0.2}_{-0.2}$ & $3.2^{+2.2}_{-1.4}$ \\ 
$8.41$ &$10.3^{+0.8}_{-0.6}$ & $7.8^{+0.8}_{-0.8}$ & $1.2^{+0.2}_{-0.1}$ & $1.3^{+0.1}_{-0.1}$ & $2.2^{+1.1}_{-0.8}$ \\ 
$11.74$ &$5.8^{+0.4}_{-0.3}$ & $6.7^{+0.6}_{-0.6}$ & $1.0^{+0.1}_{-0.1}$ & $1.2^{+0.1}_{-0.1}$ & $4.8^{+2.0}_{-1.7}$ \\ 
$12.68$ &$4.6^{+0.4}_{-0.3}$ & $5.2^{+0.5}_{-0.5}$ & $1.2^{+0.2}_{-0.1}$ & $1.0^{+0.1}_{-0.1}$ & $2.7^{+1.4}_{-1.0}$ \\ 
\enddata
\end{deluxetable}

\begin{figure}
    \centering
    \includegraphics[width=\linewidth]{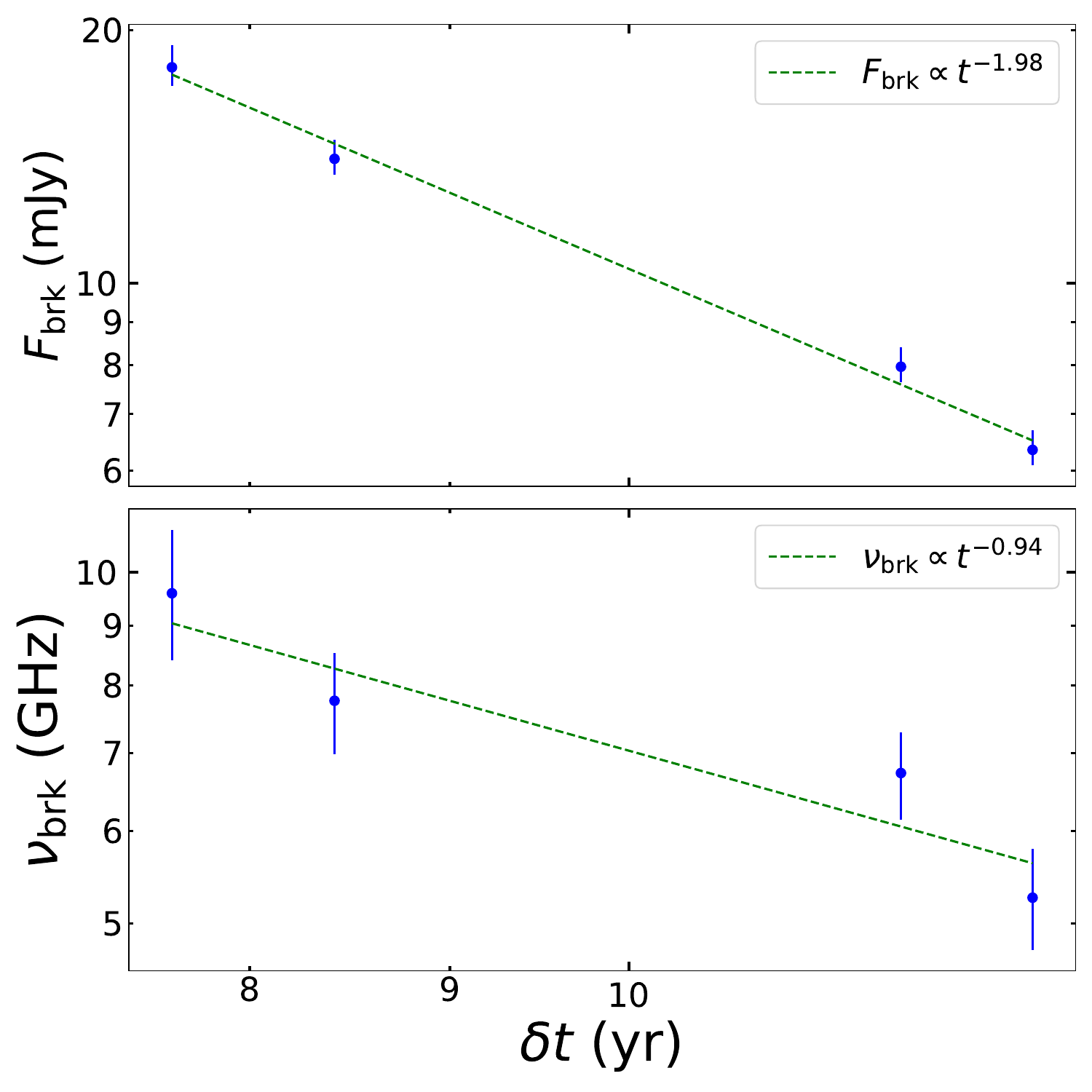}
    \caption{The best fitting values for $F_{\nu,\rm{brk}}$ and $\nu_{\rm{brk}}$ of the high frequency component as a function of time since explosion for the model-agnostic fit of \S\ref{SubSec:modelagnostic}. Both $F_{\nu,\rm{brk}}$ and $\nu_{\rm{brk}}$ follow consistent power law decays with time. }
    \label{fig:model_agnostic_F_nu}
\end{figure}

\label{SubSec:inferences}
Using the best-fitting values for the radio spectral peaks shown in Fig. \ref{fig:model_agnostic_F_nu}, and assuming the peaks are the result of synchrotron self-absorption, we can infer some of the physical properties of the high-frequency component. Independently from the details of its astrophysical nature, we can apply Eq. \ref{eq:B_field}, \ref{eq:F_brk}, and \ref{eq:nu_brk} to determine the emitting radius, magnetic field and CSM density for each epoch (Table \ref{Tab:modelagnostic}). We note that these values are based on assumed microphysical parameters $\epsilon_e = \epsilon_B = 0.1$; however, the inferred $B$ fields imply values of $\nu_c$ that violate our assumed ordering $\nu_{\rm{m}} < \nu_{\rm{sa}}<\nu_{\rm{c}}$. For our physical models of the next sections, we will require deviation from equipartition to ensure self-consistency. That being said, we can make several statements based on our model agnostic fit that have very weak dependencies on microphysical parameters:

\begin{enumerate}[(i)]
  \item The high-frequency radio component is associated with a small emitting volume, corresponding to an isotropic equivalent radius (defined in Eq. \ref{eq:equivalent volume}) of  $\approx  (f_V\frac{\epsilon_B}{\epsilon_e} )^{\frac{1}{16.2}} 10^{16}$\,cm. This radius is much smaller than the inferred radius of the initial SN shock at this epoch $\approx10^{18}$\,cm (\S\ref{subsec:12aulatetime}),
thus providing additional evidence that this is an astrophysically distinct component. Importantly, this inference is consistent with the direct VLBI size constraints by \cite{Lazda_VLBI}.
  \item The small radius of the emitting region implies a low average expansion velocity  $\approx\,500\,\rm{km\,s^{-1}}$ \emph{if} interpreted as a shock radius, and a large number density of electrons ($\approx (\frac{\epsilon_B}{0.1})^{-0.6} (\frac{\epsilon_e}{0.1})^{0.4} 10^7$ -- $10^8\, \rm{cm}^{-3}$), the range being dependent on the smoothing parameter.
  \item The lack of detected soft and hard X-ray emission at $\delta t\approx 8.4\,$yrs implies that the high-frequency radio component is either photoelectrically absorbed, or that no X-ray emission is produced. Given the low shock velocities, diffusive shock acceleration may not produce electrons of high enough energy to extend to the X-rays. 
  \item If a distinct component is responsible for the high-frequency emission at late times, then it must be  subdominant to the shock emission  at $\delta t \leq 190$\,d.
  \item While the $B$ field shows a clear, monotonic evolution with time ($B \propto t^{-0.7}$), the other quantities in Table \ref{Tab:modelagnostic} do  not: the emitting radius is consistent with being constant within uncertainties, implying an expansion velocity $\lesssim$ a few hundred $\rm{km\,s^{-1}}$.
  \item If the ordering of characteristic frequencies is the typical $\nu_{\rm{m}} < \nu_{\rm{sa}}<\nu_{\rm{c}}$, we require a deviation from equipartition of at least $\frac{\epsilon_B}{\epsilon_B} \leq 0.1$.

\end{enumerate}

Any physical model proposed to explain the observations of \sn{\,} must meet all the criteria above. In  \S\ref{Sec:CSMinteraction} and \S\ref{Sec:PWN}, we explore some of such models. 

\begin{figure*}
    \centering
    \includegraphics[trim={0 8cm 0 5cm},width=0.9\linewidth]{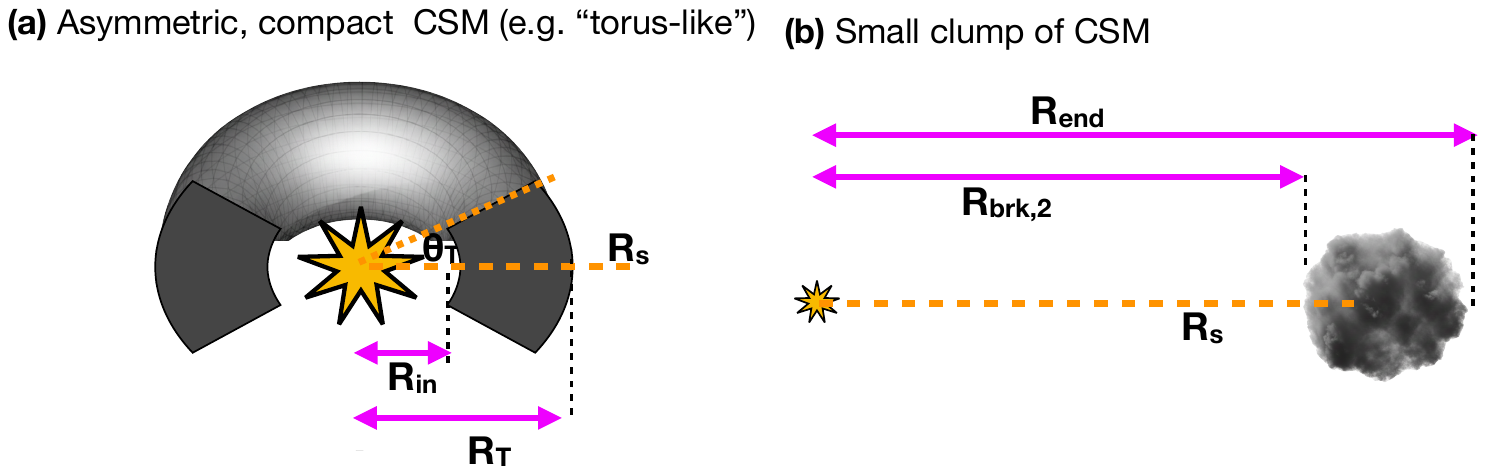}
    \caption{ Cartoon showing two different CSM configurations: (a) Spherically asymmetric torus-like CSM extending from a distance $R=R_{\rm in}$ to $R=R_T$ from the explosion site, with the shock interacting with a less dense medium at $r>R_T$. The half opening angle of the torus $\theta_T$ is related to the volumetric filling factor in our modeling as $f_V =3f_R\rm{sin}\theta_T$, where $f_R$ is the thickness of the shocked region. 
    (b) Clump of dense CSM at a distance $R_{brk,2}$ from the explosion site, extending to a distance of $R_{\rm end}$. 
    We do not assume a particular geometry for this scenario and instead 
    fit for the projected area filling factor, $f_A$, across a range of volumetric filling factors, $f_V$.}
    \label{fig:CSMcartoon}
\end{figure*}

\section{CSM Interaction Models}  \label{Sec:CSMinteraction}
We first consider the possibility that the high-frequency radio component is the result of the SN shock interaction with an aspherical CSM.
We explore two scenarios, shown in Fig. \ref{fig:CSMcartoon}: (a) dense CSM subtending a small solid angle (e.g., ``torus-like'' CSM, \S\ref{SubSec:CSMtorus}); and (b) a small clump of dense CSM, placed at some large radius from the explosion site (\S\ref{SubSec:CSMsmallclump}).

\subsection{A ``torus-like'' dense CSM}
\label{SubSec:CSMtorus}
We consider a scenario where high-density CSM is confined to a small solid angle, a geometry motivated by the equatorial torii of material frequently observed around massive stars. The solid angle $\Omega_T$ subtended by the torus is related to the half opening angle $\theta_T$ of Fig. \ref{fig:CSMcartoon} as $\Omega_T = 4\pi \rm{sin}(\theta_T)$. In this configuration, the early radio emission from \sn{}\, would result from the shock interaction with the lower-density material (for example, along the poles), with density as inferred in \S\ref{Subsub:12auearlytime}, while the late-time radio emission originates from the interaction with the high-density CSM in the torus.
Qualitatively, the dense CSM of the torus could decelerate the shock significantly, explaining the small inferred emitting volume, the low observed shock velocity, and the late emergence of the radio emission. Because of the high density, we expect the early-time emission from this component to be suppressed by free-free absorption, which would explain why it did not appear in the initial dataset. We model the free-free optical depth as in \cite{Rybicki&Lightman}:
\begin{equation}
\label{eq:tau_ff}
    \tau_{\rm{ff}} (R, \nu) = \int_R^{\infty} 0.018T_e^{-3/2}Z\nu^{-2}\bar g_{\rm ff} n_e(r)n_i(r)dr, 
\end{equation}
where all quantities are in cgs units. We assume an electron temperature, $T_e = 10^5$\,K, a Gaunt factor $\bar g_{\rm ff} = 5$, a hydrogen CSM composition with $Z=1$, and full ionization such that $n_e$, the number density of electrons, is equal to $n_i$ the number density of ions: $n_e = n_i$. A similar scenario was invoked to explain the broad-band X-ray and radio emission from the strongly interacting SN\,2014C \citep{SN2014C,Thomas22}.
We quantitatively explore this model below. 

We consider the late-time radio epochs at $\delta t >6$\,yrs, selecting data at $\nu \geq 5$\,GHz to minimize the contribution from the low-frequency component,
 while still having sufficient spectral coverage to characterize the spectral peak at each epoch. 
 We model the shock dynamics as described in \S\ref{SubSec:thinshellmodel}, adopting a broken power-law model for the dense CSM with break radii, $R_{\rm in}$ and $R_{\rm T}$, describing the inner radius and transition radius of the dense torus respectively.
 At $r<R_{\rm in}$, we assume that the early-time density profile inferred in Section \ref{Subsub:12auearlytime} ($\rho_{\rm CSM,0}(r)$, Equation \ref{eq:rhoCSMBPL}) applies. The profile at $R_{\rm in}< r < R_{\rm T}$ is poorly constrained, 
 with the only constraint being that this component is not detected at $\delta t < 6$ months. 
 Consequently, the CSM density index and normalization at $r < R_{\rm T}$ are degenerate. We thus assume an inner density profile that is wind-like without any loss of generality of our argument. 
 We also find that the density required to decelerate the ejecta to slow velocities of a few hundreds of $\rm{km\, s^{-1}}$ far exceeds the density required to produce the late-time radio emission, so we allow for a  discontinuity at $R_{\rm T}$, where the density can drop significantly. The \emph{torus} density profile is therefore described as: 

\begin{equation} 
\label{eq:slice_density_profile}
\rho_{\textrm{T}}(r) = 
\left\{
    \begin{array}{lr}
        \rho_{\rm CSM,0}(r), & r < R_{\rm in} \\
        \rho_{\rm 1} \left(\frac{r}{R_{\rm{T}}}\right)^{-2}, &  R_{\rm in}<r < R_{\rm T}\\
        \rho_{\rm 2} \left(\frac{r}{R_{\rm T}}\right)^{-s}, &  r > R_{\rm T}
    \end{array}
\right.
\end{equation}

The initial velocity of the ejecta is also poorly constrained, thus we adopt
$M_{\rm{ej}} = 5 M_\odot$ and $E_{\rm{k}} = 10^{52.5}$ erg.\footnote{We note that adopting initial values to within a factor of a few  from those assumed does not change the major conclusions from this section.} From these initial conditions and the torus CSM density profile in Eq. \ref{eq:slice_density_profile}, we  calculate the radius $R_{\rm {s}}(t)$ and velocity $v_{\rm{s}}(t)$ of the shock as it plows through the torus from numerical integration of Eq. \ref{eq:thin_shell}, and $B(t)$ through Eq. \ref{eq:B_field}. We fix $\epsilon_e = 0.1$ and fit for $\epsilon_B$ and $p$. We do not assume any specific ordering of the synchrotron break frequencies, and instead we 
self-consistently calculate the characteristic break frequencies and their corresponding SEDs using the method described in Appendix C of \cite{Itai_tvd}. Because we are considering values of $p < 2$, we need to specify a maximum value for the accelerated electron energy, $\gamma_{\rm max}$, to prevent divergent solutions. In this model, we fix $\gamma_{\rm max} = 10^7$.\footnote{Our solutions are not sensitive to this choice of $\gamma_{\rm max}$}

In our modeling, we adopt a "wedge-like" cross section for the CSM (as shown in Figure \ref{fig:CSMcartoon}) for simplicity.  
We account for the small solid angle by prescribing a volumetric filling factor $f_V$ and projected area filling factor $f_A$. To calculate these values, we define the parameter $f_R \equiv \frac{\Delta R}{R_s}$, where $\Delta R$ is the thickness of the shell of shocked material. Under this prescription, we can approximate $f_V = 3f_R \rm{sin}(\theta_T)$, which is valid for small angles of $\theta_T$. We assume a value of $f_R = 0.1$, in line with typical values derived from self-similar solutions \citep{Chevalier1982a}. The value of $f_A$ is viewing angle dependent, so we fit models for the two extreme cases: (i) viewing the torus edge-on, such that the torus appears as a thin strip, (ii) viewing the torus face-on, such that the torus appears as a ring.  

\subsubsection{Edge-on torus viewing angle} \label{subsub:edge_torus}
For an edge-on torus viewing angle, we consider a projected area filling factor $f_A = \frac{4}{\pi}\rm{sin}(\theta_T)$. At this orientation, the early time emission from the torus is not detected because it is entirely free-free absorbed. Without any constraint on $R_{\rm in}$, we assume that the torus begins very close to the surface of the progenitor, taking $R_{\rm in} = R_{\rm star}$, where $R_{\rm star} \approx R_\odot$ is the assumed radius of the progenitor at time of explosion (as appropriate for WR stars).

\begin{figure}
    \centering
    \includegraphics[width=\linewidth]{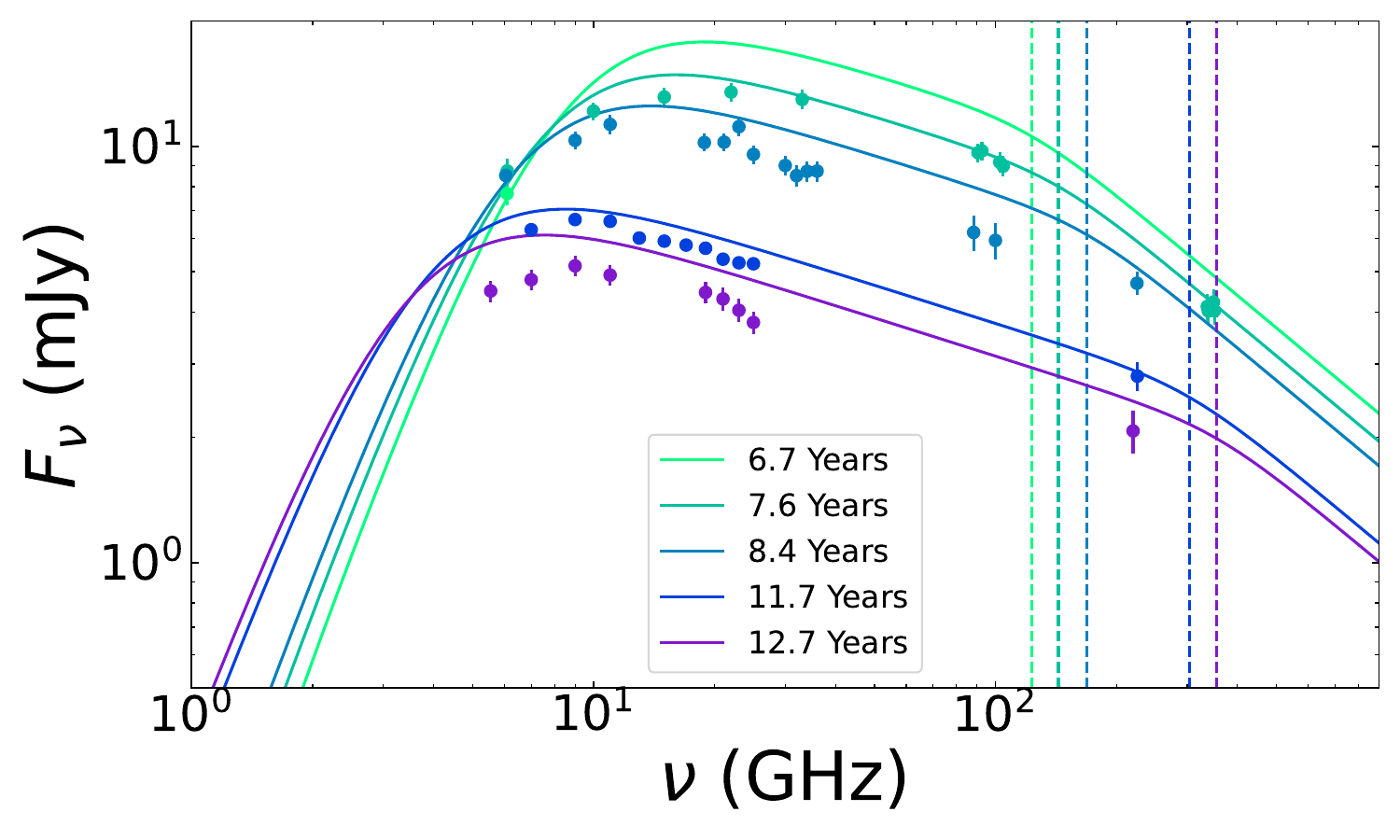}
    \caption{Best-fitting model for a torus-like dense CSM viewed edge-on. While the model captures the overall SED behavior at $\delta t = 8.4$ yr and $\delta t = 11.7$ yr reasonably well, it under-predicts the $\delta t = 7.6$ yr epoch, while over-predicting the $\delta t = 12.7$ yr epoch. The cooling breaks, visualizes as dashed lines, line up well with the breaks required in our model-agnostic fits at $\delta t = 7.6$ yr and $\delta t = 8.4$ yr.}
    \label{fig:CSM_Torus_SED}
\end{figure}

We compute the best-fitting values for the parameters that characterize the density profile: $\rho_1$, $\rho_2$, $R_T$, and $s$, in addition to the following parameters that impact the emission calculation: $\epsilon_B$, $p$, and $\theta_T$. We employ MCMC to fit, with 50 walkers iterating the model calculations 10000 times and discarding the first 1000 iterations for each walker.  The corner plots for this fits can be found in Appendix \ref{app:CSM_interaction}. We plot the best-fitting model  in Figure \ref{fig:CSM_Torus_SED}. The model fits the data well at $\delta t = 8.4$ yr and $\delta t = 11.7$ yr, while under-predicting and over-predicting the data at $\delta t = 7.6$ yr and $\delta t = 12.7$ yr, respectively. The cooling break frequencies, represented as dashed vertical lines in Figure \ref{fig:CSM_Torus_SED}, agree with those predicted in \S\ref{SubSec:modelagnostic}.

\begin{figure}
    \centering
    \includegraphics[width=\linewidth]{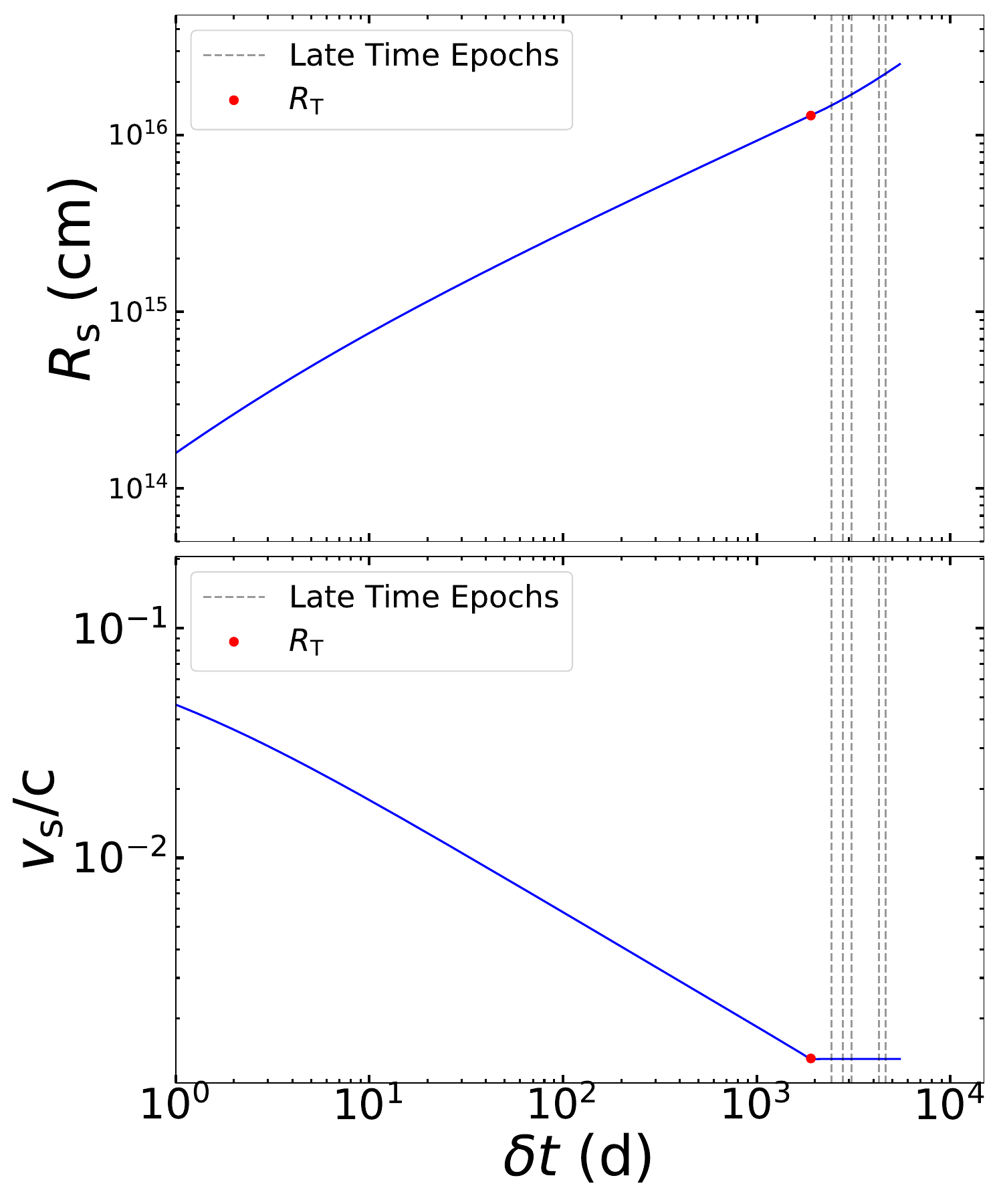}
    \caption{ Dynamical evolution of the shock (radius and velocity) expanding in a ``torus-like'' dense-CSM scenario (\S\ref{SubSec:CSMtorus}). The vertical gray dashed lines mark the epochs of our late-time radio observations. The red dot in each curve marks the transition radius, $R_T$.}
    \label{fig:CSM_slice_Rv}
\end{figure}

The resulting best-fitting shock dynamics are shown in Fig. \ref{fig:CSM_slice_Rv}. At $r > R_{\rm T}$  the shock velocity plateaus to a roughly constant value $\approx$400-500$\rm{\,km\,s^{-1}}$, as the shock expands past the dense region of the torus. We note that our best-fits indicate steep values of $s > 3$ , which should result in shock acceleration \citep{Waxman1993}. Instead we assume a constant shock velocity. We acknowledge that  this is a current limitation of our  model; 
however, we do not expect this simplified dynamics to change the major conclusions from this study.

\begin{figure}
    \centering
    \includegraphics[width=\linewidth]{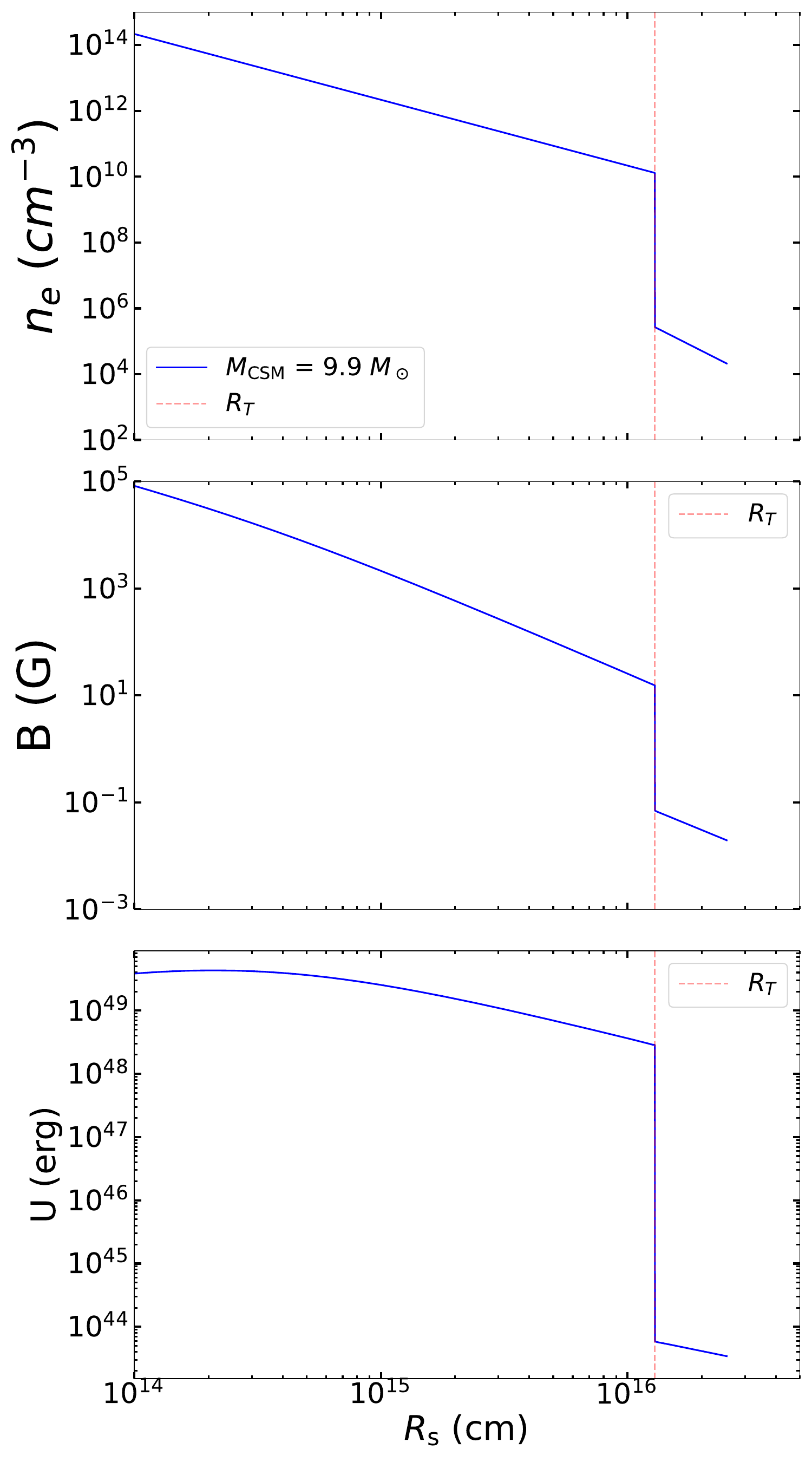}
    \caption{  Best-fitting torus CSM density ($n_e$) and shock properties ($B$ and $U$) as a function of $R_{\rm{s}}$ in the context of the ``torus-like'' dense CSM scenario viewed edge-on (\S\ref{subsub:edge_torus}).
    The sharp discontinuities in each curve correspond to when the shock reaches  the discontinuity in the CSM density profile at $R_{\rm{T}}$.  The requirement of a small emitting region translates into extremely large densities and magnetic field strengths. We infer a total CSM mass in the torus of $M_{\rm CSM} \approx 10\, \rm M_\odot$, based on integration of the displayed profile up to the time of the most recent radio observation (Eq. \ref{Eq:MassCSMtorus}).}
    \label{fig:CSM_slice_nBU}
\end{figure} 

In Fig. \ref{fig:CSM_slice_nBU}, we plot the best-fitting torus CSM number density of electrons ($n_{\rm{e}}$), the magnetic field strength ($B$), and the internal energy of the shock ($U$) as a function of shock radius, where $U$ is
calculated as: 
\begin{equation}
    U = \frac{B^2}{8\pi\epsilon_{\rm B}}  \Big (\frac{4}{3}\pi R_s^3f_V \Big )
\end{equation}
We emphasize that all these quantities apply to the shock interaction with the torus (i.e., the shock propagating along the low-density poles will have markedly different physical properties). Notably, the magnetic field strength shown in Figure \ref{fig:CSM_slice_nBU} differs significantly from those in the model-agnostic fit (reported in Table \ref{tab:agnostic_fit}). This is a result of free-free absorption affecting the observed fits in this model compared to the assumption of synchrotron self-absorption in the model-agnostic fit. In the edge-on torus case, the emission is free-free absorbed and the \emph{intrinsic SSA} peak is located at a lower $\nu_{\rm pk}$. 
While these inferences depend on the shock microphysical parameters, they are primarily dependent on their \emph{ratio}, i.e., $\epsilon_e/\epsilon_B$. Since we are fitting for $\epsilon_B$, a different choice of fixed $\epsilon_e$ would have little impact on inferred CSM mass.

A few considerations follow. First, 
very large torus densities are required at $r<R_{\rm{T}}$ to sufficiently decelerate the shock, corresponding to effective mass-loss rates 
$\dot M\gtrapprox 2.4\,M_\odot\,\rm{yr^{-1}}$ for wind velocity $v_{\rm{w}} = 1000$ km/s  (Fig. \ref{fig:CSM_clump_nBU}). These large densities also strongly free-free absorb the early radio flux, consistent with the early time radio observations. 
The associated  $B>100\,\rm{G}$ would also be problematic. We note that by setting $R_{\rm in}$ as close as possible to the explosion center, we are minimizing the required densities for deceleration and they are still extremely high. 
At $r>R_T$, we find that the CSM density must decline much more steeply than a wind-like scenario, with $s= 3.8$. This differs from the model-agnostic fit because this density profile has to account for the effects of free-free absorption, in addition to the effects of density on the total amount of emitting electrons. 

Second, the shock must propagate through a large column density of CSM in order to decelerate to the low inferred velocities. The total torus CSM mass is calculated as:
\begin{equation}\label{Eq:MassCSMtorus}
    M_{\rm{CSM}} = \int_{R_{\rm{star}}}^{R_{\rm{s}}(12.7 \rm{yr})} f_V \times \rho_T(r)\times 4\pi r^2 dr.
\end{equation}     
The CSM mass for our best fitting parameters is $M_{\rm CSM} = 10\,{\rm M}_\odot$. This CSM material is within a small radius $R \approx 10^{16}$\,cm, therefore requiring that a large fraction of progenitor mass was lost within a short time period prior to explosion. The small inferred filling factor $\log_{10}f_V = -1.48^{+0.48}_{-0.50}$ implies a very small half-opening angle of $\theta_T\le$5 deg. There is no clear physical mechanism for such collimated deposition of mass into the CSM.

Third, for the model we compute the torus CSM column density \emph{beyond} the shock radius at 8.4\,yrs, and we express the result in terms of neutral hydrogen equivalent column density for Solar-like and ejecta-like composition defined as in \S\ref{SubSec:modelagnostic}. We find that this model is in tension with the lack of X-ray detections for a Solar-like composition. The best-fitting model for CSM density predicts $\rm{NH_{int}} = 5.8\times 10^{20}\,\rm{cm^{-2}}$. This prediction is lower than the minimum $\rm{NH_{int}}$ needed to suppress the emission derived in \S\ref{SubSec:modelagnostic} -  $\rm{NH_{int}}=2.1\times 10^{22}\,\rm{cm^{-2}}$ for a Solar-like composition, but consistent with $\rm{NH_{\rm{int}}}>5\times 10^{20}\,\rm{cm^{-2}}$ for an ejecta-like composition. We note, however, that this result is limited by our 1D treatment of the shock dynamics. In a mutli-dimensional treatment, fast moving ejecta close to the torus would wrap around it, increasing the density beyond the shock radius and potentially providing enough interceding material to suppress the X-rays.

\subsubsection{Face-on torus viewing angle} \label{subsub:face_on_torus}
For a face-on torus viewing angle, we define $f_A = 2f_R$. In this scenario, the medium between the emitting region and the observer is defined by the inferred density profile in Section \ref{Subsub:12auearlytime}. This inferred profile is much less dense than the torus itself and will not result in any significant free-free absorption at early-times. In order to maintain consistency with the non-detections of this component in the early-time data, we set $R_{\rm in} = 10^{17.1}$\,cm. This is the smallest radius we can prescribe without observing interaction with the dense torus at our last early-time epoch ($\delta t = 190$\,d). Such a structure of a ring of CSM has been observed directly in SN 1987a \citep{Plait1995}. 

We perform an MCMC fit, with the same hyper-parameters as in Section \ref{subsub:edge_torus}. This model fails to converge to any solution that well-characterizes both the optically thick and optically thin sides of the SEDs. We attempt fits fixing $p=1.6$, varying between $f_R = 0.1$ and $f_R = 0.01$, and varying across different fixed $f_V$ values, but consistently only find solutions that match the data on the optically thin side. One such solution is plotted in Appendix \ref{app:CSM_interaction}. Without strong free-free absorption shifting the observed peak to a lower $F_{\rm pk}$ and higher $\nu_{\rm pk}$ (which is present in the edge-on case), there does not appear to be a density profile for the torus that can result in the observed emission.

We note that, for the range of solutions between the two extremes of viewing angle (edge-on and face-on), the suppression of the emission by free-free absorption will lie between the highly absorbed case of the edge-on viewing angle and the lack of significant absorption in the face-on case. For these intermediate viewing angles, if a solution does exist, the intrinsic SSA peak will be at lower $F_{\rm pk}$ and higher $\nu_{\rm pk}$ than the edge-on case, resulting in even more extreme values of $B$ field strength and $n_e$ at $r < R_T$.

 In summary, a high-density CSM with a very small net cross-section could explain the radio emission we observe in \sn, but the physical setup for this scenario appears highly contrived.  The required filling factor of $f _V\leq 4\times10^{-2}$, corresponds to a solid angle of $\approx 2700\, \rm{deg}^2$ or a half-opening angle of $\theta_T \approx 5$ deg.  Within that small solid angle must reside CSM of extremely high density to decelerate the shock containing solar masses of material within $R_{T}\sim$ $10^{16}\,\rm{cm}$, equivalent to far greater mass-loss than typically observed in SNe, yet leaving no imprint on the early SN optical spectroscopy. Furthermore, even in this contrived scenario, the inferred density profile is potentially inconsistent with the X-ray non-detections at $\delta t = 8.4$\,yr.

\subsection{Small Clump of Dense CSM}
\label{SubSec:CSMsmallclump}
The small cross-section scenario above requires such extreme densities at $r<R_{\rm{T}}$ because we are requiring the shock radius to be small for an extended period of time. However, the observed $F_{\rm{pk}}$ and $\nu_{\rm{pk}}$ of the high-frequency component in the late-time SEDs only imply a small \textit{emitting volume}, not a small shock radius. The shock could propagate out to a large radius before encountering a clump of over-dense CSM, where strong interaction occurs. In this scenario, this component would be radio-faint at early times not because of absorption, but by the lack of radiating electrons (i.e.,  the shock has not yet reached a dense region of CSM).

To model this scenario, we assume that the shock at early times follows the dynamics inferred in \S\ref{Subsub:12auearlytime}. As in \S\ref{Subsub:12auearlytime}, we use $M_{\rm{ej}} = 5 \,\rm{M}_\odot$ and we adopt the inferred  $E_{\rm{k}} = 10^{52.5}$ erg. At $\delta t < 190$\,d, we fix the density profile to the broken power law of Equation \ref{eq:rhoCSMBPL} with the best-fitting parameters in Table \ref{tab:early_best_fit}. The CSM density extends to larger radii as  $\rho_{\rm{CSM,0}}(r)\propto r^{-s_2}$ until it reaches a second break radius $R_{\rm{brk,}2}$, where the shock encounters a clump of uniform density, $\rho_{\rm{c}}$. The clump covers a small fraction of a sphere of radius $R_{\rm{brk,}2}$, and it is described by a volumetric filling factor $f_V$, and a projected area filling factor, $f_A$. In order to avoid assuming a particular geometry for this clump, we fit for values of $f_A$ over a range of values for $f_V$. We also fit for the radius at which the shock stops interacting with the clump, $R_{\rm{end}}$. This end radius is not important for the dynamics of the model, but will determine the contribution of free-free and photoelectric absorption to the radio and X-rays, respectively, from material in front of the shock.\footnote{We expect that since the interaction region resides at larger radii, the effects of ejecta wrapping around the clump will be minor in a multi-dimensional treatment.}

For this scenario Equation \ref{eq:tau_ff} takes the form:
\begin{equation}
    \tau_{\rm{ff}} = 0.018 T_{\rm e}^{-3/2} Zn_e n_i \nu^{-2}\bar{g}_{ff} (R_{\rm{end}}- R_s)
\end{equation}
For simplicity, we do not consider any density contribution beyond $R_{\rm{end}}$. The equation above also assumes that the radiation produced at $R_s$ travels through a distance of the order $\approx (R_{\rm{end}}-R_s)$ within the clump. The exact value clearly depends on the observer's line of sight and the actual 3D geometry of the clump.
 
As before, we fit the late-time observations at $\nu > 5$ GHz, where no significant contributions from the low frequency component are present. We are seeking clumps located at a large shock radius ($\approx 10^{18}$ cm), but that still occupy a small emitting volume. In this setup, the interaction will be delayed as the shock takes time to reach the clump from the explosion center, but the number of emitting electrons is still small. We find that to satisfy these requirements we are forced to consider a range of very small volumetric filling factors: $f_V = [5\times10^{-6},5\times10^{-7},5\times10^{-8}]$. These filling factors quantify the portion of the sphere at $R_{brk,2}$ that is covered by the clump, and thus describe the region where the shock interacts with the clump; instead, the early-time filling factors before the interaction with the clump remain as in Section \ref{SubSec:thinshellmodel}. In order to limit our number of free parameters, we fix $p = 1.6$, the value observed in all other fits. As in the torus scenario, we fix $\gamma_{\rm max} = 10^7$.

\begin{figure}
    \centering
    \includegraphics[width=\linewidth]{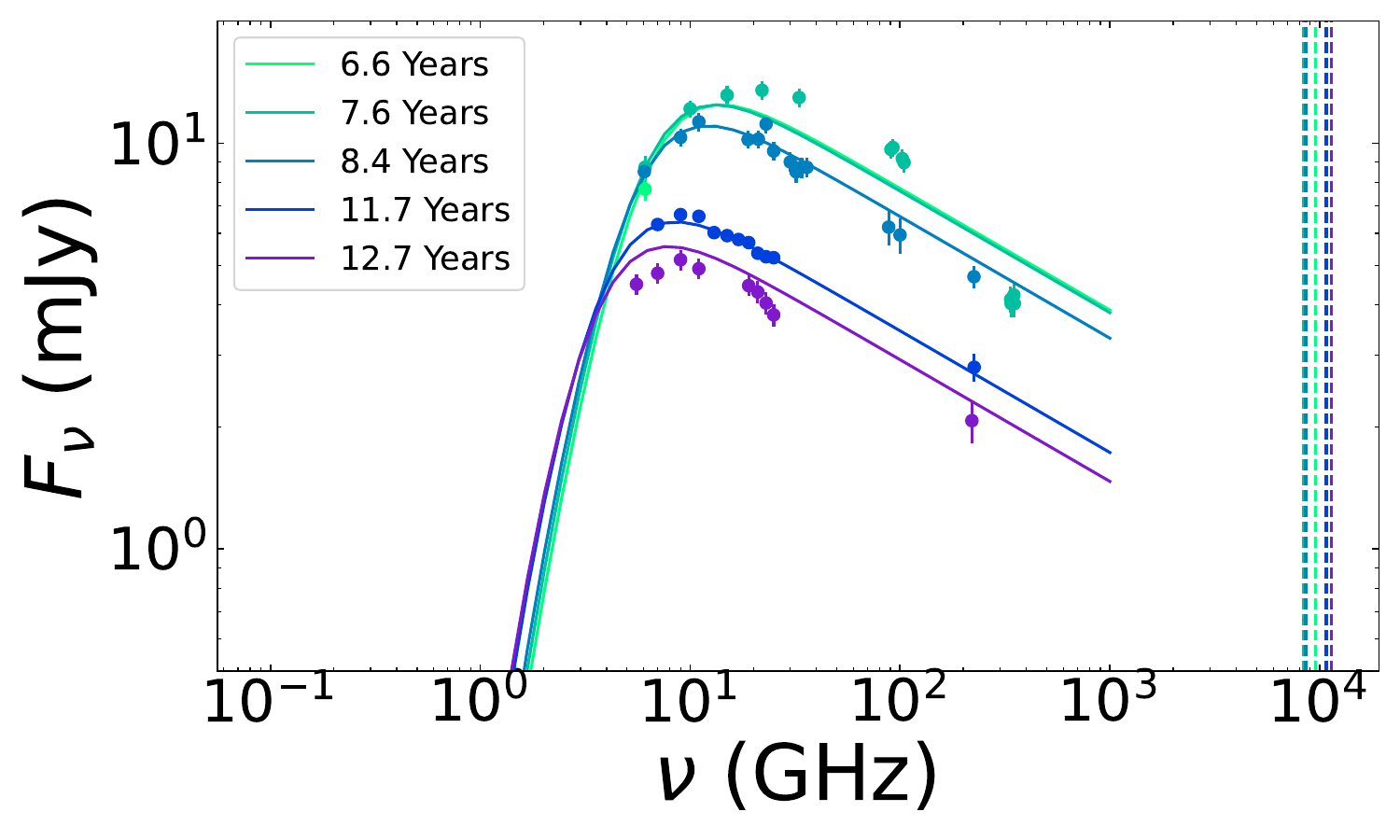}
    \caption{Best-fitting models for a small clump of dense CSM with filling factor $f_V=5\times10^{-8}$ at $R_{brk,2}= 4.7\times10^{17}$\,cm through $R_{\rm end} = 1.2\times10^{18}$\, cm. Like the edge-on torus model, the data are well fit at  $\delta t = 8.4$ yr and $\delta t = 11.7$ yr, while being under-predicted at $\delta t = 7.6$ yr epoch and over-predicted at $\delta t = 12.7$ yr. Unlike in the torus model, the cooling breaks, represented by dashed vertical lines, lie outside the range of observed frequencies, failing to reproduce the observed spectral break at $\delta t = 7.6$ yr.}
    \label{fig:CSM_SED}
\end{figure}

We again employ MCMC, using 25 walkers iterating the model calculations 1000 times and discarding the first 200 iterations for each walker. The resulting corner plots can be found in Appendix \ref{app:CSM_interaction}. An example of one of the fits is shown in Figure \ref{fig:CSM_SED} for $f_V=5\times10^{-8}$, which is the model with the lowest $\nu_c$ (i.e. the closest to the data) in this class. While much of the broad evolution is well captured in these models, this class of models fail to self-consistently account for the location of $\nu_{c}$ (dashed vertical lines in the Figure).

\begin{figure}
    \centering
    \includegraphics[width=\linewidth]{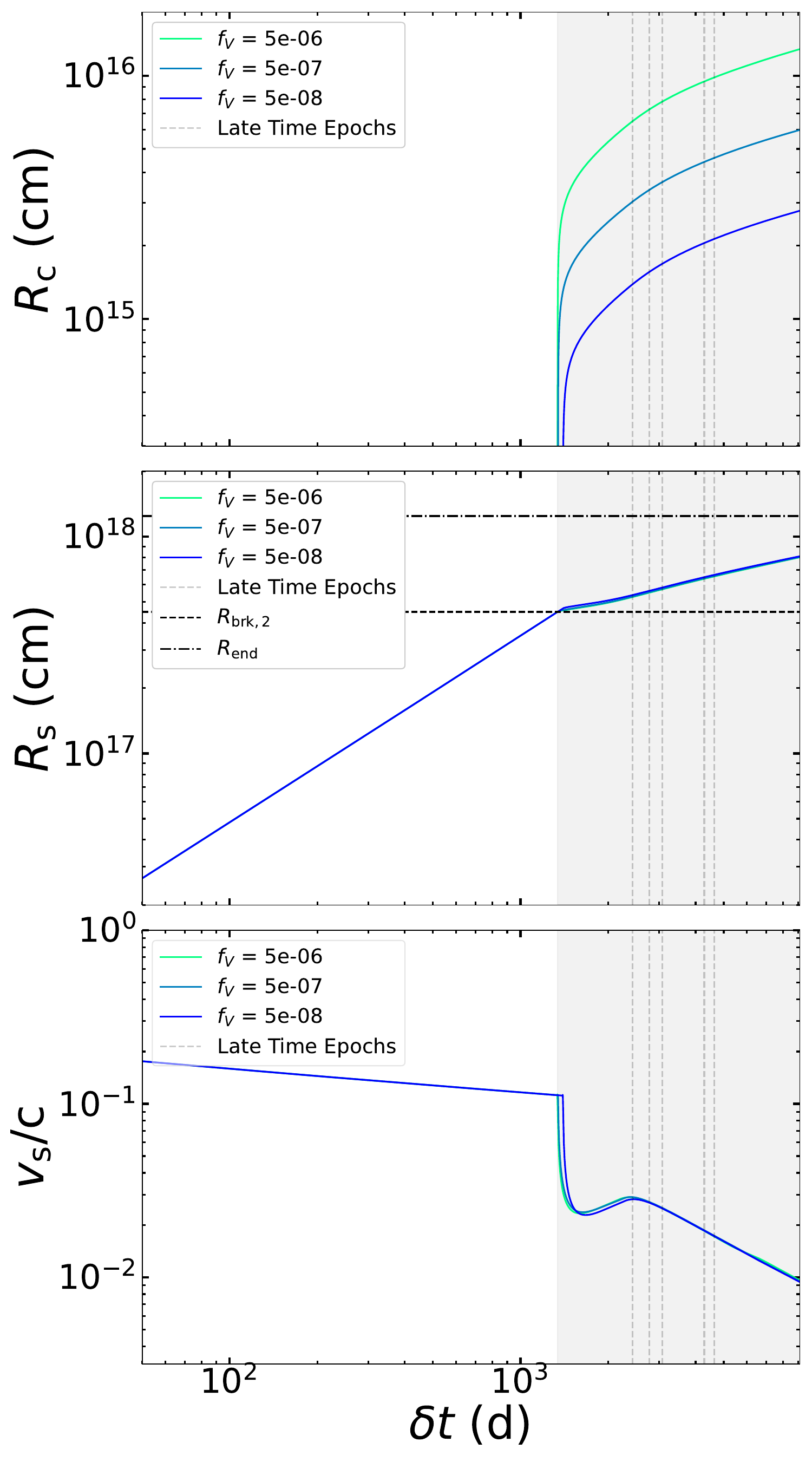}
    \caption{The dynamical evolution of the shock radius ($R_{\rm{s}}$), shock velocity ($v_{\rm{s}}$), and radius of the emitting region ($R_{\rm{c}}$) for the  CSM clump scenario. The different colored lines correspond to different volumetric filling factors of the dense material. The gray dashed lines correspond to the epochs at which we have late-time observations of the event. The dashed and dotted-dashed black horizontal lines in the $R_{\rm{s}}$  plot mark $R_{\rm brk,2}$ and $R_{\rm end}$, respectively. For the times after the shock encounters the clump at $R_{\rm brk,2}$ (marked as the gray region), the dynamics reported only reflect the portion of the shock interacting with the clump. This portion does not reach $10^{18}$ cm at the times of our observations like we expect for the extrapolations of the early shock because it is significantly decelerated by the dense clump. }
    \label{fig:CSM_clump_Rv}
\end{figure}

The best-fitting dynamical evolution for each filling factor can be found in Figure \ref{fig:CSM_clump_Rv}. We plot the evolution of the shock radius $R_{\rm{s}}$, as before, but also include $R_{\rm{c}}$, which represents the radius of a sphere with the same volume as the emitting volume, computed as:
\begin{equation}
    V \equiv \frac{4}{3}\pi R_c^3 = \frac{4}{3}\pi f_V (R_s^3 - R_{\rm{brk,}2}^3) 
\end{equation}
which implicitly defines
\begin{equation}
    R_c \equiv f_V^{1/3} (R_s^3 - R_{\rm{brk,}2}^3)^{1/3} 
\end{equation}
The 1-D shock dynamics for each filling factor are almost exactly the same. Higher filling factors result in larger emitting volumes, but these larger volumes are precisely offset by lower values of $\epsilon_{\rm{B}}$. Notably, this precise balance is only possible because these models \emph{violate} the 
spectral break constraint at $\nu_c$. Instead, in the torus-like model of \S\ref{SubSec:CSMtorus}, $\epsilon_{\rm{B}}$ cannot compensate for the filling factor because each model requires approximately the same magnetic field strength at $\delta t = 7.6$ yr to account for the same cooling break frequency at that epoch. 

With $R_{\rm{c}}< 10^{16}$ cm, clump models satisfy the compact emitting region requirement by requiring large deviations from equipartition: $\epsilon_B \ll\epsilon_e$. After encountering the clump, the velocity of the shock inside the clump sharply decreases, leveling out as the denser parts of the ejecta profile reach the clump, then smoothly decreases.

\begin{figure}
    \centering    
    \includegraphics[width=\linewidth]{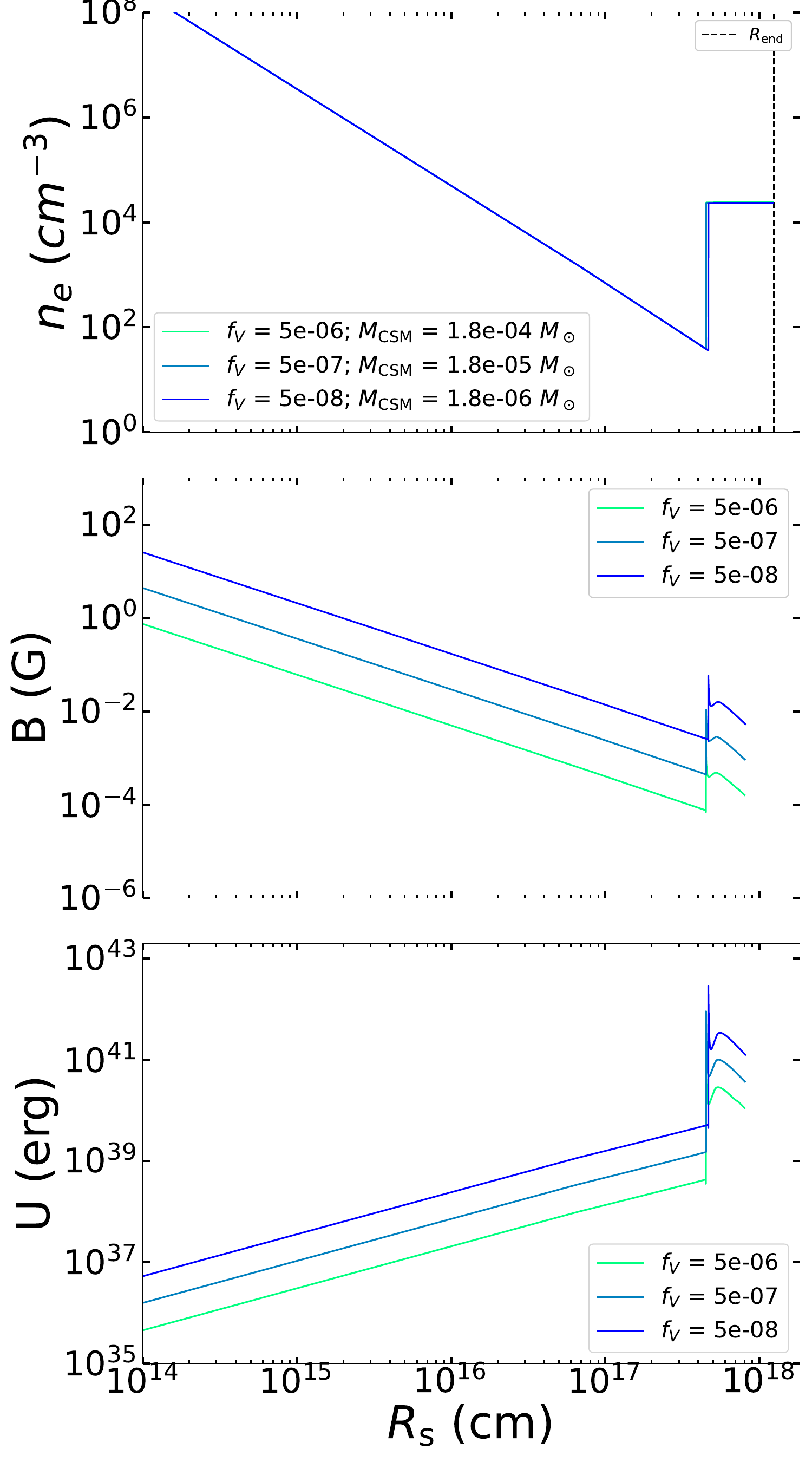}
    \caption{The CSM density ($n_e$) and shock properties ($B$ and $U$) as a function of $R_{\rm{s}}$ for the dense CSM clump scenario. The different colored lines correspond to different volumetric filling factors composed of this dense material. }
    \label{fig:CSM_clump_nBU}
\end{figure}

For each  $f_V$, we also plot the clump $n_e$, $B$, and $U$ as a function of $R_{\rm{s}}$ in Figure \ref{fig:CSM_clump_nBU}. Since the 1-D dynamics for each scenario are nearly identical, the density profiles are all almost identical as well, reaching the clump at $R_{brk,2}\approx 10^{17.7}\,$cm with density $\rho_c\approx 10^{-19.4}\,\rm{g/cm^3}$ ($n_e = 2.3\times10^4\, \rm{cm}^{-3}$). The models all converge to $10^{-5}<f_A < 10^{-4}$, with $f_A$ increasing for decreasing $f_V$. The required CSM masses reported are all $ < 2\times10^{-4} M_\odot$, notably much lower than the CSM mass required in the edge-on torus scenario. The large CSM mass in the torus case was required to slow the shock down and keep it contained within a small radius. In the clump scenario, this requirement is bypassed by having the interaction occur at a separated distance from the explosion site. A large CSM mass is not required for the interaction that produces the observed emission.

The models all account for the same level of free-free absorption, modeling by an end radius of $\log\big (\frac{R_{\rm{end}}}{\rm cm} \big) = 18.09^{+0.22}_{-0.19}$. Any larger $R_{\rm{end}}$ would result in absorption of the observed flux at lower frequencies.
The magnetic field strength and shock energy are both largest for the smallest filling factor, in order to match the observed luminosity within a smaller emitting volume. These two values also exhibit a spike at the density discontinuity; this should be regarded as a numerical artifact and not physical.  For these models, $R_{\rm{end}}\approx10^{18.1}\,\rm{cm}$ 
implies $\rm{NH_{int}}\ge 2\times 10^{22}\,\rm{cm^{-2}}$, which is consistent with the requirements to suppress the X-ray emission from synchrotron radiation (\S\ref{SubSec:modelagnostic}) even in the case of solar composition.

In summary, a small, disconnected clump of dense CSM can match some of the observed late-times properties of \sn. However, it fails to properly account for the observed spectral cooling break at $\nu_c$ in our $\delta t = 7.6$ yr epoch, and overall struggles to reproduce the flux temporal evolution. Furthermore, we also lack a strong astrophysical motivation for the existence of dense clump (which would have a density contrast of a factor $\approx 10^3$ compared to its surroundings), especially one that does not show strong hydrogen features in the optical spectra \citep{Milisavljevic2018}.

We note that while keeping $f_A$ as a free parameter does allow us to not make any assumptions about the geometry of the clump or how it is distributed, we still assume that the material is of uniform density and placed at a specific distribution of radii. This problem could instead be treated as inhomogeneities in the CSM with variations in density, distribution of relativistic electron, and magnetic field strength as described in \cite{Bjjornsson2013,Bjornsson2017,Bjornsson2024}. CSM inhomogeneities have been suggested as explanations for broad synchrotron SED peaks, as observed in \sn{} \citep{De_Colle_inprep}. These inhomogeneities can be modeled by assuming a probability distribution of magnetic field strength, as done for SN 2019oys \citep{Sfaradi2024} and PSNJ1404 \citep{Chandra_2019}. Such a treatment is beyond the scope of this paper, but we note that, based on the radio data, we cannot entirely rule out the possibility of a inhomogeneous CSM for \sn{}.

\section{A Nascent Pulsar Wind Nebula}
\label{Sec:PWN}
Motivated by the lack of a clearly viable CSM-interaction interpretation and the conclusions of \cite{Milisavljevic2018}, \cite{Pandey21}, and \cite{Lazda_VLBI}, we now consider the possibility that the high-frequency radio component is the result of an emerging pulsar wind nebula (PWN). In a PWN, a central rapidly rotating neutron star injects energy into the surrounding ejecta via magnetic winds of charged particles. This central engine has often been proposed as a powering mechanism for superluminous SNe (SLSNe; e.g., \citealt{Kasen10,Dessart2012,Metzger2015,Kashiyama2016, Murase2016,Sukhbold2016,Nicholl_2017,Dessart2018,Murase2021,Eiden2026}), but could also be relevant in broad-line and regular H-stripped core-collapse SNe (e.g., \citealt{Maeda2007,Woosley2010,Kashiymama2016,Taddia2019, Lin_2020,Rodriguez2024}) because of the large inferred $^{56}\rm{Ni}$ mass and explosion energy compared to their H-rich explosion analogs.

Because of degeneracies in the modeling of optical SN light curves and the faintness of late-time optical spectra, the non-thermal emission can be a better probe of the suspected central engine (e.g., \citealt{Kotera13,Metzger14,Murase15,Murase2016,Kashiymama2016,Omand2018}). PWN synchrotron emission results from wind interaction with the innermost ejecta, offering a natural explanation of the small inferred emitting volume and large density of electrons and positrons of \sn{}. The high-density ejecta surrounding the PWN may also explain the lack of observed radio from this component at early times and the continued lack of X-ray detections until late times (see \S\ref{SubSubSec:PWNRadio} and \S\ref{SubSubSec:PWNXrays}). 
In particular, high-frequency radio and millimeter emission in the $\sim 10$--$100$ GHz range has been proposed as a powerful probe of nascent PWN emission from pulsar- or magnetar-powered supernovae, because synchrotron radiation from freshly injected and accumulated electron-positron pairs can emerge once synchrotron self-absorption and free-free absorption in the ejecta become sufficiently weak \citep{Murase2016,Omand2018,Murase2021}. Parameter inference with a full, realistic PWN model would involve time-dependent numerical treatments that are beyond the scope of this paper. Instead, in this section, we investigate the feasibility of a PWN model through (i) observational comparisons to known PWNe, (ii) comparison to existing numerical models, and (iii) analytic estimates based on known properties of \sn.

\subsection{Observational comparisons}
We begin by considering observational comparisons to known PWNe. Typically, the radio spectra of PWNe are characterized by flat spectral indices: $0.0 < \alpha < -0.3$ (e.g., \citealt{Slane2017,GaenslerSlane06}). The observed optically thin radio spectral index of \sn{} ($F_\nu \propto \nu^{-0.31\pm0.02}$) falls within this range of PWNe observations. Explosive transients, on the other hand, are expected to have $\alpha < -0.5$, as diffusive shock acceleration (DSA) models predict $\alpha = -0.5$ and non-linear corrections only steepen the spectrum \citep{Urosevic2019}. The broadband spectra of PWNe typically exhibit a spectral steepening in the millimeter or optical bands \citep{Slane2017}. This steepening is generally greater than $\Delta \alpha = -0.5$, suggesting that a simple cooling break applied to a single power law distribution of electrons is insufficient to describe the emission \citep{GaenslerSlane06}. This steepening is also observed in our data at $\delta t = 8.4$\, yr (see Fig. \ref{fig:model_agnostic_fit}).

\begin{figure*}
    \centering
    \hskip -0.4 cm
    \includegraphics[width=0.8\linewidth]{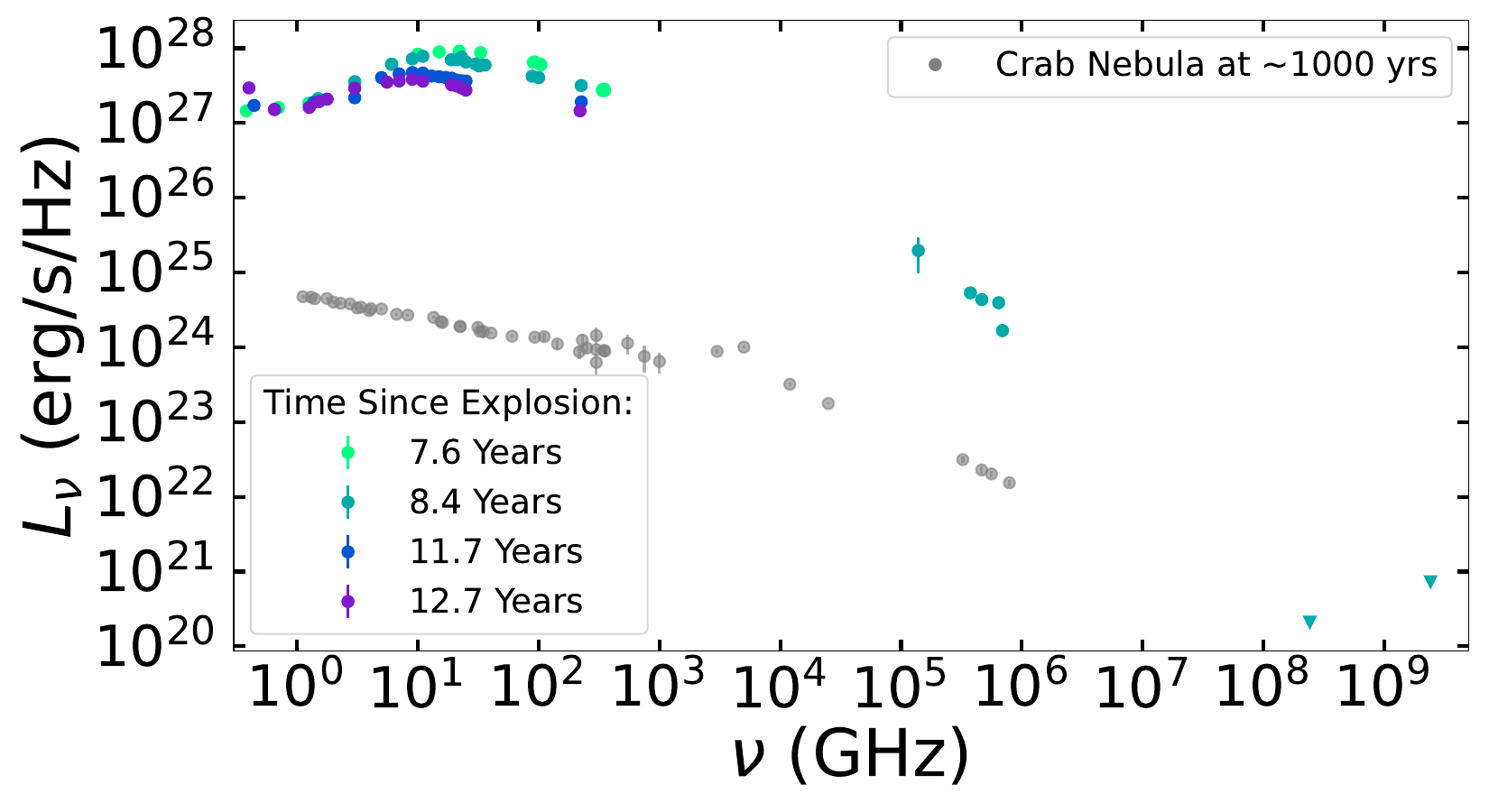}
    \vspace{-15pts} 
    \caption{Late-time ($\delta t > 6$\, yr) broadband SED of \sn{}, compared to the Crab Nebula at $\delta t \sim 1000$\, yr (using data compiled in \citealt{Crab_SED}). Both objects exhibit a shallow spectral index at radio frequencies, which steepens at some point in the mm-optical regime. The Crab Nebula is significantly less luminous than \sn{}, as expected for a much older source.}
    \label{fig:crab_comp}
\end{figure*}

In Figure \ref{fig:crab_comp}, we plot \sn{}, alongside the Crab Nebula, the best-observed PWN. The Crab Nebula exhibits a similarly shallow spectral index in the radio regime, which steepens in the infrared. Notably, the Crab is 2--3 orders of magnitude less luminous than \sn{}; however, this would be unsurprising given the Crab is a factor of $\approx$ 100 times older than \sn{} and PWNe are expected to fade over time \citep{GaenslerSlane06}. Theoretical work also predicts that young PWNe will have luminosities 10--1000 times greater than the present day Crab Nebula \citep{Bandiera1984}, in line with our observations of \sn{}.

While direct observations can provide useful comparisons, we lack any confirmed detections of PWNe as young as our object. The youngest confirmed Galactic PWN  has an inferred age 70--500\,yr  \citep{Smeaton2024}. The emergence of radiation from a PWN has been proposed in some  SNe: 1979C \citep{Patnaude2011,Ahlvind2026,Lundqvist2026}, 1986J (\citealt{Bietenholz_2017,Tanaka2023}), 1987A \citep{Fransson2024}, PTF10hgi \citep{Eftekhari2019},  and VT 1137–0337 \citep{Dong_2023}. Among these: SN 1979C has been observed to have progressively steepening spectral indicies, in contradiction with expectations for a central engine \citep{Lundqvist2026}; SN 1986J has a central radio-emitting component that is consistent with shock interaction with a dense and highly structured CSM \citep{Bietenholz_2017}; PTF10hgi has shown evidence for late-time appearance of hydrogen in the spectra, suggesting that CSM interaction is a more viable explanation of its late-emerging radio emission \citep{Quimby_2018}; for VT 1137–0337, a CSM interaction model cannot be ruled out \citep{Dong_2023}. SN\,1987A has the strongest evidence for a PWN, with observation of accompanying neutrinos at the time of explosion indicating compact object formation and infrared spectroscopy revealing emission features consistent with PWN photoionization (\citealt{Fransson2024}).\footnote{See \cite{Greco21,Greco22} for the association of hard X-ray emission and a PWN in SN\,1987A.} However, SN 1987A lacks a comparable radio signal\footnote{The compact, faint ALMA  radio source is possibly the only evidence for a PWN in the radio bands \citep{Cigan19}.} and originates from a lower energy explosion ($\sim10^{51}$\, erg; \citealt{Woosley1988}) of a completely different progenitor --- a blue supergiant star. Given this sample, we lack any direct observational analogs for \sn{} at early times. 

\sn{}, however, clearly presents many of the observational traits of more-evolved, observed PWNe. In order to compare to early-time PWN emission, we now turn to the predictions of analytical and numerical models. 
\subsection{Comparison to Numerical Models}
\cite{Gelfand09} present a semi-analytic model of the dynamical and radiative evolution of PWNe that explores epochs as early as 10 yr post-explosion. The dynamics are computed assuming that the PWN is surrounded by a thin shell of material, and the set up is applicable to the pre-radiative evolutionary phases of the object's evolution (we are fully within this regime at $\delta t \sim 10$\, yr). The PWN is powered by the NS rotational energy, and 
a simple power-law injection of particle energies is assumed. The radio spectral index is found to be $\alpha \sim -0.3$ for the PWN evolution at late times (10--10000\,yr). Furthermore, these authors find a PWN radius at $\delta t \approx 10$\, yr of $R_{\rm{pwn}} \approx 10^{16} \rm{cm}$ and ejecta velocities on the order of $10^2\,\rm{km\,s^{-1}}$, comparable to our inferred radius and velocity for a spherically symmetric geometry in equipartition in \S\ref{SubSec:inferences}. We caution, however, that the PWN radius, the swept-up shell radius, and the equipartition radius inferred from the radio SED need not be identical.
Overall, the observed and inferred properties of \sn{} are order-of-magnitude consistent with numerical estimates for a young PWN simulated by \cite{Gelfand09}, although the \emph{visibility} of early-time radio emission through the young SN ejecta is more directly affected by the radio transparency (discussed in \S\ref{SubSubSec:PWNRadio}).

The possibility of detecting non-thermal radio and mm emission from a nascent pulsar or magnetar wind nebula embedded in SN ejecta has been further explored in the context of pulsar-driven SNe, SLSNe, and FRB-related scenarios. \citet{Murase2016} showed that persistent synchrotron emission from electron-positron pairs in young PWNe can become detectable once the surrounding ejecta becomes sufficiently transparent, with high-frequency radio and mm bands being especially promising at early epochs. The radio spectral index transits from $\alpha \approx -0.8$ in the fast cooling case to $\alpha \approx -0.3$ in the slow cooling case. This framework was further developed for embryonic SLSN remnants by \citet{Omand2018}, and was later compared with ALMA and NOEMA observations of SLSNe by \citet{Murase2021}.

\cite{Omand2023} utilize this model for \sn{} with the goal of explaining the optical spectra and light curves of \cite{Milisavljevic2013,Milisavljevic2018}, with non-local thermodynamic equilibrium (NLTE) spectral synthesis computations. The inferred values of surface dipole $B$ field ($B_* = 3-4\times10^{14}$\, G) and initial pulsar period ($P_0 = 15$\, ms) are comparable to those of \cite{Pandey21} ($P_0 = 18$\,ms, $B_* = 8 \times 10^{14}$\, G) and fall within our allowed region (see  \S\ref{subsubsec:PWNpara}, Figure \ref{fig:P_Pdot}). \cite{Omand2023} also offer a prediction of the radio evolution of \sn{} based on the best-fitting parameters from their optical spectra modeling. Comparing to our data, we find that the predicted optically thin radio emission at $\delta t = 10\,$yr is a good representation of the observations  both in terms of flux density and spectral index. 
However, the predicted peak frequency is significantly lower than observed, by a factor of $\approx 5$. The theoretical optically thick spectrum is also very steep, (presumably dominated by free-free absorption), 
compared to the observed $F_\nu \propto \nu^{0.5}$. This discrepancy can be partially attributed to the additional contribution of the low frequency component in the observed spectrum. As noted in \cite{Omand2023}, clumping in the ejecta may also have significant, unaccounted for effects on the spectrum. Nevertheless, based on these numerical models a PWN explanation remains plausible for our observations. In Section \ref{subsubsec:PWNpara}, we will compare the PWN parameters inferred from these models to constraints we place based on the radio data. 

We end by noting that the large $P_0$ inferred by \cite{Pandey21} and \cite{Omand2023} imply small rotational kinetic energy $\approx 6\times 10^{50}$\,erg, which is $\ll$ the $E_{\rm{k}}$ carried by the ejecta. A different energy source is needed to explain the large $E_{\rm{k}}$ of \sn{} in these models. 

\subsection{Analytical Estimates and Considerations}

\subsubsection{\texorpdfstring{Radio detectability of the PWN at $\delta t\approx 10$\,yr}{Radio detectability of the PWN at delta t approx 10 yr}}\label{SubSubSec:PWNRadio}
We first investigate whether the radio  emission from a central PWN would be detectable on our observed timescales. To compute the timescale for our object, we employ the analytic formalism used in the transparency calculation of \cite{Metzger_2017}. Following their Equation 11, we expect oxygen-dominated (Z = 8, A = 16) ejecta to be transparent to free-free absorption at the following time post explosion:
\begin{equation}
\begin{split}
t_{\rm{ff}} &= 6.25\, \bar g_{\rm{ff}}^{1/5} f_{\rm{ion}}^{2/5}
\bigg(\frac{T_{\rm{ej}}}{10^4 \rm{K}}\bigg)^{-3/10}
\bigg(\frac{M_{\rm{ej}}}{5\,{\rm M}_\odot}\bigg)^{2/5} \\
&\quad \times
\bigg(\frac{v_{\rm{ej}}}{0.1c}\bigg)^{-1}
\bigg(\frac{\nu}{1\,\rm{GHz}}\bigg)^{-2/5} \rm{yr},
\end{split}
\end{equation}

\noindent
where $\bar g_{\rm{ff}} = 5$ is the Gaunt factor at radio frequencies,  $f_{\rm{ion}}$ is the ionization fraction of the ejecta, $T_{\rm{ej}}$ is the ejecta electron temperature, $M_{\rm{ej}}$ is the ejecta mass, 
and $v_{\rm{ej}}$ is the ejecta velocity. 
For \sn{}, we assume  $T_{\rm{ej}}=10^{4}\,$K, $M_{\rm{ej}}=5\,{\rm M}_\odot$,
and use the inferred total kinetic energy from Section \ref{Subsub:12auearlytime}, $E_{\rm k} = 10^{52.5}$\,erg, to calculate the ejecta velocity ($E_{\rm k} = \frac{1}{2} M_{\rm{ej}}v_{\rm{ej}}^2$).  
We first verify that PWN emission would be completely free-free absorbed during the early-time observations ($\delta t \leq 190$\,d). To do this, we calculate $t_{\rm{ff}}$ over a range of ionization fractions $f_{\rm{ion}} = [0.1,1]$ 
for $\nu = 37$ GHz, the maximum observed frequency at early times (i.e., the frequency that will see the smaller free-free absorption). We find that $t_{\rm{ff}} > 190$\,d over the range of computed ionization fractions. While emission at this frequency may not be absorbed for $f_{\rm{ion}} < 0.1$, we know that there is some degree of ionization because of the ionized emission features in the early time optical spectra \citep{Milisavljevic2013}. Higher values of ejecta temperature may also lower $t_{\rm{ff}}$; however, at $T_{\rm{ej}} = 10^5$\, K, $t_{\rm{ff}} > 190$\,d for all $f_{\rm{ion}} > 0.25$. 
Next, we calculate whether the ejecta are transparent to free-free absorption during the late-time observations ($\delta t \geq 6.7$\,yr). We compute $t_{\rm{ff}}$ over the same range of ionization fractions for $\nu = 16$ GHz, the observing frequency of the earliest late-time epoch. We find that $t_{\rm{ff}} < 6.7$\, yr across all ionization fractions, even for a low ejecta temperature of $T_{\rm{ej}} = 10^3$\, K. We therefore conclude that the emergence of the high frequency component in radio on our observed timescales is consistent with a PWN interpretation.

Our conclusion is strengthened by previous theoretical and numerical treatments of nascent PWN radio emission. \citet{Murase2016} calculated the attenuation of radio waves by synchrotron self-absorption in the PWN and free-free absorption in the surrounding ejecta, and showed that higher-frequency radio waves can escape more readily than GHz emission and can be used as signatures of magnetar-driven SNe including SLSNe. \citet{Omand2018} find a timescale of $\sim 10-100$ yr for transparency to 1\,GHz emission for SLSNe while the ejecta becomes transparent earlier at higher frequencies, on timescales of order $\sim 1-10$ yr at 100 GHz \citep[see also][]{Murase2021}. \cite{Margalit_2018_CLOUDY} present a numerical calculation of the ionization states of the SN ejecta using the photoionization code \texttt{CLOUDY}. They also find that for a sample of SLSNe, the expected timescale for ejecta transparency to free-free absorption is on the order $\approx$10--100\,yr. These SNe have typical explosion energies of $E_{\rm k} \sim 10^{52}$\, erg, comparable to \sn{}, while having larger $M_{\rm ej}\sim10M_\odot$. This larger $M_{\rm ej}$ will only increase $t_{\rm ff}$, making these values an overestimate for our source.

\subsubsection{\texorpdfstring{X-ray suppression of the PWN  at $\delta t\approx 10$\,yr}{X-ray suppression of the PWN  at delta t approx 10 yr}} \label{SubSubSec:PWNXrays}

In the PWN scenario, the lack of detectable X-ray emission at late times can be due to (i) photoelectric absorption by the SN ejecta, or (ii) intrinsic faintness, i.e., a spectral break between the optical and the X-ray regime, which is often observed in PWNe and the nature of which is still a  matter of debate (e.g., \citealt{GaenslerSlane06}). In the following we discuss the X-ray absorption scenario. Radio transparency does not necessarily imply X-ray transparency, because free-free absorption depends on the ionized gas density whereas soft X-ray photoelectric absorption is controlled by the ejecta column density and ionization state of the ejecta. Thus, a PWN can become visible at the high-frequency radio band while its soft X-ray emission remains attenuated by the ejecta.

From \S\ref{SubSec:modelagnostic}, for ejecta-like composition, a neutral hydrogen column density $\rm{NH_{\rm{int}}}>6.2\times 10^{20}\,\rm{cm^{-2}}$ is needed to suppress the X-ray emission below detectability, which is easily achieved by \emph{neutral} SN ejecta at $\approx 3000$\,d. However, the pulsar wind can lead to high-energy X-ray photons (e.g., the wind-injected electron-positron pairs cool via synchrotron and inverse Compton emission) that can ionize the SN ejecta from the inside.\footnote{We note that the SN ejecta will be eventually collisionally ionized by the passage of the reverse shock. However, this happens on a Sedov-Taylor time scale of hundreds of years  (e.g., \citealt{Reynolds_Chevalier1984,Truelove99}) that is much larger than the timescales considered in this paper, and thus not relevant.} The result is a time- and spatially-dependent ionization state of the ejecta, which directly impacts the X-ray opacity  (with ionized ejecta having lower X-ray opacity). The problem has been treated by different authors in the literature (\citealt{Kotera13,Murase15,Metzger14}) and an ionization breakout has been invoked to explain the late UV brightening of ASASSN-15lh \citep{Margutti17-15lh}. Here we use the results from the numerical simulations of \cite{Margalit_2018_CLOUDY}, that self-consistently track the ejecta ionization state with time as the spin-down power declines.

\cite{Margalit_2018_CLOUDY} concluded that for oxygen-dominated composition and ionizing luminosity $\propto t^{-2}$ that is relevant to our problem (see \S\ref{subsubsec:PWNpara}), the ejecta is either ionized during a spin-down time scale $\tau_0$, or never achieve a state of complete ionization (and hence transparency to the X-rays). For efficient deposition of energy into the ejecta 
$\tau_0$ needs to be comparable to the time scale of evolution of the SN (e.g., $\tau_0\approx t_{\rm{rise}}$), 
and thus $\ll 10$\,yr. This is consistent with the value inferred from the optical light curves in \cite{Pandey21}: $\tau_0 = 24.49 \pm 2.12$\,d. 

For a ratio of absorptive to scattering opacity in the ejecta $\eta_{\rm thr} >1$ that applies here (Eq. A7 of \citealt{Metzger14}), the ionization time scale of the SN ejecta is:
\begin{align}
\label{Eq:tion}
t_{\rm ion} \approx 340\,{\rm d}
&\left ( \frac{M_{\rm ej}}{5\,\rm{M_{\odot}}}\right )
\left ( \frac{v_{\rm ej}}{10000\,\rm{km~s^{-1}}}\right )^{-3/2}
\left (\frac{T}{10^5\,\rm{K}}\right)^{-0.4} \nonumber \\
&\times \left(\frac{X_{Z}}{0.35}\right)^{1/2}
\left(\frac{E_{\rm{ion}} }{10^{52}\,\rm{erg}}\right)^{-1/2} \nonumber  \times Z_8^{3/2}
\end{align}
where $T$ is the temperature of electrons in the recombination layer; $X_Z$ is the mass fraction  of elements with atomic number $Z = 8Z_8$ in the ejecta expanding with velocity $v_{ej}$. The calculation above suggests that $t_{\rm ion}> \tau_0$, even assuming an ionizing energy $E_{\rm{ion}}= L_{\rm{ion}}\times t$  comparable to the total ejecta kinetic energy (where $L_{\rm{ion}}$ is some fraction of the spin-down luminosity).  It is thus unlikely to achieve complete ionization of the SN ejecta on a spin-down time scale, from which we conclude that it is plausible that the lack of X-ray detection at $\approx 10\,\rm{yr}$ is due to photo-electric absorption.  

\subsubsection{Analytical Inferences on the PWN Parameters}
\label{subsubsec:PWNpara}
\begin{figure*}
    \centering
    \includegraphics[width=0.75\linewidth]{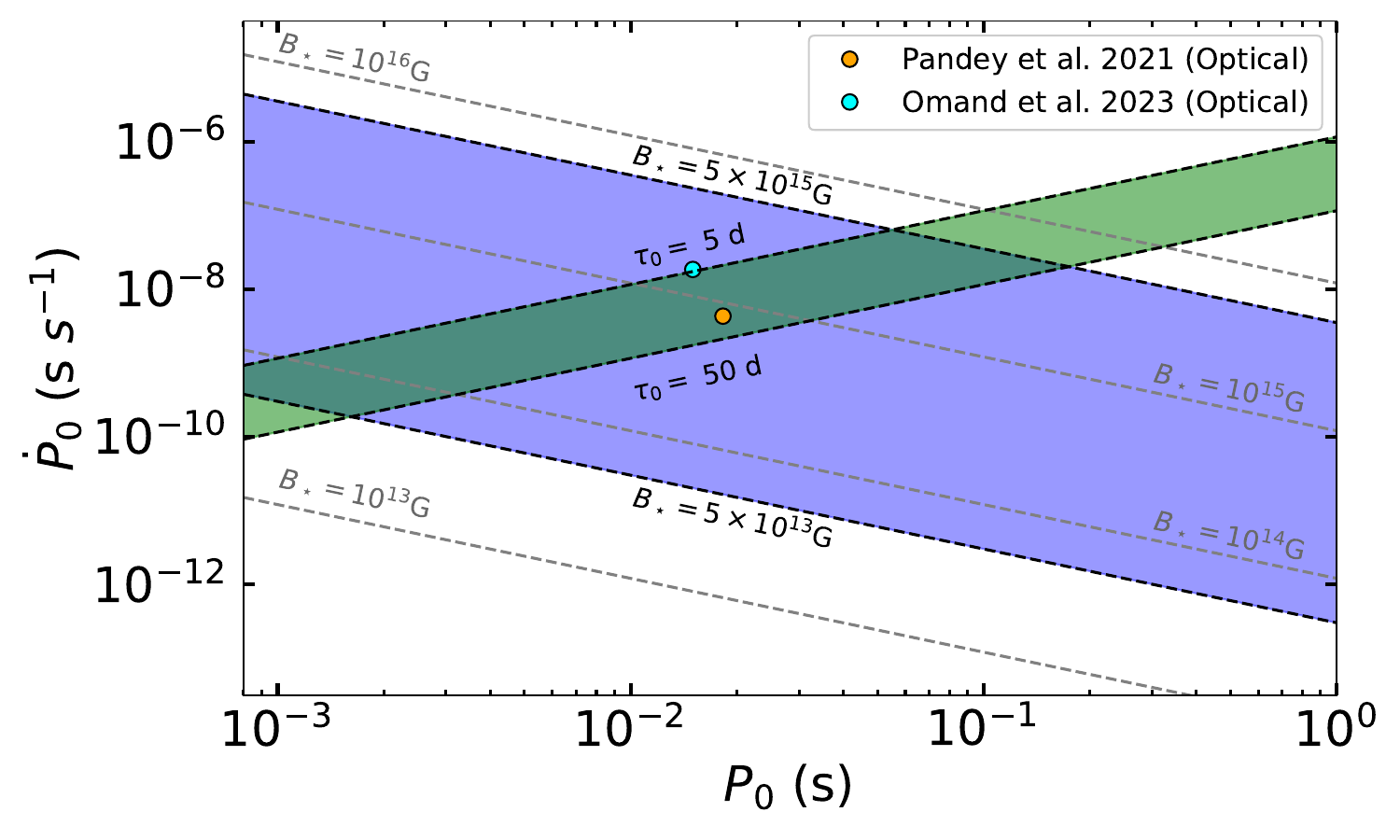}
    \caption{
    Motivated parameter space for the initial spin period, \(P_0\), and period derivative, \(\dot P_0\), in the PWN emission scenario. The green region shows the reference range motivated by the optical rise time if the central engine contributed significantly to the optical luminosity. The blue region shows the parameter space inferred from the observed radio luminosity using an effective radio efficiency range $\eta_{\rm r,eff}=[1,10^{-4}]$. These regions should be interpreted as illustrative rather than model-independent constraints. The parameters inferred in the optical studies by \cite{Pandey21} and \cite{Omand2023} are also plotted as an orange and cyan point, respectively. The gray dashed lines represent lines of constant surface dipole magnetic field strength, $B_\star$.
    The allowed region (i.e., overlap) includes many combinations of $P_0$ and $\dot P_0$ that correspond to magnetar engines. 
     }
    \label{fig:P_Pdot}
\end{figure*}

Energy injection from a central neutron star may come from two sources: rotational energy, in the form of magnetized winds; or magnetic energy from the interior magnetic field, in the form of magnetar giant flares. In this analysis, we focus on rotationally-powered energy injection because of the larger energy budget (up to  a few $\sim 10^{52}$\,erg for ms period magnetars,  compared to the more modest $\sim 10^{50}$\,erg of magnetic energy for a strong B field on the order $\sim10^{16}$\,G) and because this scenario is expected to dominate at early times \citep{Margalit_2018_CLOUDY}.  In this case, 
the PWN luminosity is related to the evolution of the spin-down luminosity of the pulsar, which evolves as \citep{GaenslerSlane06}: 
\begin{equation}
\label{eq:L_sd}
    L_{\rm{sd}} = \dot E_0 \bigg(1 + \frac{t}{\tau_0}\bigg)^{-\frac{n+1}{n-1}},
\end{equation}
where $\dot E_0$ is the initial spin-down luminosity at time $t_0$, $n$ is the braking index of the pulsar with angular frequency $\Omega$ (such that $\dot \Omega \propto \Omega^{-n}$), and $\tau_0$ is the initial spin down timescale \citep{GaenslerSlane06}: 
\begin{equation}
\label{eq:tau0}
    \tau_0 = \frac{P_0}{(n-1)\dot P_0}, 
\end{equation}
where $P_0$ is the initial pulsar period, and $\dot P_0$ is the initial period derivative. For $t \ll \tau_0$, $L_{\rm sd}$ is constant in time, whereas for $t \gg \tau_0$,
\begin{equation}
\label{eq:L_sd_asymptotic}
        L_{\rm{sd}} \approx \dot E_0 \bigg(\frac{t}{\tau_0}\bigg)^{-\frac{n+1}{n-1}}.
\end{equation}
The spin-down timescale can be related to the surface dipole magnetic field strength, $B_*$, as in \cite{Kasen10}:
\begin{equation}
\label{eq:B_*}
\tau_0 = 1.3 \bigg(\frac{B_{*}}{10^{14}\,{\rm G}}\bigg)^{-2} \bigg(\frac{P_0}{10\, \rm {ms}}\bigg)^{2} \rm{yr},
\end{equation}
where it is assumed  $\rm{sin}^2\alpha =1/2$ ($\alpha$ being the angle between the rotational axis and the magnetic dipole axis), $B_{*}$ is the NS dipole  magnetic  field at the poles, the NS moment of inertia  is $I\approx 10^{45}  \rm{g\,cm^{2}}$  and the NS radius is $R_{\rm{NS}}=10$ km. $B_\star$ differs from the magnetic field that powers the synchrotron emission of the nebula $B_n$, which was probed in our model agnostic fit (\S\ref{SubSec:modelagnostic}).  From \S\ref{SubSec:modelagnostic} we find that $B_n$ decreases monotonically with time, as expected from the adiabatic expansion of the nebula and/or the rapid decay of injected energy with time. The resulting synchrotron emission can either trace in time the energy injection, or not.  We therefore distinguish between \(B_\star\), which controls the spin-down power of the neutron star, and \(B_n\), which controls the synchrotron emission from the nebula. The two quantities are not directly interchangeable, because \(B_n\) depends on the magnetization of the wind, the integrated injection
history, and the subsequent expansion of the nebula.

We compute the optically thin luminosity, $\nu L_\nu$, and find that it scales approximately as $L(t)\propto t^{-2}$ thus implying that, if the observed luminosity reflects $L_{\rm sd}$, we are in the regime $t \gg \tau_0$ and $n \approx 3$ (Equation \ref{eq:L_sd_asymptotic}).
The $n=3$ braking index is typically used in the literature for magnetar spin-down \citep{Ostriker1969,Kasen10}, as it corresponds to the expected breaking in the case of magnetic dipole radiation \citep{Ostriker1969,Spitkovsky2006}. This reasoning is consistent with (but cannot be considered a proof of) 
our assumption that the PWN is primarily rotation powered as the observed radio luminosity does \emph{not} have to mirror the spin-down luminosity (see below). 

For an independent probe of $\tau_0$, we turn to the observed optical evolution of the source. If the same central engine contributed significantly to the bright optical luminosity, one expects \(\tau_0\) to be comparable to the photon diffusion time,
\(t_{\rm diff}\), which is observationally related to the SN rise time. From \cite{Milisavljevic2013},  $t_{\rm{rise}}=16.5 \pm 1.0$\, d, which is $\ll$ than all our late-time epochs - independently indicating that we are in the $t \gg \tau_0$ regime.

For comparison with simple spin-down estimates, we define an effective radio efficiency, \(\eta_{\rm r,eff}\equiv L_{\rm obs}/L_{\rm sd}\). 
This quantity should not be interpreted as a constant bolometric conversion efficiency, because in time-dependent PWN calculations, the observed radio emission depends on the accumulated population of
relativistic pairs, the nebular magnetic field, adiabatic and radiative losses, synchrotron self-absorption, and free-free absorption in the surrounding ejecta (e.g., \citealt{Murase2016,Omand2018,Murase2021,Vurm_2021}). 
This is particularly relevant when most of the rotational energy is injected at \(t\ll t_{\rm obs}\), as expected for the regime of interest \(\tau_0\ll t_{\rm obs}\).
In this regime, the nebular magnetic energy may evolve approximately as \(\epsilon_B \Delta {\mathcal E}_{\rm rot} \sim V_n B_n^2/8\pi\), where \(\Delta {\mathcal E}_{\rm rot}\sim \dot E_0 \tau_0\) is the total energy injected at early times and \(V_n\propto R_n^3\). For approximately homologous expansion with \(R_n\propto t\), this gives
\(B_n\propto t^{-3/2}\), ignoring more details such as decay. Then, if the number of radio-emitting pairs is approximately conserved (see \S\ref{SubSubSec:sourcepairs}), the optically thin synchrotron flux scales as \(F_\nu\propto N_e B_n^{(p+1)/2}\), which can naturally produce a decline close to \(t^{-2}\) for \(p=1.6\), as observed.

When $t \gg \tau_0$ we cannot place valuable constraints on $P_0$ and $\dot P_0$ based purely on the evolution of luminosity, as the normalization of $L_{\rm sd}$ constrains a product of $P_0$, $\dot P_0$ and $\dot E_0$, but not the individual terms. To break this degeneracy, we conservatively adopt \(5\,{\rm d}<\tau_0<50\,{\rm d}\) as a motivated reference range based on the rise time from \cite{Milisavljevic2013},
rather than as a model-independent constraint.
This $\tau_0$ range translates  into an allowed range of $P_0$/$\dot P_0$ ratios via Equation \ref{eq:tau0}  (green-shaded region in Figure \ref{fig:P_Pdot}).

These parameters are further constrained by comparing $L_{\rm sd}$ with the observed luminosity $L_{\rm obs}$. Since in our physical scenario  \citep{GaenslerSlane06}:
\begin{equation}
\label{eq:Erot}
    L_{\rm{sd}} = \dot E_{\rm{rot}} = 4\pi^2I\frac{\dot P}{P^3},
\end{equation}
(where $I$ is the moment of inertia of the pulsar and $I \approx 10^{45}$ g\,$\rm{cm}^{-2}$ for standard values  of $M = 1.4\, \rm{M_\odot}$ and $R = 10$\, km), and $ L_{\rm obs} \equiv \eta_{\rm r,eff} L_{\rm sd}$ with $\eta<1$, it follows that for $n\approx 3$ and $t \gg \tau_0$:
\begin{equation}
\label{eq:L_obs}
    L_{\rm obs} \equiv \eta_{\rm r,eff} L_{\rm sd} \approx \pi^2 I \eta (P_0 \dot P_0)^{-1} t^{-2}
\end{equation}

Following \cite{Dong_2023}, a lower limit on $\eta_{\rm r,eff}$ can be placed by observational surveys. Radio observations of several 1000s of years old PWNe \citep{Frail_1997} imply $\eta_{\rm r,eff}^{\rm{min}} \sim 10^{-4}$ among the population of detected pulsars. Since $\eta_{\rm r,eff}$ is expected to decrease over time, we adopt 
$\eta_{\rm r,eff}^{\rm{min}} > 10^{-4}$, which leads us to conservatively consider the range of efficiencies $1 \geq \eta \geq 10^{-4}$.
The observed radio luminosity\footnote{The radio luminosity is calculated by integrating the higher frequency component of \S\ref{SubSec:modelagnostic} over the typical range of 100 MHz to 100 GHz \citep{GaenslerSlane06}.} and Equation \ref{eq:L_obs} constrain the product of $P_0 \dot P_0$ as in Figure \ref{fig:P_Pdot}, blue-shaded region. Lines of constant $B_\star$ are depicted in gray, computed by combining Equations \ref{eq:tau0} and \ref{eq:B_*}.
Interestingly, the PWN parameters inferred by \cite{Pandey21} and \cite{Omand2023} (orange and cyan filled circle, respectively, in Fig. \ref{fig:P_Pdot}) fall within the overlapping region of allowed parameter space. With $B_\star = (0.5-50)\times 10^{14}$\,G, our central source likely qualifies as a magnetar. 
 We note that this result does differ from the corresponding figure in \cite{Lazda_VLBI}, which instead focuses on the scenario where $t < \tau_0$. Our allowed regions fall within the ``Unconstrained'' region of their Figure 4, which they do not exclude, but rather defer to future works. 

\begin{figure}[t!]
    \centering
    \includegraphics[width=\linewidth]{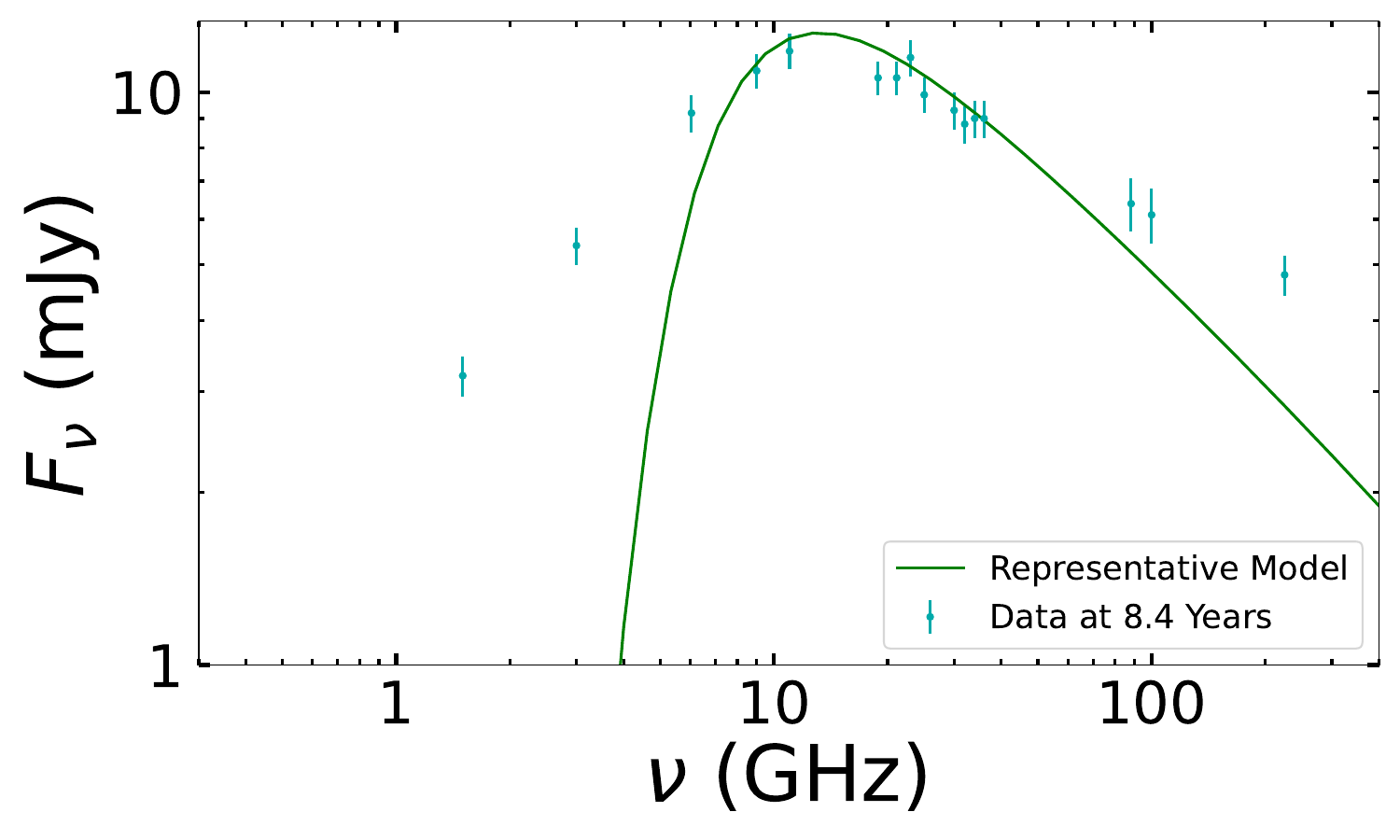}
    \caption{Observed late-time radio SED of \sn{} compared with a representative nascent PWN synchrotron model \citep{Murase2016}. The model is not a formal fit, but demonstrates that high-frequency radio emission from a young PWN can reach the observed luminosity scale once absorption in the nebula and ejecta becomes sufficiently weak. This model utilizes a pulsar with $P_0 = 15 $\,ms and $B_*=10^{14}$\,G and explosion parameters of $M_{\rm ej}=7 \rm \,M_\odot$ and $E = 10^{52}$\,erg with $p=1.6$, $\gamma_b=3\times{10}^4$, $\epsilon_B=0.1$ and $f_{\rm ion}=0.5$.}
    \label{fig:PWN}
\end{figure}

To place our source in the context of detailed nascent PWN calculations, we compare the observed radio SED with a representative time-dependent PWN model based on the framework of \citet{Murase2016,Omand2018,Murase2021}. A selected model with pulsar parameters  $P_0 = 20 $\,ms and $B_*=10^{14}$\,G and explosion parameters of $M_{\rm ej}=7 \rm \,M_\odot$ and $E = 10^{52}$\,erg with $p=1.6$, $\gamma_b=3\times{10}^4$, $\epsilon_B=0.1$ and $f_{\rm ion}=0.5$ is plotted in Figure \ref{fig:PWN}. The comparison is not intended as a formal parameter fit, but it illustrates that a young PWN can reproduce high-frequency radio emission of the observed data. However, we emphasize that the representative PWN model does not explain the entire radio spectrum, especially at low frequencies. This is not surprising because the low-frequency side is particularly sensitive to synchrotron self-absorption, free-free absorption, ejecta clumping, asphericity, inhomogeneous ionization, and possible contamination from a separate CSM-interaction component. These effects can shift or broaden the apparent spectral turnover, while leaving the optically thin high-frequency component largely consistent with a nascent PWN interpretation. A quantitative fit including these effects is beyond the scope of this work, and we therefore use
the comparison only to demonstrate that a young PWN can plausibly produce the observed high-frequency radio emission.\footnote{We similarly attempt to find a representative time-dependent model for an electron-ion nebula, using the formalism of \cite{Margalit+Metzger_PWN}. We find that while we are able to match the evolution of the peak flux and frequency, both the optically thick and optically thin slopes are far too steep to match the data.}

\subsubsection{Source of Injected Electron-Positron Pairs}
\label{SubSubSec:sourcepairs}

A key element of the PWN model to address is what process produces the radio-emitting electrons (or electron-positron pairs).
We start with an estimate of the total number of emitting leptons in the nebula  $N_e$ at the time of our observations. 
Using the equation for optically thin synchrotron flux from 
\cite{chevalier98}:
\begin{equation}
\label{eq:F_thin}
    F_\nu = \frac{4\pi f_VR^3}{3D^2}c_5 n_0 B^\frac{p+1}{2}\bigg(\frac{\nu}{2c_1}\bigg)^{\frac{1-p}{2}},
\end{equation}
where $B\equiv B_n$ is the nebula field, $n_0$ is a normalization on the density of  relativistic electrons per unit energy, such that: 
\begin{equation}
\label{eq:n_0E}
    \frac{dn}{dE} = n_0E^{-p}
\end{equation}
extending between $E_{\rm min}$ and $E_{\rm max}$.
It follows that:
\begin{equation}
   N_e = \frac{F_\nu D^2}{c_5(p)} B_n^{-(1+p)/2} \bigg(\frac{2c_1}{\nu}\bigg)^{(1-p)/2} \frac{ (E_{\rm max}^{1-p} - E_{\rm min}^{1-p})}{1-p}
\end{equation}
Considering a range of energies $(1-10^7)m_ec^2$
and selecting a representative optically thin spectrum at $\delta t = 8.4$\,yr, $F_\nu(\nu =30 \rm\, GHz) = 9.3\,mJy$ with $p=1.6$, we infer:

\begin{equation}
\label{eq:N_e_obs}
        N_e \approx 7\times10^{52} \bigg(\frac{B_n}{0.1\rm G}\bigg)^{-1.3}, 
\end{equation}
corresponding to a number density of radiating leptons: 
\begin{equation}
        n_e \approx 3\times10^4 \bigg(\frac{f_V}{0.5}\bigg)^{-1}\bigg(\frac{R_n}{10^{16}\rm cm}\bigg)^{-3} \bigg(\frac{B_n}{0.1\rm G}\bigg)^{-1.3} \rm{cm}^{-3}
\end{equation}

We compare these estimates to the ``classic'' picture of direct pulsar lepton injection at multiplicity $\mu_{\pm}$, i.e.,  the Goldreich-Julian (GJ) injection rate $\dot N_{GJ}$ \citep{GoldreichJulian1969}, where $\mu_{\pm}$ is the number of leptons per primary GJ particle. 
Assuming an outflow speed of $\approx c$ of GJ particles from a polar cap area $A_{\rm pc}$, we estimate: 
\begin{equation}
    \dot N_{\rm GJ} = \frac{\rho_{\rm GJ}}{e}A_{\rm pc} c\approxeq \frac{B_*R_{\rm NS}^3 \Omega^2}{2ec},
\end{equation}
where $e$ is the electron charge. We can relate this quantity to $L_{\rm sd}$ expressed in terms of magnetic dipole luminosity as in \cite{Kasen10}: 
\begin{equation}
    L_{\rm sd} = \frac{B^2R_{\rm NS}^6\Omega^4}{12c^3},
\end{equation}
therefore,
\begin{equation}
\label{eq:N_GJ_dot}
    \dot N_{\rm GJ}  \approxeq L_{\rm sd}^{1/2} \frac{\sqrt{3}c^{1/2}}{e}.
\end{equation}
$\dot N_{\rm GJ}$ is related to the lepton injection rate $\dot N_{\pm}^{\rm inj}$ by the multiplicity factor, $\mu_\pm$, such that:
\begin{equation}
\label{eq:N_e,inj}
    \dot N_{\rm \pm}^{\rm inj} = \mu_\pm \dot N_{\rm GJ}
\end{equation}

Over time, $\dot N_{\pm}^{\rm inj}$ will evolve like $ \dot N_{\pm}^{\rm inj} \propto L_{\rm sd}^{1/2}$:
\begin{equation}
\label{eq:Ne_inj(t)}
        \dot N_{\pm}^{\rm inj}(t)= \dot N_{\rm \pm,0}^{\rm inj} \bigg(1 + \frac{t}{\tau_0}\bigg)^{-1}.
\end{equation}

We compute the maximum  $E_{0}$ allowed in the overlapping region of Figure \ref{fig:P_Pdot} and find $\dot E_{0}^{\rm max} \approx 1\times10^{46}$\,erg/s, from which we derive:  
\begin{equation}
    \dot N_{\pm,0}^{\rm inj} \le 6\times10^{42} \Big ( \frac{\mu_\pm}{10^5}\Big )\, \rm s^{-1}.
\end{equation}
We normalize $\mu_\pm = 10^5$, the order of magnitude of the theoretical maximum multiplicity \citep{TimokhinHarding2019} and observational measurements of the Crab Pulsar \citep{Hibschman_2001}.
Integrating Equation \ref{eq:Ne_inj(t)} from $t=0$ to $t = 8.4$\,yr for $\tau_0 \approx 24.5$\,d, the total number of injected leptons is:
\begin{equation}
         N_{\pm}^{\rm inj} \le 6\times10^{49} \Big (\frac{\mu_\pm}{10^5}\Big ),
\end{equation}
We conclude that  even when we maximize the $L_{\rm sd}$ and $\mu_\pm$, the directly injected particle count is a factor $\sim 10^3$ below the $N_e\approx 10^{53}$ radiating leptons number required by observations (Eq. \ref{eq:N_e_obs}). We thus consider alternative scenarios, starting with giant magnetar flares.

A magnetar giant flare has an associated baryonic mass ejection into the nebula that could supply the needed electrons. Relativistic hydrodynamical simulations imply that these flares eject crustal material characterized by an electron fraction of $Y_e \approx 0.4$ \citep{Jakub2024}. Recent work has inferred a baryonic ejecta mass of $M_b \sim 10^{-6} \rm \,M_\odot$ from the SGR 1806-20 magnetar giant flare \citep{Patel_2025}. Normalizing based on these parameter, the total number of electrons from a magnetar flare is:
\begin{equation}
    N_{\rm e,flare} \approx5\times10^{50} \bigg(\frac{Y_e}{0.4}\bigg)\bigg(\frac{M_b}{10^{-6} \rm \,M_\odot}\bigg),
\end{equation}
therefore, giant flares appear to be insufficient to account for the total number of required electrons. 

Next we consider the scenario proposed in \cite{MetzgerPiro14} (see also \citealt{Vurm_2021}), whereby the enormous energy density in the magnetic fields and radiation in the early-time nebula generates high-energy photons that produce a high abundance of electron-positron pairs. A key parameter is the pair multiplicity factor, $Y \equiv L_{\rm sd}/(m_ec^{2})$, defined as the fraction of the spin-down power placed into electron/positron rest mass.  We take $Y = 0.1$, corresponding to an estimate of maximum pair yield for a ``saturated'' cascade \cite{Svensson1987}. We begin by estimating a freeze-out timescale, $t_{\rm fo}$ after which the number of pairs created through this cascade ceases to change significantly as a result of the nebula expansion. At $t_{\rm fo}$ the equilibrium timescale of pair annihilation and creation ($t_{\rm eq}$) is comparable to the nebula expansion timescale: $t_{\rm eq}(t_{\rm{fo}}) \approx R_{\rm fo}/v$.  Following \cite{MetzgerPiro14},  the optical depth of pairs through the nebula at freeze-out is:
\begin{equation}
\label{eq:tau_es(t_fo)}
    \tau_{\rm es}^n(t_{\rm fo}) = \frac{16}{3} \beta,
\end{equation}
where $\beta \equiv v/c$ and
$ \tau_{\rm es}^n(t)$ is determined by the balance between pair  creation and annihilation (\citealt{MetzgerPiro14}, their Eq. 14).  
We consider $\tau_{\rm es}^n(t)$ in the limit $t \gg t_{\rm sd}$ as $t_{\rm fo} > t_{\rm sd}$ for our relevant range of parameters. Setting Equation \ref{eq:tau_es(t_fo)} equal to Equation 14 in \cite{MetzgerPiro14}\footnote{With normalization revised for self-consistency with our spin-down time-scale definition of Equation \ref{eq:B_*}} we find:
\begin{equation}
    t_{\rm fo} = 760\,\rm d \bigg(\frac{\beta}{1.67\times10^{-3}}\bigg)^{-1}\bigg(\frac{B_*}{5\times10^{14} \rm G}\bigg)^{-2/3},
\end{equation}
where we normalize to our average value of $B_*$ and a nebula expansion velocity of $v = 500$\,km/s as inferred from observations.
The total number of pairs at freeze out is then: 
\begin{equation}
    N_\pm^{\rm fo} = n_\pm^{\rm fo} V^{\rm fo} = \frac{\tau_{\rm es}^n (t_{\rm fo})}{\sigma_T R_{\rm fo}} \frac{4\pi R_{\rm fo}^3}{3},
\end{equation}
Using $R_{\rm fo} = \beta c t_{\rm fo}$, leads to:
\begin{equation}
    N_\pm^{\rm fo} \approx 6\times10^{53} \bigg(\frac{\beta}{1.67\times10^{-3}}\bigg)\bigg(\frac{B_*}{5\times10^{14} \rm \,G}\bigg)^{-4/3},
\end{equation}
which is comparable  with the observational constraint of Equation  \ref{eq:N_e_obs}. While we acknowledge that residual $e^{+}-e^{-}$ annihilation might still play a role after formal freeze out, the early pair cascade thus provides a plausible source of radiating leptons in the PWN scenario.

We conclude that  the late-time radio observations are consistent with most of the emitting electrons originating from pair creation in the first $\approx2$\,years post-SN-explosion. In this scenario, the observed decline in radio luminosity is accounted for by the adiabatic expansion of the nebula, rather than mirroring the spin-down luminosity of the pulsar. We note that this is a simple treatment, that only explains the total number of emitting leptons and not the lepton energies needed to produce the observed emission.

\section{Conclusion}
\label{Sec:Conclusion}
In this work, we have presented the broadband evolution of \sn{} across 13 years of evolution. While the early radio and X-ray emission of \sn{} at $\delta t<190$\,d are consistent with a standard model of a shock propagating through a wind-like 
 density medium, the late-time evolution  ($\delta t \ge 6.7$\,yr) markedly deviates from this model. Our monitoring campaign reveals the onset of a major  radio re-brightening that dominates at $\nu\gtrsim 5$\,GHz. This luminous emission component is observationally characterized by a broad and rapidly evolving spectral peak, with a shallow optically thin slope ($F_\nu \propto \nu^{-0.31\pm0.02}$) and an optically thick spectral slope that is shallower than SSA from a single population of electrons. 
 Modeling the broad-band X-ray to radio spectrum we infer the following properties: 
\begin{itemize}
    \item The late-time mm-to-radio data are consistent with a two-component synchrotron emission model, with the lower frequency component being roughly consistent with extrapolations of the emission expected from the initially observed shock.
    \item The component responsible for the higher frequency emission must be subdominant to the initial shock emission at $\delta t<190$\,d. 
    \item The higher frequency component is associated with a small emitting volume (with an isotropic equivalent radius $R\approx 10^{16}$\,cm), low average expansion velocities ($v\approx500$\,km/s), and high number densities of electrons ($n_e \geq 10^4 \rm\,cm^{-3}$). 
    \item The higher frequency component is characterized by a hard electron power law index of $p \approx 1.6$.
    \item No high-energy emission is detected at any epoch. The late-time soft and hard X-ray emission is either photoelectrically absorbed or intrinsically faint.    
\end{itemize}
We interpret these findings in the context of two families of models: shock-CSM interaction with an asymmetric medium, and the emergence of radiation from a PWN. Through our analysis of these models, we have found that: 
\begin{itemize}
    \item A ``torus-like'' CSM interaction geometry is able to reproduce many of the observational features of the late-time SEDs; however, the parameters required are extreme (e.g., a large CSM mass of $10\,\rm M_\odot$, within a radius of $\sim10^{16}$\,cm with half-opening angle of $\theta < 5^{\circ}$) and lack viable stellar evolutionary motivation. 
    \item Models invoking the shock interaction with an isolated small clump of CSM at a large radius can account for the delayed onset, but fail to reproduce some key observational features of the late-time SEDs, like the observed spectral cooling breaks. While the parameters in this model are not unphysical, the astrophysical setup is highly contrived. 
    \item The late-time broadband SEDs of \sn{} bear many similarities to those of evolved PWNe. In particular, our observed optically thin slope is consistent with the flat radio spectral indices of PWNe and inconsistent with standard predictions of diffusive shock acceleration in an explosive event. Additionally, we find that the high-energy limits are naturally accounted for by photo-electric absorption within the expanding, oxygen-dominated  ejecta. By contrast, the ejecta are transparent to radio photons  because of the significantly lower free-free optical depth, allowing the emergence of the new radio component \citep{Murase2016}. Within this model we find that the radio evolution is consistent with the adiabatic expansion of a relic saturated electron-positron pair-cascade plasma injected by a central engine posited by \cite{MetzgerPiro14}, providing new insights into young pulsar physics.
     \footnote{We note that numerical modeling applied to the radio data can reproduce the high frequency emission, while discrepancies at lower frequencies can be explained by inhomogeneities, ejecta clumping, and contamination from the separate CSM-interaction component.} 
\end{itemize}

Based on this evidence, we conclude that a PWN represents the most likely explanation of the remarkable radio re-brightening of \sn{}.
While this conclusion is based purely on the data presented in this paper,  this model is further supported by the late-time optical/NIR data presented in \cite{Milisavljevic2018} and \cite{Pandey21}, the VLBI data of \cite{Lazda_VLBI}, the spectropolarimetric observations of \cite{DeSoto_2025}, and the modeling efforts of \cite{Pandey21}, \cite{Omand2023}, and \cite{Dessart2024}. Taken together these studies establish \sn{} as the most compelling candidate for an unprecedentedly young extragalactic PWN. As described in \cite{Lazda_VLBI}, planned future VLBI observations will allow us to conclusively rule out CSM interaction scenarios and directly probe the physical properties of an emerging PWN. 


\section*{Acknowledgments}
E.~W. would like to thank his PhD qualifying exam committee: profs. Wenbin Lu, Daniel Kasen, and Joshua Bloom for helpful comments and conversations that directed impacted the content of this paper.
The National Radio Astronomy Observatory (NRAO) is a facility of the National Science Foundation operated under cooperative agreement by Associated Universities, Inc. We thank the NRAO for carrying out the Karl G. Jansky Very Large Array (VLA).
This paper makes use of the following ALMA data: 2019.A.00006.T. ALMA is a partnership of ESO (representing its member states), NSF (USA) and NINS (Japan), together with NRC (Canada), NSTC and ASIAA (Taiwan), and KASI (Republic of Korea), in cooperation with the Republic of Chile. The Joint ALMA Observatory is operated by ESO, AUI/NRAO and NAOJ.
We thank the staff of the GMRT that made these observations possible. The GMRT is run by the National Centre for Radio Astrophysics of the Tata Institute of Fundamental Research.
The Submillimeter Array is a joint project between the Smithsonian Astrophysical Observatory and the Academia Sinica Institute of Astronomy and Astrophysics and is funded by the Smithsonian Institution and the Academia Sinica. We recognize that Maunakea is a culturally important site for the indigenous Hawaiian people; we are privileged to study the cosmos from its summit.
This work is based on observations carried out under project number D20AE with the IRAM NOEMA Interferometer. IRAM is supported by INSU/CNRS (France), MPG (Germany) and IGN (Spain).
This research has made use of the NuSTAR Data Analysis Software (NuSTARDAS) jointly developed by the ASI Space Science Data Center (SSDC, Italy) and the California Institute of Technology (Caltech, USA).
This paper employs a list of Chandra datasets, obtained by the Chandra X-ray Observatory, contained in the Chandra Data Collection (CDC) 607~\dataset[doi:10.25574/cdc.607]{https://doi.org/10.25574/cdc.607}
Some of the data presented herein were obtained at Keck Observatory, which is a private 501(c)3 non-profit organization operated as a scientific partnership among the California Institute of Technology, the University of California, and the National Aeronautics and Space Administration. The Observatory was made possible by the generous financial support of the W. M. Keck Foundation. The authors wish to recognize and acknowledge the very significant cultural role and reverence that the summit of Maunakea has always had within the Native Hawaiian community. We are most fortunate to have the opportunity to conduct observations from this mountain.
 
R.~M. acknowledges partial support from the National Science Foundation (grant number AST-2221789 and AST-2224255), by NASA through  80NSSC22K0991,  80NSSC22K1292, and by SAO through GO3-24048X. 
D.M. acknowledges support from the National Science Foundation through grants PHY-2209451 and AST-2206532.
FDC acknowledges support from the DGAPA/PAPIIT grant
 IN113424.
M.~L. acknowledges the support of the Natural Sciences and Engineering Research Council of Canada (NSERC-CGSD).
L.~R. acknowledges funding from the  Trottier Space Institute Fellowship, the Natural Sciences and Engineering Research Council of Canada (NSERC) Arthur B. McDonald Fellowship and Discovery Grant programs, the Canada Research Chairs (CRC) program, the Fondes de Recherche Nature et Technologies (FRQNT), the Centre de recherche en astrophysique du Québec (un regroupement stratégique du FRQNT), and the AstroFlash research group. 

\facilities{VLA, GMRT, ALMA, SMA, NOEMA, Chandra, NuSTAR, Keck:I(LRIS)}
\software{
astropy \citep{Astropy}, 
\texttt{emcee} \citep{Foreman_Mackey_2013},
CASA \citep{McMullin07},
AIPS \citep{AIPS},
IRAF \citep{Tody1986IRAF},
SNOoPY \citep{SNOoPY},
DAOPHOT \citep{Stetson1987}
NuSTARDAS \citep{Harrison_2013}
CIAO \citep{CIAO}
}

\bibliography{BibliographyFile}{}
\bibliographystyle{aasjournal}

\appendix

\section{Early-Time Fit Corner Plot}
Here we present the corner plot for the early-time radio data fit described in Section \ref{Subsub:12auearlytime}.
\label{app:early_corner}
\begin{figure}
	\centering 
    \includegraphics[width=0.9\columnwidth]{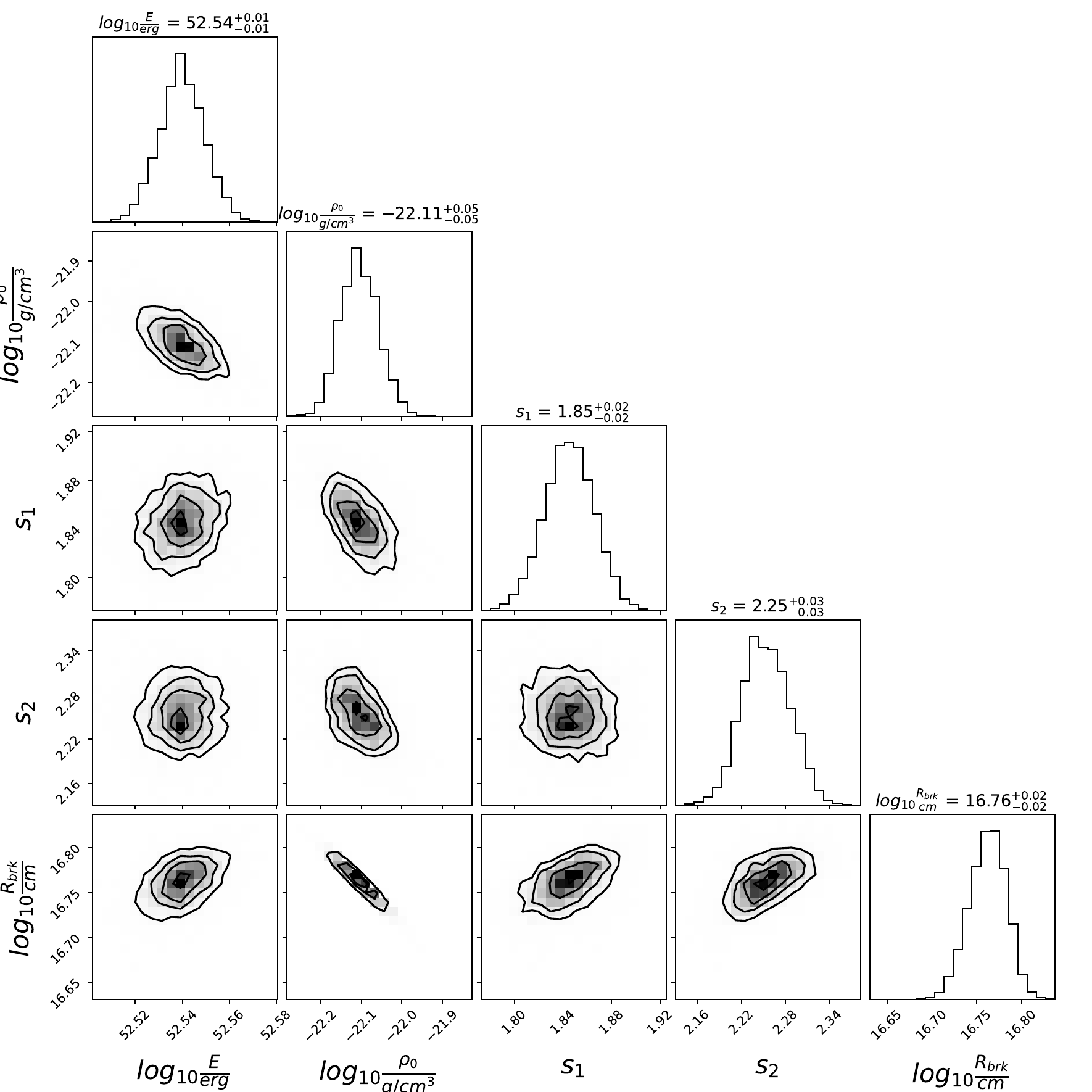}
    \caption{Corner plots for the posteriors of the fit to the early time data $\delta t\le 190$\,d described in \S\ref{Subsub:12auearlytime}. The data are well modeled by synchrotron emission originating from a blast wave propagating into a broken power-law density CSM, with a  change in density indexes from $s_1=1.85\pm{0.02}$, to $s_2=2.25\pm{0.03}$. Chains have converged with all parameters' ESS $>300$ and $\hat R \approx 1$.} 
    \label{fig:early_shock_fit}
\end{figure}

\section{Model-Agnostic Fit Corner Plot}
Here we present the corner plot for the model-agnostic 2-component fit described in Section \ref{tab:agnostic_fit}.

\begin{figure}
\label{fig:agnostic_corner}
\centering 
   \includegraphics[width=\columnwidth]{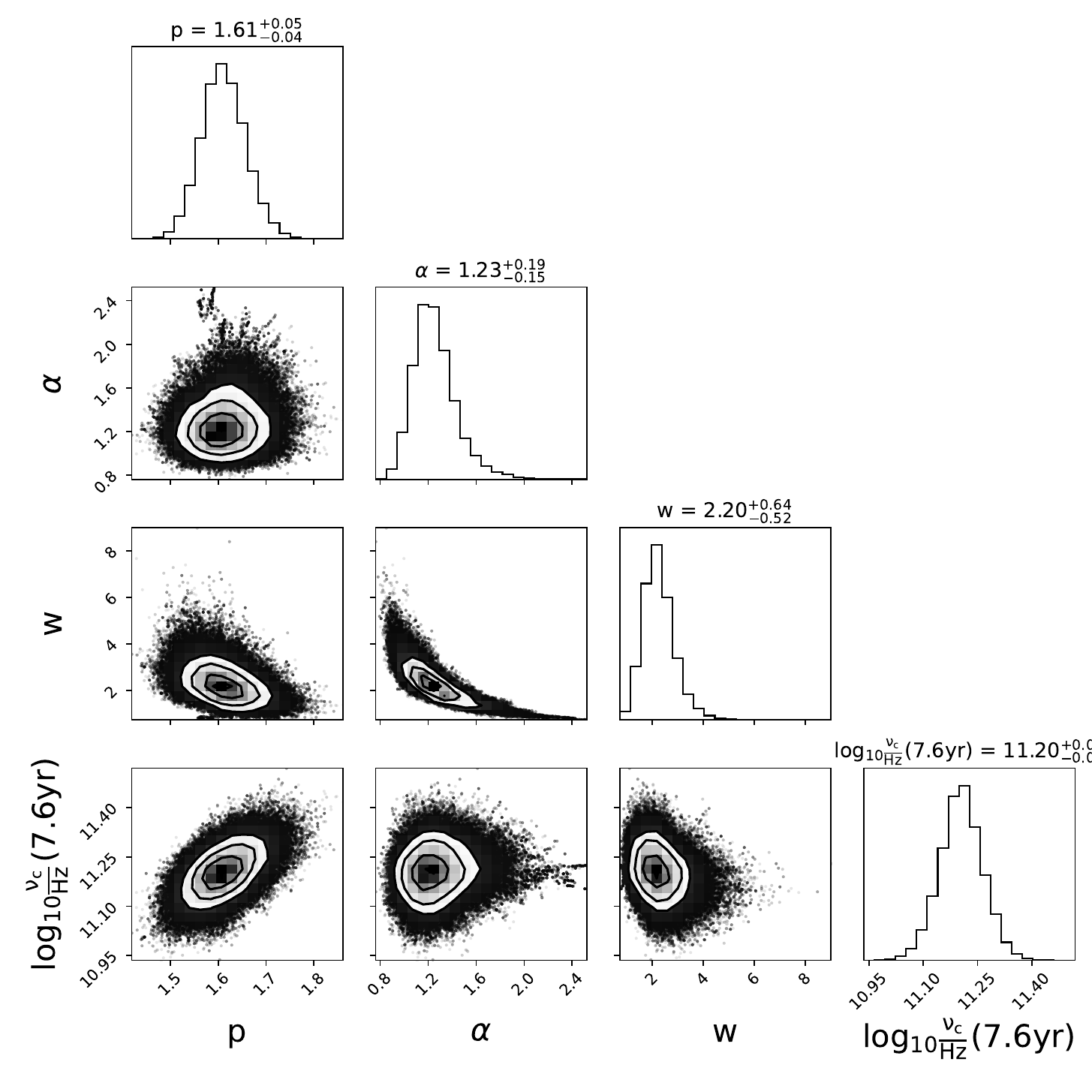}
    \caption{Corner plot for the posteriors of key parameters in the model agnostic fit of the late-time data shown in Figure \ref{fig:model_agnostic_fit} and described in \S\ref{SubSec:modelagnostic}. Additional fitted parameters include the $F_{\rm {brk}}$ and $\nu_{\rm{brk}}$, shown in Figure \ref{fig:model_agnostic_F_nu}, and the normalization on the optically thin, low frequency component. Chains have converged with all parameters' ESS $>600$ and $\hat R \approx 1$}
\end{figure}

\section{Radio and X-ray Data}
The error bars on the radio flux density measurements reported are the image rms. At the time of modeling the radio data, we add 5\% systematic flux calibration uncertainty in quadrature to this reported rms. 

\begin{longtable}{ccrrcccc}
    \caption{Radio observations of \sn{}.} \label{Tab:radio} \\
    \hline
    \hline
    Start Date$^a$ & 
    Time$^b$ & 
    Frequency & 
    Flux Density$^c$ & 
    Telescope & 
    Array Config. & 
    Project Code & 
    PI \\
    (UT) & (days) & (GHz) & (mJy) & & & & \\
    \hline
    \endfirsthead 
    
    \multicolumn{8}{c}{{\bfseries \tablename\ \thetable{} -- Continued from previous page}} \\
    \hline
    Start Date & Time & Frequency & Flux Density & Telescope & Array Config. & Project Code & PI \\
    (UT) & (days) & (GHz) & (mJy) & & & & \\
    \hline
    \endhead 
    
    \hline
    \multicolumn{8}{r}{{Continued on next page...}} \\
    \endfoot 
    
    \hline
    \hline
    \endlastfoot 
9/9/2012 & 190.0 & 1.51 & 4$\pm$0.20 & VLA & BnA & 12A-458 & Soderberg\\
9/9/2012 & 190.0 & 3.0 & 1.8$\pm$0.10 & VLA & BnA & 12A-458 & Soderberg\\
9/9/2012 & 190.0 & 6.0 & 1.02$\pm$0.06 & VLA & BnA & 12A-458 & Soderberg\\
11/5/2018 & 2438.0 & 6.1 & 7.7$\pm$0.50 & VLA & D & 18A-119 & Margutti\\
9/12/2019 & 2749.0 & 1.5 & 3.2$\pm$0.20 & VLA & A & SC1068 & Margutti\\
9/12/2019 & 2749.0 & 3.0 & 5.2$\pm$0.30 & VLA & A & SC1068 & Margutti\\
9/12/2019 & 2749.0 & 6.1 & 9.3$\pm$0.60 & VLA & A & SC1068 & Margutti\\
9/28/2019 & 2765.0 & 10.0 & 12.6$\pm$0.60 & VLA & A & SC1068 & Margutti\\
9/28/2019 & 2765.0 & 15.0 & 13.5$\pm$0.70 & VLA & A & SC1068 & Margutti\\
9/28/2019 & 2765.0 & 22.0 & 13.8$\pm$0.70 & VLA & A & SC1068 & Margutti\\
10/12/2019 & 2779.0 & 33.06 & 13.2$\pm$0.70 & VLA & A & SC1068 & Margutti\\
10/29/2019 & 2796.0 & 102.5 & 9.3$\pm$0.50 & ALMA & C43-3 & 2019.A.00006.T & Coppejans\\
10/29/2019 & 2796.0 & 104.5 & 9.1$\pm$0.50 & ALMA & C43-3 & 2019.A.00006.T & Coppejans\\
10/29/2019 & 2796.0 & 336.52 & 4.2$\pm$0.30 & ALMA & C43-3 & 2019.A.00006.T & Coppejans\\
10/29/2019 & 2796.0 & 338.42 & 4.1$\pm$0.30 & ALMA & C43-3 & 2019.A.00006.T & Coppejans\\
10/29/2019 & 2796.0 & 348.52 & 4.3$\pm$0.30 & ALMA & C43-3 & 2019.A.00006.T & Coppejans\\
10/29/2019 & 2796.0 & 350.48 & 4.1$\pm$0.30 & ALMA & C43-3 & 2019.A.00006.T & Coppejans\\
10/29/2019 & 2796.0 & 90.5 & 9.8$\pm$0.50 & ALMA & C43-3 & 2019.A.00006.T & Coppejans\\
10/29/2019 & 2796.0 & 92.5 & 9.9$\pm$0.50 & ALMA & C43-3 & 2019.A.00006.T & Coppejans\\
12/19/2019 & 2847.0 & 0.38 & 2.2$\pm$0.30 & GMRT & ---& ddtC104 & Coppejans\\
12/24/2019 & 2852.0 & 0.7 & 2.43$\pm$0.20 & GMRT  & ---& ddtC104 & Coppejans\\
12/29/2019 & 2857.0 & 0.19 & $\leq$ 2.3$\pm$0.00 & GMRT & ---& ddtC104 & Coppejans\\
12/29/2019 & 2857.0 & 1.25 & 2.79$\pm$0.20 & GMRT  & ---& ddtC104 & Coppejans\\
7/1/2020 & 3042.0 & 1.5 & 3.2$\pm$0.20 & VLA & B & SC1074 & Margutti\\
7/1/2020 & 3042.0 & 3.0 & 5.4$\pm$0.30 & VLA & B & SC1074 & Margutti\\
7/1/2020 & 3042.0 & 6.05 & 9.2$\pm$0.50 & VLA & B & SC1074 & Margutti\\
7/11/2020 & 3052.0 & 0.19 & $\leq$ 2.2$\pm$0.00 & GMRT & ---& ddtC104 & Coppejans\\
7/11/2020 & 3052.0 & 11.0 & 11.8$\pm$0.60 & VLA & B & SC1074 & Margutti\\
7/11/2020 & 3052.0 & 18.87 & 10.6$\pm$0.50 & VLA & B & SC1074 & Margutti\\
7/11/2020 & 3052.0 & 21.13 & 10.6$\pm$0.50 & VLA & B & SC1074 & Margutti\\
7/11/2020 & 3052.0 & 23.0 & 11.5$\pm$0.60 & VLA & B & SC1074 & Margutti\\
7/11/2020 & 3052.0 & 25.0 & 9.9$\pm$0.50 & VLA & B & SC1074 & Margutti\\
7/11/2020 & 3052.0 & 9.0 & 10.9$\pm$0.50 & VLA & B & SC1074 & Margutti\\
7/13/2020 & 3054.0 & 225.0 & 4.8$\pm$0.30 & SMA &  & POETS & Berger\\
9/5/2020 & 3108.0 & 30.0 & 9.3$\pm$0.50 & VLA & B & SC1079 & Margutti\\
9/5/2020 & 3108.0 & 32.0 & 8.8$\pm$0.50 & VLA & B & SC1079 & Margutti\\
9/5/2020 & 3108.0 & 34.0 & 9$\pm$0.50 & VLA & B & SC1079 & Margutti\\
9/5/2020 & 3108.0 & 36.0 & 9$\pm$0.50 & VLA & B & SC1079 & Margutti\\
11/17/2020 & 3181.0 & 88.26 & 6.39$\pm$0.60 & NOEMA & --- & ID D20AE & \\
11/17/2020 & 3181.0 & 100.0 & 6.11$\pm$0.60 & NOEMA & --- & ID D20AE & \\
10/31/2023 & 4259.0 & 3.0 & 3.29$\pm$0.44 & VLA & D & SF151070 & Margutti\\
10/31/2023 & 4259.0 & 5.0 & 6.12$\pm$0.17 & VLA & D & SF151070 & Margutti\\
10/31/2023 & 4259.0 & 7.0 & 6.85$\pm$0.10 & VLA & D & SF151070 & Margutti\\
10/31/2023 & 4259.0 & 9.0 & 7.15$\pm$0.10 & VLA & D & SF151070 & Margutti\\
10/31/2023 & 4259.0 & 11.0 & 7.04$\pm$0.08 & VLA & D & SF151070 & Margutti\\
11/16/2023 & 4275.0 & 225.5 & 2.9$\pm$0.22 & SMA & --- & POETS & Berger\\
12/15/2023 & 4304.0 & 0.65 & 2.29$\pm$0.24 & GMRT & ---& ddtC315 & Nayana AJ\\
12/19/2023 & 4308.0 & 13.0 & 6.42$\pm$0.19 & VLA & D & 23B-330 & Nayana AJ\\
12/19/2023 & 4308.0 & 15.0 & 6.29$\pm$0.15 & VLA & D & 23B-330 & Nayana AJ\\
12/19/2023 & 4308.0 & 17.0 & 6.14$\pm$0.11 & VLA & D & 23B-330 & Nayana AJ\\
12/21/2023 & 4310.0 & 1.37 & 2.85$\pm$0.28 & GMRT & ---& ddtC315 & Nayana AJ\\
12/27/2023 & 4316.0 & 0.44 & 2.6$\pm$0.26 & GMRT & ---& ddtC315 & Nayana AJ\\
12/27/2023 & 4316.0 & 19.0 & 6.02$\pm$0.12 & VLA & D & 23B-330 & Nayana AJ\\
12/27/2023 & 4316.0 & 21.0 & 5.67$\pm$0.12 & VLA & D & 23B-330 & Nayana AJ\\
12/27/2023 & 4316.0 & 23.0 & 5.55$\pm$0.13 & VLA & D & 23B-330 & Nayana AJ\\
12/27/2023 & 4316.0 & 25.0 & 5.51$\pm$0.14 & VLA & D & 23B-330 & Nayana AJ\\
10/27/2024 & 4621.5 & 3.0 & 4.421$\pm$0.03 & VLA & A & 24B-311 & Wiston\\
10/27/2024 & 4621.5 & 1.52 & 2.924$\pm$0.05 & VLA & A & 24B-311 & Wiston\\
10/27/2024 & 4621.5 & 9.0 & 5.807$\pm$0.08 & VLA & A & 24B-311 & Wiston\\
10/27/2024 & 4621.5 & 5.55 & 5.316$\pm$0.04 & VLA & A & 24B-311 & Wiston\\
10/27/2024 & 4621.5 & 7.0 & 5.514$\pm$0.05 & VLA & A & 24B-311 & Wiston\\
10/27/2024 & 4621.5 & 11.0 & 5.489$\pm$0.09 & VLA & A & 24B-311 & Wiston\\
10/27/2024 & 4621.5 & 19.0 & 4.9$\pm$0.08 & VLA & A & 24B-311 & Wiston\\
10/27/2024 & 4621.5 & 25.0 & 4.16$\pm$0.12 & VLA & A & 24B-311 & Wiston\\
10/27/2024 & 4621.5 & 21.0 & 4.72$\pm$0.14 & VLA & A & 24B-311 & Wiston\\
10/27/2024 & 4621.5 & 23.0 & 4.44$\pm$0.13 & VLA & A & 24B-311 & Wiston\\
10/27/2024 & 4621.5 & 1.26 & 2.444$\pm$0.02 & VLA & A & 24B-311 & Wiston\\
10/27/2024 & 4621.5 & 1.775 & 3.153$\pm$0.07 & VLA & A & 24B-311 & Wiston\\
11/17/2024 & 4642.5 & 0.4 & 4.45$\pm$0.19 & GMRT & ---& 47\_021 & Wiston\\
11/18/2024 & 4643.5 & 0.65 & 2.29$\pm$0.20 & GMRT & ---& 47\_021 & Wiston\\
2/13/2025 & 4730.5 & 220.0 & 2.2$\pm$0.22 & SMA & --- & POETS & Berger\\

\end{longtable}
\vspace{-0.2cm} 
\begin{minipage}{\linewidth}
    \small 
    \noindent \textsc{Note}---$^a$ Dates here are expressed in the standard American format MM/DD/YY. \par
    \noindent $^b$ Since explosion (3 March 2012). \par
    \noindent $^c$ Uncertainties are quoted at 1$\sigma$, and upper-limits are quoted at $3\sigma$.
\end{minipage}

\begin{deluxetable*}{ccccccccc}[ht]
\tablecaption{CXO and NuSTAR X-ray observations of SN\,2012au, with inferred count-rate and flux limits. }
\tablehead{
\colhead{Obs ID} & \colhead{$\delta t$} & \colhead{Count-Rate$^a$} & \colhead{Photon Index}&
\colhead{Unabsorbed Flux$^b$} & \colhead{Exposure$^{c}$} &
\colhead{Spacecraft} &
\colhead{PI}   \\
\colhead{} & \colhead{$\rm \left[ d \right]$} &\colhead{$10^{-4}\,[\rm{c\,s^{-1}}$]} &  &\colhead{$10^{-15}[\rm{erg\,s^{-1}\,cm^{-2}}$]} & \colhead{[ks]}}
\startdata 
\hline
21160&2343.2& $<1.5$ &  2$^d$ &$<2.3$  &19.8&  CXO & D. Patnaude\\
23316&3060.3 & \rdelim\}{2}{*}[\,$<1.5$ ] &2$^d$  & \rdelim\}{2}{*}[\,$<2.7$ ]& 9.8 & CXO & R. Margutti\\
23331& 3060.9 & & 2$^d$ & & 10.1 & CXO & R. Margutti\\
25187& 3676.7 & $<1.6$ &2$^d$  & $<3.1$ & 19.3 & CXO & M. Stroh\\
90601521002 & 3034.9 & $<4.1$&2$^d$ & $<24.5$& 23.8/23.6& NuSTAR & R. Margutti\\
\hline
\enddata
\tablecomments{$^a$ 3$\sigma$ limit; Background-subtracted; 0.5--8 keV for CXO, 8--20 keV for NuSTAR, both modules.  \\ $^b$  CXO: 0.3--10 keV and  $\rm{NH_{\rm{int}}}=0\,\rm{cm^{-2}}$. NuSTAR: 8--20 keV.  \\$^{c}$ For NuSTAR we report the exposure time of the A module and the B module.  \\$^{d}$ Assumed.
\label{Tab:Xraydata}}
\end{deluxetable*}

\section{Synchrotron Constants}
\label{app:constants}
Here we report the constants defined in \cite{Pacholczyk1970} that are used to compute the synchrotron emissivity and self-absorption. These constants are used in our treatment of synchrotron radiation and enter in our Equations \ref{eq:F_brk} and \ref{eq:nu_brk}. All constants are in c.g.s. units.

\begin{equation}
    c_1 =  6.27\times10^{18}
\end{equation}
\begin{equation}
    c_5 = \frac{\sqrt{3}e^3}{16\pi m_e c^2} \frac{p+7/3}{p+1}\Gamma \Big(\frac{3p+2}{12}\Big)\Gamma\Big(\frac{3p+7}{12}\Big)
\end{equation}
\begin{equation}
    c_6 = \frac{\sqrt{3}\pi}{72} em_e^5c^{10} (p+10/3) \Gamma\Big(\frac{3p+2}{12}\Big)\Gamma\Big(\frac{3p+10}{12}\Big)
\end{equation}

\section{CSM Interaction Plots}
Here we present the plots relevant to the CSM interaction models for late-time emission described in Section \ref{Sec:CSMinteraction}. 
\label{app:CSM_interaction}
\begin{figure*}[ht]
\centering
\includegraphics[width=1.8\columnwidth]{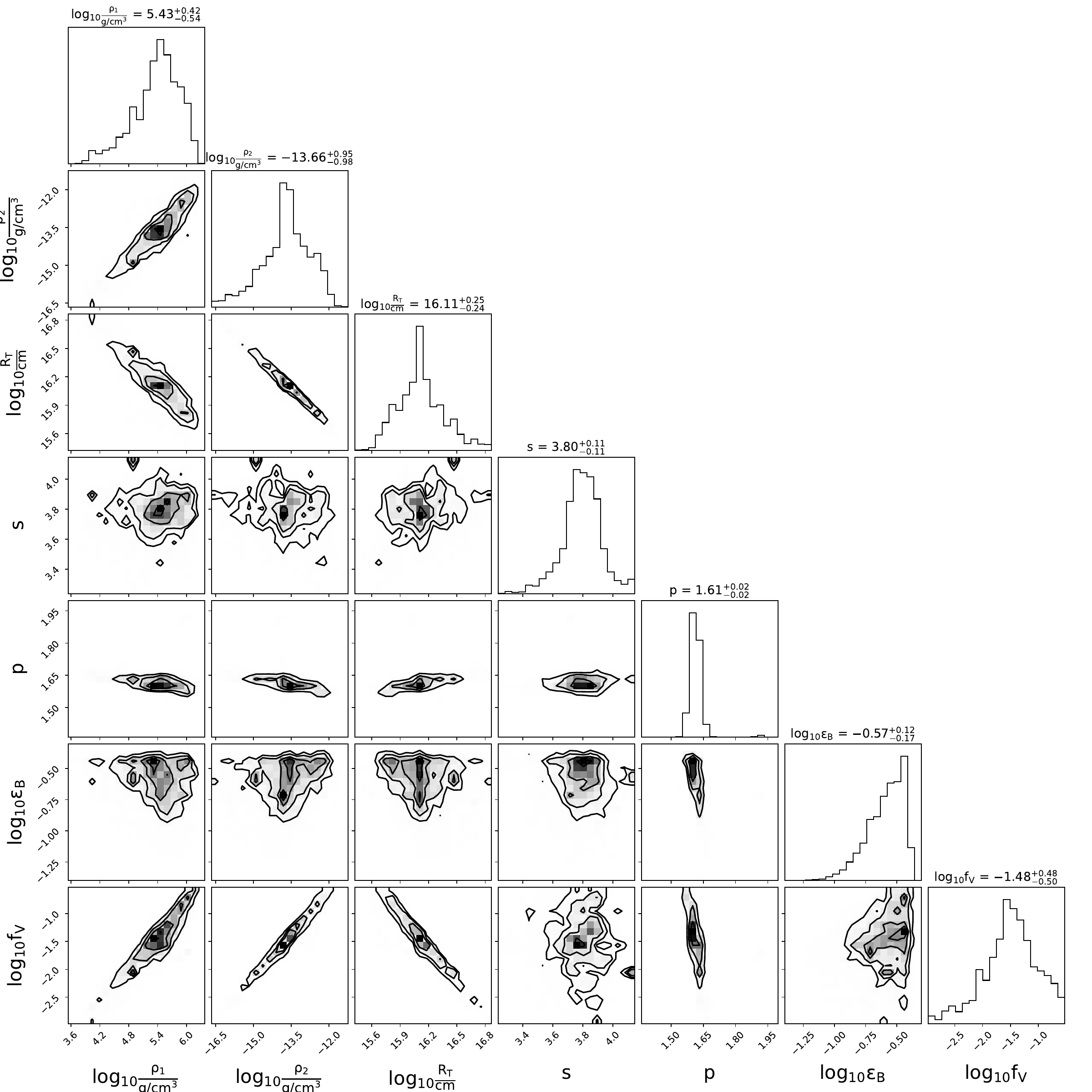}

\caption{Corner plot for the edge on torus-like dense CSM model in Section \ref{subsub:edge_torus}. The region at $R < R_T$ is characterized by extremely high densities ($\rho_1 \approx 10^5 \rm{g/cm}^3$), needed to significantly decelerate the shock. At $R = R_T$, there is a large drop in density to $\rho_2 = 10^{-12.56} \rm{g/cm}^3$ that declines in radius with a steeper than wind density profile: $\rho_{\rm CSM}(R) \propto R^{-3.29}$. All chains have converged with all parameters' ESS $>10^5$ and $\hat R \approx 1$.}
\label{fig:torus_corners}
\end{figure*}

\begin{figure*}[ht]
\centering
\includegraphics[width=1.8\columnwidth]{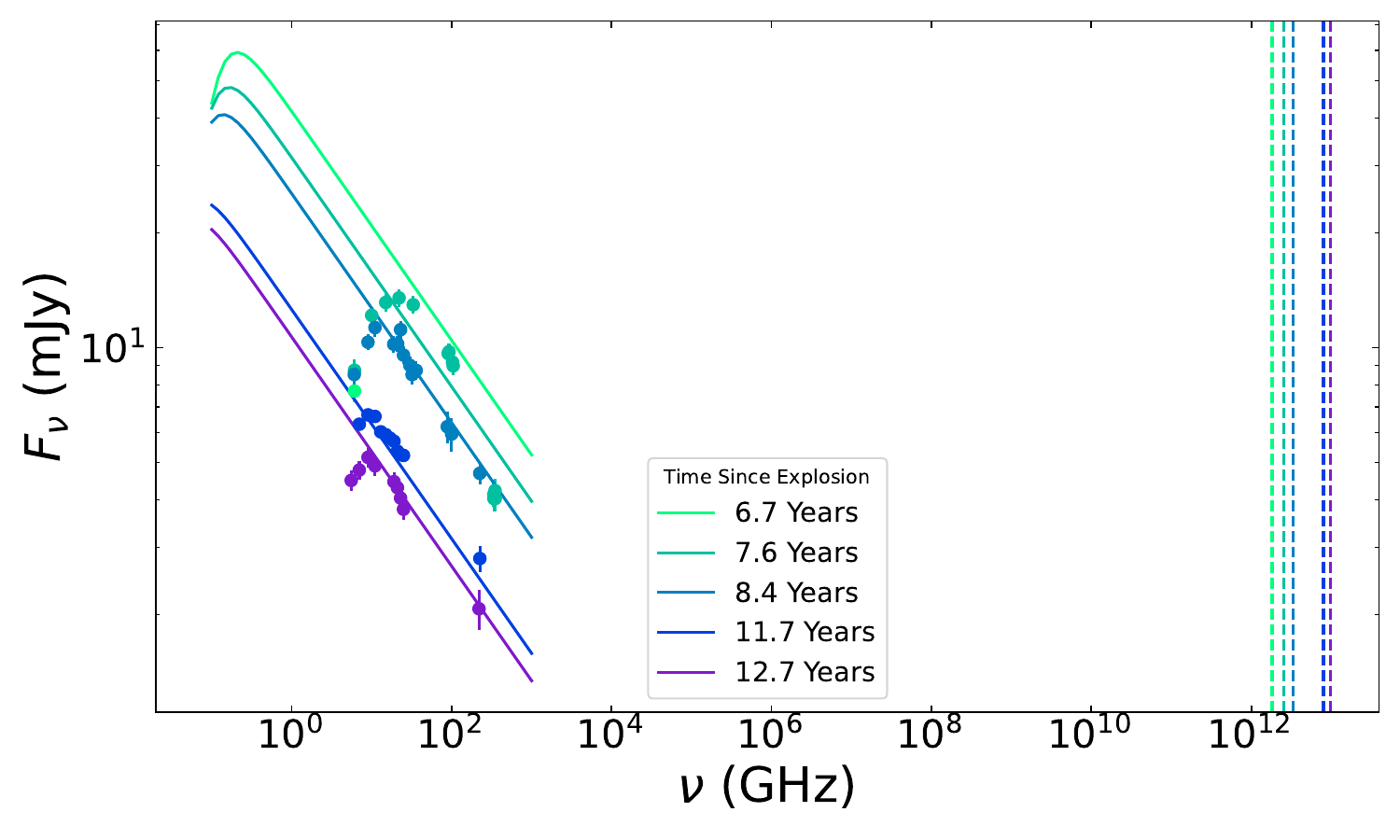}

\caption{The best model fit for the dense torus viewed from a face-on perspective, described in Section \ref{subsub:face_on_torus}. While the optically thin side of the observed SEDs is well modeled, the optically thick side is significantly overestimated by the model. In the face-on perspective, there is not significant free-free absorption, which would otherwise shift the observed peak to higher $\nu_{\rm pk}$ and lower $F_{\rm pk}$.}
\label{fig:torus_corners_face_on}
\end{figure*}

\begin{figure*}[ht]
\centering

\gridline{
  \fig{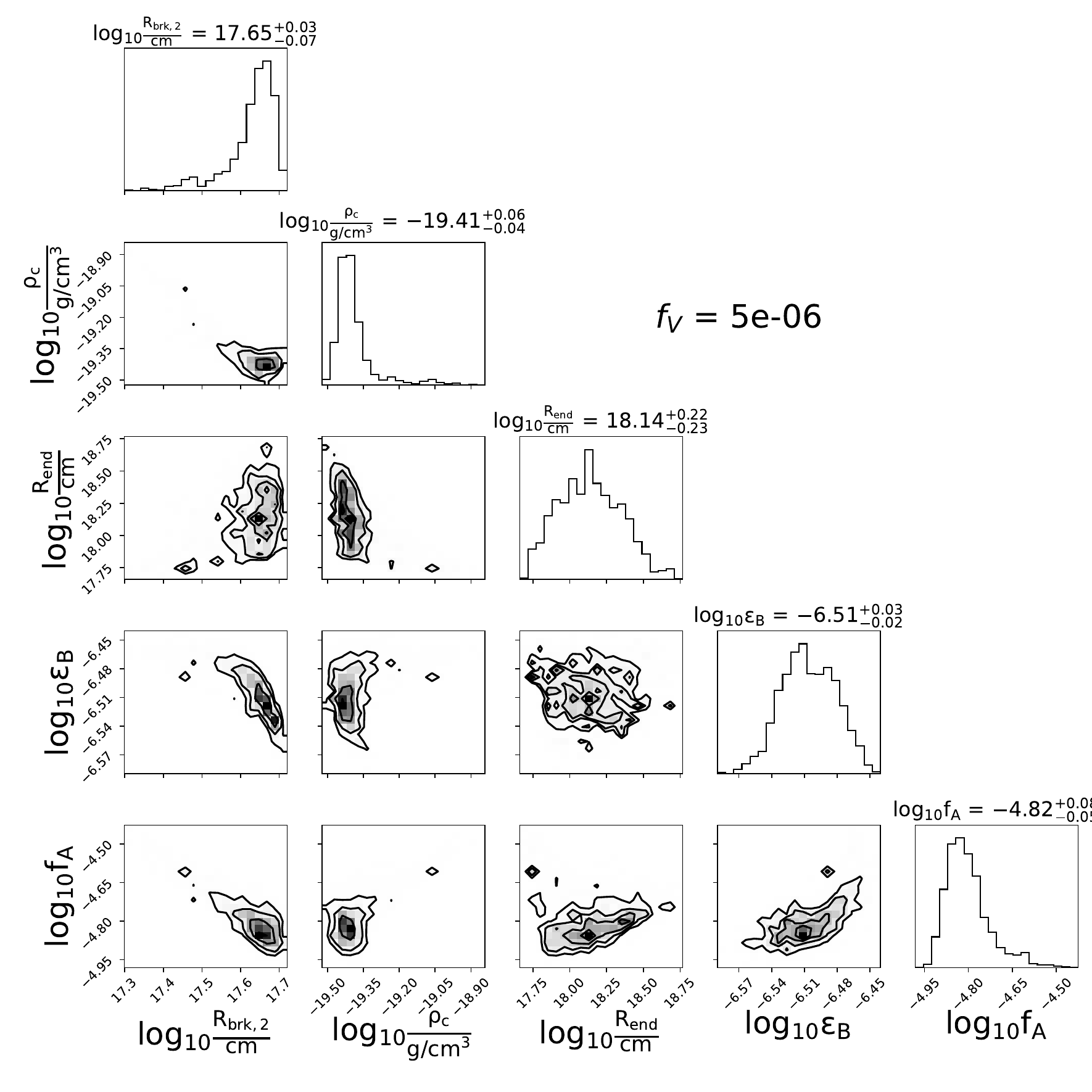}{0.31\textwidth}{(a)}
  \fig{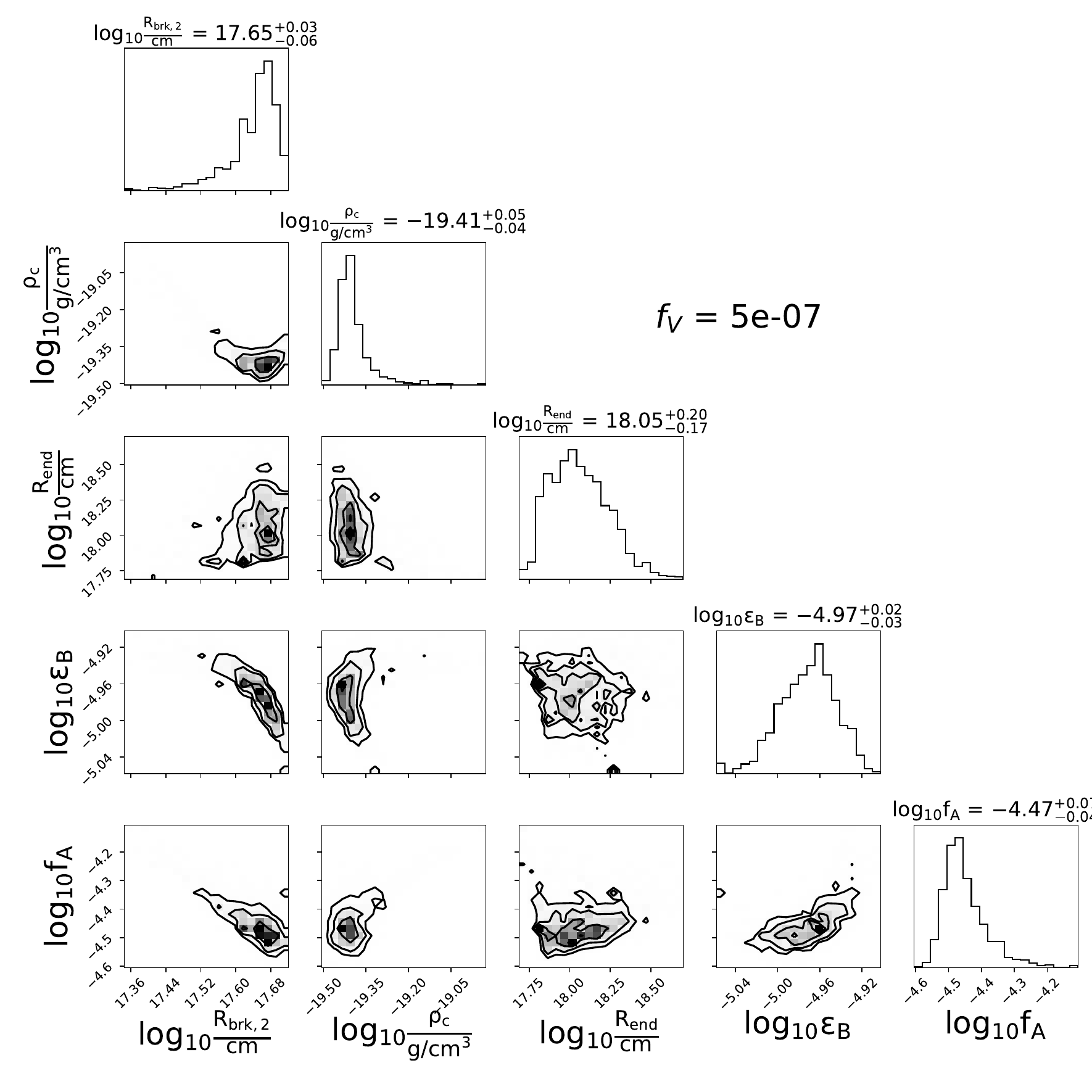}{0.31\textwidth}{(b)}
  \fig{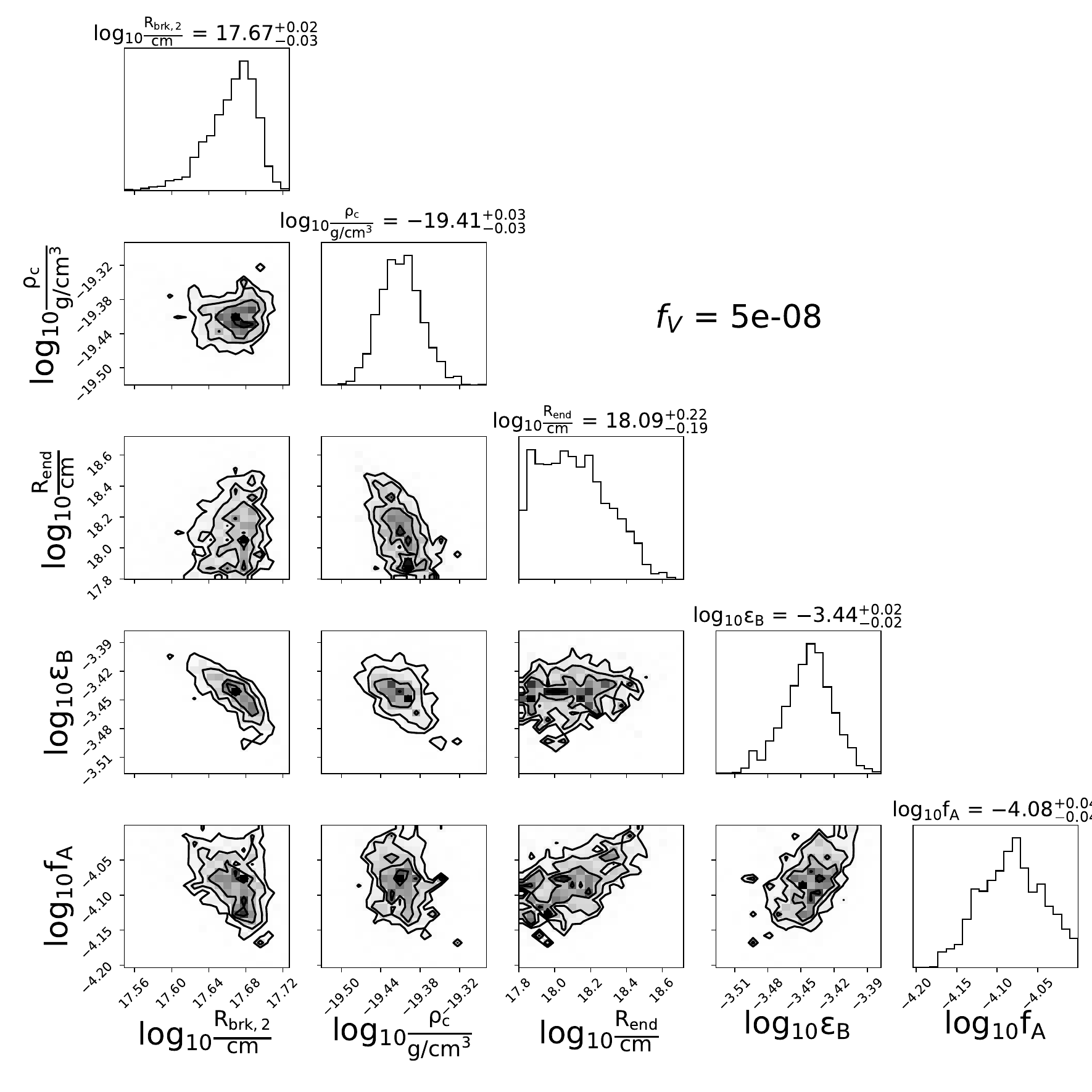}{0.31\textwidth}{(c)}
}

\caption{Corner plots for the small clump of dense CSM models models in Section \ref{SubSec:CSMsmallclump}. Plots (a) - (c) show the parameter space for $f_V=5\times10^{-6}$ through $f_V=5\times10^{-8}$ respectively. The factors related to the dynamics of the shock $R_{\rm brk,2}$, $\rho_c$, and $R_{\rm end}$ are all constant across different filling factors. The only variable that changes with considerably with $f_V$ is $\epsilon_B$, which increases with decreasing filling factor. This compensates for the lower emitting volume to maintain the same observed luminosity. These models consequently do not have the same $B$ field strength, like the torus models, because they are not matching the observed cooling break. All chains have converged with all parameters' ESS $>10^5$ and $\hat R \approx 1$.} 
\label{fig:clump_corners}
\end{figure*}

\end{document}